\documentclass[enabledeprecatedfontcommands,bibliography=totoc,listof=totoc,index=totoc,twoside=true,BCOR=12mm,DIV=12]{scrbook}

\usepackage[utf8]{inputenc}                       
\usepackage[bachelor]{mdsg}
\lmutitle{Konzeption und Umsetzung einer mobilen Applikation zur Validierung von fälschungssicheren Produktlabeln}                       
\lmustudentone{Oliver Linne}                    

\lmuprofone{Prof. Dr. Claudia Linnhoff-Popien}    
\lmuadvisorone{Markus Friedrich, M.Sc.}                    
\lmuadvisortwo{Dr. Sebastian Feld}                    
\lmuadvisorthree{Prof. Dr. Dr. Ulrich Rührmair}                  
\lmudeadline{8. Juni 2021}                      
\usepackage{cmap}                                 
\usepackage[T1]{fontenc}                          
\usepackage{lmodern}
\usepackage[english,ngerman]{babel}               
\usepackage{bibgerm}                              
\usepackage{tabularx}                             
\usepackage{booktabs}                             
\usepackage{rotating}                             
\usepackage{multirow}                             
\usepackage{amssymb,amsmath}                      
\usepackage[hyphens]{url}
\usepackage[breaklinks]{hyperref} 
\lmuhypersetup                                    
\usepackage{flafter}                              
\usepackage{pdflscape}                            
\usepackage{hyphenat}                             
\usepackage[all]{hypcap}                          
\usepackage{enumitem}                             
\usepackage{caption}
\usepackage{amsthm}
\usepackage{nicefrac}
\usepackage[german,ruled,linesnumbered]{algorithm2e}
\usepackage{physics}
\usepackage{graphicx}
\usepackage{cleveref}

\usepackage{subcaption}

\graphicspath{{./text/pictures/}}                      

\newtheorem{definition}{Definition}

\SetKwInput{Zurueck}{Rückgabe}
\SetKwFor{FuerAlle}{für alle}{tue parallel}{Ende}
\SetKwInput{usedAlgorithm}{\textbf{Verwendeter Algorithmus}}

\begin{document}
    \setlist{noitemsep}                           
    \lmufront                                     
    \newpage
    \cleardoublepage
    \cleardoubleemptypage
    \thispagestyle{empty}
    \vspace*{2cm}

\begin{center}
    \textbf{Abstract}
\end{center}

\vspace*{1cm}

\noindent 

Auf Grund weltweit steigender Zahlen der Produktpiraterie soll ein kostengünstiges Verfahren zur Verifizierung der Herkunft eines Produktes entwickelt werden.
Dafür lässt sich durch exakt messbare, einzigartige, jedoch schwer rekonstruierbare Eigenschaften spezieller physischer Objekte ein Echtheitszertifikat kreieren.
Dieses ist im Kontext der vorliegenden Arbeit ein fälschungssicheres Label, das sich in einem semi-transparenten Material aus zufällig verteilten Goldnanokügelchen oder -stäbchen zusammensetzt.
Die charakteristischen Positionierungen der Elemente des Labels lassen sich mit der Kamera eines Smartphones und zusätzlichen Technologien präzise messen.
Dadurch kann für die breite Bevölkerung ohne die Notwendigkeit einer bestehenden Netzwerkverbindung ein offline verwendbares Verifikationsverfahren erschaffen werden.
Die vorliegende Arbeit liefert einen ersten Teil des Machbarkeitsnachweises, dass ein derartiges System und insbesondere das damit einhergehende algorithmische Berechnungsverfahren in einer mobilen Applikation implementier- und effizient einsetzbar ist.
Zudem wird je eine in der Praxis geeignete Methode zur Übermittlung und Sicherung der benötigten Informationen eruiert.
Des Weiteren werden die Resultate der Validierung von fälschungssicheren Produktlabeln ausführlich analysiert und vorhandene Schwächen aufgezeigt.                       
    \thispagestyle{empty}
    \frontmatter                                  
    \tableofcontents                              
    \mainmatter                                   
%
%
    \chapter{Einleitung}\label{chapter:Einleitung}
Seit März des Jahres 2020 ist das tägliche Leben vielerorts auf Grund der Corona-Pandemie von Einschränkungen und Verboten geprägt \cite{WHOPandemieAusruf}.
Allerdings ist die Lösung schon gefunden: Impfungen.
In Deutschland genießen vollständig gegen das Corona-Virus geimpfte Personen seit Mai 2021 wieder mehr Freiheiten und entkommen damit teilweise dem festen Griff des Pandemie-Lebens \cite{VerordnungBundesregierung}.
Den Nachweis, dass eine Person geimpft ist, gibt beispielsweise ein entsprechender Eintrag im Impfpass \cite{VerordnungBundesregierung}.
Dieser begehrte, weil zu mehr Freiheit führende, Eintrag besteht für gewöhnlich aus einem Aufkleber des Impfstoffherstellers, einem Stempel und einer Unterschrift der impfenden Einrichtung.
Augenscheinlich sind diese Informationen jedoch leicht zu fälschen.
So verwundert es nicht, dass Berichten zufolge der Handel mit nachgemachten Bescheinigungen über Corona-Impfungen floriert \cite{FAZFake,BR,ZDFFake,TagesschauFake,ZeitFake}.
Impfpässe stellen im Kontext der Corona-Pandemie allerdings nicht das einzige vermehrt manipulierte Produkt dar.
Laut Weltgesundheitsorganisation (WHO), Weltzollorganisation und Interpol war mit Beginn der Corona-Krise auch eine starke Zunahme z.B. gefälschter medizinischer Artikel wahrzunehmen \cite{BBC-WHO,WCO-Weltzollorganisation,InterpolRiseInFake}.
Hierbei ist der Betrug nicht nur bei Gesichtsmasken, Handschuhen und Handdesinfektionsmitteln, sondern vor allem auch bei Medikamenten und Impfstoffen zu beobachten \cite{InterpolRiseInFake,InterpolNetworkDismantled,InterpolPublicWarning}.
Die Verabreichung dieser kann eine Gefahr für die Gesundheit bedeuten, da es sich hier oft um Verunreinigungen, minderwertige oder sogar schlicht andere Stoffe als die angegebenen handelt \cite{MedicalProductQualityReport,InterpolRiseInFake,InterpolNetworkDismantled,InterpolPublicWarning}.
Auch wenn auf Grund der erhöhten Nachfrage nach bestimmten Gütern während der Corona-Krise vermehrte Täuschungsversuche zu verzeichnen sind, stellte der Handel mit falschen, ungeprüften medizinischen Produkten bereits vor Beginn der Corona-Pandemie eine große Gefahr für die Gesundheit und Teile der Wirtschaft dar.
Untersuchungen der WHO zufolge handelte es sich zwischen den Jahren 2007 und 2016 in Ländern mit niedrigem und mittlerem Einkommen (Länderklassifizierung der Weltbank nach Einkommensniveau \cite{WorldBankCountry}) bei über $10~\%$ der untersuchten Medikamente um minderwertige Qualität, nicht zugelassene Artikel oder gar Fälschungen \cite{WHOStudyImpact,WHOStudySurveillance}.

Aber auch wohlhabendere Länder, wie zum Beispiel Deutschland, haben über das Problem der Fälschung medizinischer Produkte hinaus mit unterschiedlichsten Formen der Warenfälschung zu kämpfen \cite{WHOStudySurveillance, OECDPharmaceuticalTrade,OECDAllgemeineTrends}.
Der Organisation für wirtschaftliche Zusammenarbeit und Entwicklung (OECD) zufolge machte der Handel mit imitierten Produkten im Jahr 2016 über $3~\%$ des Welthandelsvolumens aus.
Dies entspricht einem Wert von mehr als 500 Milliarden US-Dollar. Betrachtet man einzig die Importe der Europäischen Union, betrug der Anteil der Produktfälschungen sogar über $6~\%$, was ein Volumen von 134 Milliarden US-Dollar bedeutet – mit steigender Tendenz. \cite{OECDAllgemeineTrends}

Um diese Problematik einzudämmen, wäre ein kostengünstiges und vielseitig einsetzbares Verfahren zur Authentifizierung des Herstellers eines Produktes nützlich.
Einen Lösungsansatz zur Entwicklung einer derartigen Methode stellen sogenannte fälschungssichere Label dar.
Es handelt sich dabei um einzigartige und eindeutig identifizierbare physische Objekte. Das Besondere an diesen ist, dass ihre Eigenschaften zwar leicht zu messen sind, es allerdings unmöglich ist, sie exakt zu klonen.
Auf Basis dieser Label lässt sich ein Sicherheitsmechanismus konstruieren, mittels dessen der Hersteller eines Produktes verifiziert werden kann. \cite{Ruehrmair2019,Kirovski2010}
Dazu entwickelten Rührmair und Marin im Vorfeld dieser Arbeit einen konzeptionell neuen Ansatz fälschungssicherer Label für den Produktschutz.
Ein solches Label besteht aus Goldnanokügelchen beziehungsweise -stäbchen, welche in einem durchsichtigen Medium, zum Beispiel einem Plastikplättchen, zufällig verteilt sind.
Da die Anordnung und exakte Positionierung der Kügelchen respektive Stäbchen willkürlich erfolgt, ist jedes Label einzigartig. \cite{Ruehrmair2021,ZachMarin2021}

Mithilfe spezieller Technologien lässt sich die dreidimensionale Lage der in einem Label enthaltenen Objekte bis auf wenige Nanometer genau bestimmen und als Daten in einer 3D-Punktwolke extrahieren.
Dabei wird für auf Goldnanokügelchen basierende Label jedes dieser Kügelchen durch einen Punkt repräsentiert.
Bei Label auf Goldnanostäbchen-Basis werden Anfangs- und Endpunkt eines Stäbchens jeweils durch einen Punkt modelliert, so dass ein Stäbchen in der Punktwolke also durch zwei einzelne Punkte dargestellt wird. \cite{Ruehrmair2021,ZachMarin2021}

Verbindet man die digitalisierte Version, also die Punktwolke mit dem physischen Label zu einer logischen Einheit, so ist diese unzertrennbar.
Denn würde eine der beiden Komponenten verändert werden, würde bei einer Überprüfung konstatierbar sein, dass physische und digitalisierte Version des Labels nicht zueinander passen. \cite{Ruehrmair2021}

Dieser Umstand kann nun für den Produktschutz genutzt werden.
Zur eindeutigen Identifizierung bringt der Hersteller eines Produktes auf diesem ein fälschungssicheres Label an, vermisst dieses und fügt als Referenz die erhaltene digitalisierte Version in Kombination mit einer individuellen, nicht fälschbaren digitalen Signatur hinzu.
Die digitale Signatur wird mit Hilfe eines privaten, nur dem Hersteller bekannten Schlüssels in Abhängigkeit der weiteren Daten erzeugt.
Ein Anwender vollführt dann zur Authentifizierung der Herkunft des Produktes selbständig eine Messung des Labels und vergleicht die dadurch erhaltene Punktwolke mit der Referenz des Herstellers.
Stimmen beide – bei gleichzeitig korrekter Signatur - überein, ist der Nachweis für die Herkunft des Produktes erbracht.
Andernfalls wird die Ware bei Nonkonformität als Imitat entlarvt. \cite{Ruehrmair2021}

Die Validierung eines fälschungssicheren Labels und damit die Authentifizierung eines Produktes besteht im Kern folglich aus dem Vergleich zweier 3D-Punktwolken.
Jede Messung eines Labels unterliegt in der Praxis allerdings gewissen Ungenauigkeiten und Fehlern.
Je nach Betrachtungswinkel verändert sich beispielsweise die relative Lage der im Label enthaltenen Objekte.
Auch tauchen in der Punktwolke standardmäßig Punkte auf, die keines der in der Realität im Label enthaltenen Objekte repräsentieren.
Zudem gehen bei der Messung und damit verbundenen Erstellung der Punktwolke naturgemäß Punkte verloren.
Es werden also nicht alle Objekte des Labels als Teil der Punktwolke modelliert.
Des Weiteren kommt es auf Grund von Verrauschungen dazu, dass die Positionierung eines Punktes, der stets dasselbe Objekt repräsentiert, bei mehrmaligen Messungen variiert.
Jedoch befinden sich die unterschiedlich positionierten Punkte jeweils mit hoher Wahrscheinlichkeit innerhalb des sogenannten Fehlerbereichs des Objekts.
Dieser ist im dreidimensionalen Raum ein Quader, aufgespannt durch die drei sogenannten Fehlerradien (einer je Dimension) mit der Position des Punktes als Mittelpunkt. 
Zwei Messungen eines Labels sind in der Praxis folglich höchstwahrscheinlich nicht identisch, sondern maximal ähnlich zueinander. \cite{Ruehrmair2021,ZachMarin2021,Lankheit2020}

Der Vergleich der zwei Punktwolken stellt eine komplexe Aufgabe dar.
Zur Lösung dieser lassen sich in der Literatur zahlreiche potenziell einsetzbare algorithmische Verfahren mit unterschiedlichen zugrundeliegenden Fragestellungen finden.
Um für den vorliegenden Spezialfall das geeignetste Verfahren zu bestimmen, wurde in einer dieser Arbeit vorangegangenen Untersuchung bereits der Vergleich mehrerer Algorithmen vollzogen \cite{Lankheit2020}.
Daraus resultierend wurde eine Empfehlung gegeben, welcher für den weiteren Einsatz praktikabel erscheint.

\begin{figure}
  \begin{subfigure}{0.3\textwidth}
  \centering
    \includegraphics[width=1\textwidth]{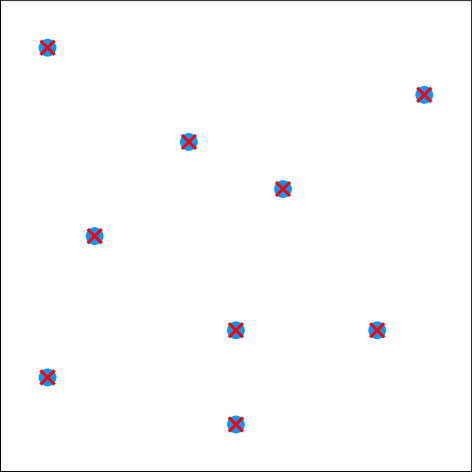}
    \subcaption{Optimalfall: Die beiden Punktwolken sind identisch.}
    \label{figure:PunktwolkenA}
  \end{subfigure}
  \hspace{0.03\textwidth}
  \begin{subfigure}{0.3\textwidth}
    \centering
    \includegraphics[width=1\textwidth]{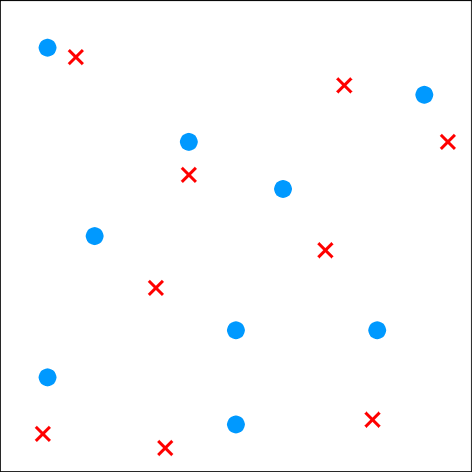}
    \subcaption{Praxisfall: Rotation und Messungenauigkeiten treten auf.}
    \label{figure:PunktwolkenB}
  \end{subfigure}
  \hspace{0.03\textwidth}
  \begin{subfigure}{0.3\textwidth}
    \centering
    \includegraphics[width=1\textwidth]{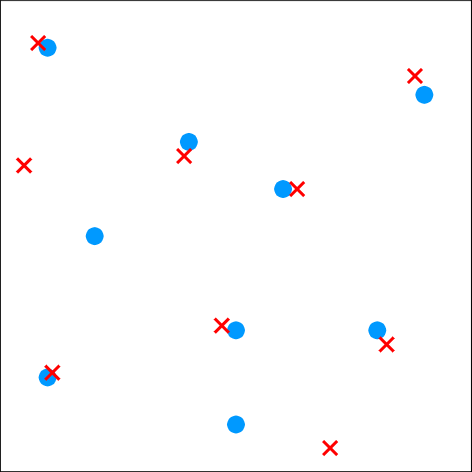}
    \subcaption{Algorithmische Lösung: Annäherung der beiden Punktwolken.}
    \label{figure:PunktwolkenC}
  \end{subfigure}
  \caption{Die drei Bilder zeigen jeweils zwei Punktwolken (eine dargestellt durch blaue Kreise, die andere durch rote Kreuze). (a) zeigt den theoretischen Optimalfall, (b) den realistischen Praxisfall und (c) eine mögliche Annäherung als Lösung des Algorithmus.
  }
  \label{figure:dreimalPunktwolken}
\end{figure}

Der Algorithmus berechnet eine Transformation, mit der eine Punktwolke möglichst präzise auf die andere abgebildet werden kann.
Anhand des Resultats lässt sich bestimmen, ob die zwei Punktwolken gleich oder ungleich sind.
Zur Veranschaulichung des beschriebenen Sachverhalts dient symbolisch Abbildung \ref{figure:dreimalPunktwolken}.
Hierbei handelt es sich aus Gründen der Anschaulichkeit um zweidimensionale Punktwolken.
Darstellungen von 3D-Punktwolken finden sich im Anhang dieser Arbeit (Abbildungen \ref{figure:3D-Gut} und \ref{figure:3D-Schlecht}).
In der theoretischen Überlegung liefert die zweimalige Messung desselben Labels zwei identische Punktwolken (\ref{figure:PunktwolkenA}).
In der Praxis treten allerdings Messungenauigkeiten und Fehler auf (\ref{figure:PunktwolkenB}).
Als Lösung liefert der eingesetzte Algorithmus eine Annäherung beider Punktwolken (\ref{figure:PunktwolkenC}).

Um das beschriebene Verfahren der Produktauthentifizierung der breiten Bevölkerung zugänglich zu machen, soll zukünftig eine offline verwendbare, also ohne die Notwendigkeit einer bestehenden Netzwerkverbindung, mobile Applikation für Android bereitgestellt werden.
Mithilfe dieser soll sowohl ein Hersteller durch den Prozess der Sicherung einer Ware vor Imitation als auch ein Prüfer durch den Prozess der Validierung eines Produktes geleitet werden.
Das hauptsächliche Ziel der vorliegenden Arbeit ist es, den Nachweis für die Umsetzbarkeit einer solchen mobilen Applikation zu erbringen.

Es stellt sich unter anderem die Frage, wie die Daten der Referenzmessung des Herstellers kostengünstig, mit minimalem Aufwand, auf kleiner Fläche und für den Anwender offline auslesbar neben dem Label angebracht werden können.
Da außer der Referenz auch eine digitale Signatur und weitere benötigte Produktinformationen, wie beispielsweise die Seriennummer, die exakte Bezeichnung des Artikels und der Name des Herstellers, übermittelt werden müssen, ist es ein weiteres Ziel dieser Arbeit, dafür eine praktisch einsetzbare und direkt in der mobilen Applikation integrierbare Methode zu bestimmen. 

Ferner soll in der vorliegenden Arbeit ein Verfahren der Nicht-Manipulierbarkeit der dem Label beigefügten Daten ermittelt werden, welches trotz kurzer Signaturen auch eine hohe Sicherheit bietet.
Da die vom Hersteller erzeugte digitale Signatur Teil der Informationen ist, die dem Anwender übermittelt werden, sollte diese aus Gründen der praktischen Verwendbarkeit eine möglichst kurze Länge aufweisen.
Der Wunsch nach hoher Sicherheit liegt auf der Hand: Sobald die Daten manipulierbar sind, ist die Unfälschbarkeit der Produkte nicht mehr gegeben.
Die Entwicklung der für den praktischen Einsatz digitaler Signaturen zusätzlich benötigten umfangreichen Infrastruktur ist jedoch nicht Teil dieser Arbeit.

Des Weiteren soll der für den Punktwolkenvergleich geeignetste Algorithmus als Teil der mobilen Applikation implementiert werden.
Im Vordergrund steht das Bestreben, dem Anwender nach weniger als einer Sekunde Wartezeit das Resultat der Validierung des Produktlabels zu präsentieren.
Gemeint ist die Zeit, die von der App für die Berechnung des Punktwolkenvergleichs und dessen Auswertung benötigt wird.
In der dieser Arbeit vorangegangenen Untersuchung wurde das Zeitlimit allerdings nicht eingehalten, obwohl die Berechnungen dabei mit einem Computer und nicht einem vergleichsweise leistungsärmeren Smartphone durchgeführt wurden \cite{Lankheit2020}.
Zu eruieren ist, ob das Ergebnis der Produktauthentifizierung von einer mobilen Applikation auf einem Smartphone in weniger als einer Sekunde berechnet werden kann.

Die tatsächliche Messung physischer Produktlabel und die Einbindung der dafür vorgesehenen Technologien ist nicht Teil der vorliegenden Arbeit.
Denn weder reale Produktlabel noch die besagten Technologien zur Messung dieser stehen bisher zur Verfügung.
Entsprechend wird in dieser Arbeit auf synthetische Datensätze zurückgegriffen.

Da im Vergleich zu der dieser Arbeit vorangegangenen Untersuchung erstens inzwischen neue Annahmen über die fälschungssicheren Produktlabel getroffen wurden, die zu veränderten Punktwolken und damit neuartigen Datensätzen führen, und zweitens detailliertere Erkenntnisse zu den Resultaten des Punktwolkenvergleichs von Interesse sind, ist es ein weiteres Ziel, Aufschluss über das in der Praxis zu erwartende Validierungs-Verhalten zu geben.
Dabei sollen jedoch ausschließlich die Ergebnisse des eingesetzten Algorithmus untersucht werden und insbesondere kein erneuter Vergleich mit weiteren potenziell geeigneten algorithmischen Verfahren vollzogen werden. 

Der Aufbau der vorliegenden Arbeit gliedert sich wie folgt:
Zunächst wird in Kapitel \ref{chapter:Validierung} die Produktauthentifizierung detaillierter beschrieben.
Dazu werden in Abschnitt \ref{section:verwandteArbeiten} die wesentlichen Resultate verwandter Arbeiten dargestellt, in \ref{section:Grundlagen} die benötigten Grundlagen vermittelt und  in \ref{section:EinordnungErgebnisse} die Ergebnisse der vorangegangenen Untersuchung in den Kontext der vorliegenden Arbeit eingeordnet.
Anschließend werden in Kapitel \ref{chapter:unfälschbarkeitProduktInfos} digitale Signaturen und die dafür eingesetzten algorithmischen Verfahren vorgestellt.
Zur Übermittlung der Referenzdaten, Produktinformationen und digitalen Signatur vom Hersteller an den Prüfer wird in Kapitel \ref{chapter:auswahlBarcode} eine kostengünstige Methode präsentiert.
Dabei zeigt sich die Verwendung von Barcodes im Allgemeinen und QR-Codes im speziellen als besonders geeignet.
Kapitel \ref{chapter:EntwicklungDerApp} umfasst sowohl eine Erläuterung zur Implementierung der mobilen Applikation und die Klärung der Frage, welche Funktionalitäten dabei ausgelassen werden, als auch eine Erklärung zur praktischen Verwendung dieser. 
Die Vorstellung der Ergebnisse und ihre Evaluation werden in Kapitel \ref{chapter:Evaluation} vollzogen.
Hierbei wird insbesondere auf die Resultate des primären Punktwolkenvergleichs, dem Kern der Produktauthentifizierung, eingegangen.
Die Barcodes und digitalen Signaturen werden in diesem Kapitel nicht evaluiert, da sie sowohl in sich geschlossene Themen als auch ausschließlich sekundäre, ergänzende Technologien darstellen.
Abgerundet wird die vorliegende Arbeit durch ein Fazit in Kapitel \ref{chapter:Fazit}.
    \chapter{Produktauthentifizierung mithilfe fälschungssicherer Label}\label{chapter:Validierung} 
Die digitalisierte Version des physischen Labels, die durch eine dreidimensionale Vermessung entsteht, ist im Kontext dieser Arbeit eine 3D-Punktwolke.
Bei einem Label, welches Goldnanokügelchen beinhaltet, stellt jedes einen der dreidimensionalen Punkte dar.
Bei Goldnanostäbchen werden dagegen die beiden Endpunkte eines Stäbchens als ein jeweils eigenständiger Punkt aufgefasst.
Somit können sowohl auf Kügelchen als auch auf Stäbchen basierende Label in ihrer digitalisierten Form als eine 3D-Punktwolke aufgefasst werden.
Dadurch lässt sich für beide Label-Varianten die nachfolgende Bearbeitung mit dem gleichen Verfahren fortsetzten.
Aus Sicht der Algorithmen bedarf es keiner weiteren Unterscheidung der unterschiedlichen Typen.
Erfolgt im Verlauf der vorliegenden Arbeit der Hinweis, dass eine Punktwolke aus $n \in \mathbb{N}$ Punkten besteht, dann würde das zugrundeliegende Label entweder $n$ Goldnanokügelchen oder $\lfloor\frac{n}{2}\rfloor$ Goldnanostäbchen enthalten.

Im Mittelpunkt des Verfahrens der Validierung von fälschungssicheren Produktlabeln und damit der gesamten Produktauthentifizierung befindet sich der Vergleich zweier Punktwolken.
Dieser mag möglicherweise trivial erscheinen, erweist sich in der Praxis allerdings als anspruchsvoll.
Für den Vergleich bedarf es eines Algorithmus, welcher den Grad an Gleichheit der beiden Punktwolken ermittelt.
Vorschläge zur Implementierung derartiger Berechnungsverfahren lassen sich in Vielzahl im Gebiet der Computer-Vision finden.
Hierbei handelt es sich um einen Bereich, der sich mit der Extraktion und Verarbeitung verschiedenster Inhalte aus Bilddaten befasst \cite{szeliski2010computer}.
Methoden der Computer-Vision, die wiederum versuchen, Punktwolken möglichst präzise aufeinander abzubilden, werden unter dem Begriff der Point-Set-Registration zusammengefasst \cite{zhu2019review}.
Damit liefern die Algorithmen der Point-Set-Registration prinzipiell eine Lösung für das dieser Arbeit zugrundeliegende Problem.
Allerdings liegen ihrer Entwicklung unterschiedliche Problemstellungen und Zielsetzungen zugrunde, sodass kein Algorithmus existiert, der eine allgemeingültig optimale Lösung bietet.
Deshalb musste der Einleitung weiterer Entwicklungsschritte eine Evaluation, welches Verfahren für das spezielle Szenario im Kontext der vorliegenden Arbeit am geeignetsten ist, vorangehen.
Dazu wurden ausgewählte Algorithmen der Point-Set-Registration implementiert, miteinander unter den gegebenen Bedingungen verglichen und als beste Lösung für die weitere Verwendung der von Myronenko und Song entwickelte \textbf{Coherent Point Drift (CPD)} \cite{Myronenko2010} vorgeschlagen \cite{Lankheit2020}.

Ehe in Abschnitt \ref{section:EinordnungErgebnisse} detailliert auf die Ergebnisse der Vorarbeit eingegangen wird und insbesondere die Differenzen der grundsätzlich getroffenen Annahmen im Vergleich zur vorliegenden Arbeit aufgezeigt werden, liefert Kapitel \ref{section:verwandteArbeiten} einen Überblick verwandter Arbeiten, die sich ebenfalls mit der Unfälschbarkeit und Authentifizierung von Produkten auseinandersetzen.
Zudem wird in Kapitel \ref{section:Grundlagen} elementares Wissen über das zugrundeliegende Optimierungsproblem und die eingesetzten Algorithmen vermittelt sowie auf die Nomenklatur für die weitere Arbeit eingegangen.

\section{Verwandte Arbeiten}\label{section:verwandteArbeiten}
Bereits in den 1980er-Jahren entstand die Idee, bestimmte physikalische Eigenschaften eines Objekts für den Echtheitsnachweis zu verwenden.
Ein derartiges System zur Authentifizierung basiert auf speziellen, einzigartigen und meist zufälligen Phänomenen und Charaktereigenschaften des Gegenstands, welche leicht messbar, aber nicht praktikabel reproduzierbar sind.
Goldman \cite{goldman1985non} nutzte hierfür die charakteristische Lichtdurchlässigkeit von Papier und erkannte, ähnlich zur vorliegenden Arbeit, die potenziellen Vorteile der Speicherung einer Referenzmessung auf dem Artikel.
Auch Simmons \cite{Simmons1991} und Bauder \cite{Bauder1983} begriffen den Nutzen, eine Vergleichsmessung des einzigartigen Objektes mit auf diesem anzubringen, und verwendeten dafür erstmals digitale Signaturen.
Dabei stellten in semi-transparentem Material eingelassene Fasern aufgrund ihrer zufälligen Verteilung die Unfälschbarkeit des Objektes dar.
Die Authentifizierung wurde durch die Messung der Anfangs- und Endpunkte der Fasern vollzogen.
Damit ist sowohl die Problemstellung des Verfahrens als auch der Grund für die Verwendung digitaler Signaturen vergleichbar mit dieser Arbeit.

Eine weitere Eigenschaft, welche sich für die Verwendung der Produktauthentifizierung eignet, ist die einzigartige Oberflächenbeschaffenheit bestimmter Materialien.
Hierbei lassen sich aus den zufälligen und natürlich vorkommenden Unvollkommenheiten der Oberflächen von zum Beispiel Produktverpackungen, CDs, Plastikkarten und Papier geeignete Charakteristika für die Unfälschbarkeit von Produkten ableiten. \cite{buchanan2005fingerprinting,hammouri2009cds}
Beispielsweise haben Clarkson et al. \cite{Clarkson2009Fingerprinting} die Oberfläche von Papierdokumenten dreidimensional vermessen und so eine eindeutige Referenz des Dokuments zur späteren Verifikation erstellt.
Zwar auch auf Grundlage der Oberfläche von Papier haben Haist und Tiziani \cite{haist1998optical} ein Verfahren zur Kontrolle der Echtheit von Banknoten entwickelt, allerdings haben sie dazu die zufällige mikroskopische Faserstruktur des Papiers als das eindeutige Merkmal verwendet.
Im Gegensatz zur vorliegenden Arbeit wird bei diesen Methoden lediglich die Oberfläche vermessen.
Nichtsdestotrotz sind die detaillierte Untersuchung der Objekte und der Vergleich mit einer Referenzmessung bei diesen Varianten im Kern analog zur vorliegenden Arbeit.

Ein anderes Beispiel für ein fälschungssicheres Identifikationsschema ist das von Pappu et al. \cite{pappu2001physical,pappu2002physical} entwickelte Verfahren zur Erstellung und Messung von sogenannten Sprenkel-Mustern (engl. \textit{speckle patterns}).
Hierfür wird während des Produktionsprozesses eine große Zahl kleiner Glaskügelchen in einem transparenten quaderförmigen Kunststoffplättchen zufällig verteilt.
Richtet man einen Laserstrahl auf das Plättchen, wird das Licht mehrfach gestreut und es entsteht ein einzigartiges Sprenkel-Muster.
Zwar wird bei diesem Verfahren nicht das Kunststoffplättchen selbst vermessen, aber das für die Einzigartigkeit verantwortliche Objekt - in Kunststoff eingelassene, zufällig angeordnete Kügelchen - ist verwandt dem unfälschbaren Produktlabel der vorliegenden Arbeit.

Bei allen bisher vorgestellten Techniken wird allerdings spezielle zusätzliche Hardware zum Extrahieren und Vergleichen der einzigartigen Eigenschaften benötigt.
Dabei handelt es sich beispielsweise um einen Warenscanner \cite{Clarkson2009Fingerprinting}, ein Mikroskop \cite{haist1998optical} oder einen Laser \cite{pappu2001physical,pappu2002physical}.
Da der Authentizitätsnachweis in der vorliegenden Arbeit jedoch mit einem mobilen Telefon erfolgen soll, eignen sich diese Methoden hierfür nicht.

2020 haben Leem et al. \cite{leem2020edible} einen neuen Ansatz entwickelt, der dem übergeordnetem Konzept dieser Arbeit nahekommt.
Mithilfe der Kamera eines mobilen Telefons wird dabei das charakteristische Emissionsverhalten fluoreszierender Partikel gemessen.
Leider erfolgt die Echtheitsverifikation nicht offline, also ohne die Notwendigkeit einer bestehenden Internetverbindung, sondern mit online hinterlegten Referenzdaten.

Nach bestem Wissen existiert keine mobile Applikation, mithilfe derer offline die Authentizität eines fälschungssicheren Labels nachweisbar ist.
Wie das Konzept einer App zur Validierung von fälschungssicheren Produktlabeln aussehen kann und welche zusätzlichen Verfahren eingebunden werden müssen, wird in weiten Teilen dieser Arbeit erörtert.

\section{Abgrenzung zu Physical Unclonable Functions}

Ein zu dieser Arbeit nahe benachbartes Forschungsfeld sind sogenannte Physical Unclonable Functions (PUFs).   Sie sollen zur Abgrenzung der vorliegenden Arbeit in einem kurzen Überblick umrissen und von den Anstrengungen dieser Arbeit differenziert werden;  dabei wird in Teilen \cite{Lankheit2020} gefolgt.  Die PUF-Foschung wurde durch eine frühe Vorläuferarbeit von Lofstrom et al.\ \cite{lofstrom2000ic} sowie durch die bereits teilweise oben angesprochenen Veröffentlichungen zu optischen PUFs \cite{pappu2002physical} und elektrischen PUFs \cite{gassend2002silicon} um die Jahrtausendwende herum begründet, also vor ungefähr zwei Jahrzehnten.  Die beiden genannten optischen und elektrischen PUFs ordnet man heute rückblickend als sogenannte ``Strong PUFs'' ein:  Diese sind eine PUF-Unterart, die sehr viele verschiedene Möglichkeiten der externen Anregung besitzt, oder auch sehr viele sogenannte ``Challenges'' in der Sprechweise des Gebiets \cite{ruhrmair2022secret,ruhrmair2014pufs,ruhrmair2012security}.  Sogenannte ``Weak PUFs'', die nur wenige Challenges, aber dennoch breite praktische Einsetzbarkeit besitzen, wurden erst später eingeführt \cite{holcomb2008power,simons2012buskeeper,kim2018dram,tehranipoor2016dram,guajardo2007fpga,kumar2008butterfly,rahman2014aro,xiao2014bit,vskoric2005robust,maes2012pufky,cao2015low,orosa2022spyhammer}.  Sowohl Weak PUFs als auch Strong PUFs haben bestimmte Übereinstimmungen und Überschneidungen mit den im Zuge dieser Arbeit untersuchten Produktabels:  Beispielsweise weisen sie ebenfalls zufällige und unkontrollierbare Fabrikationsschwankungen auf, die für ihre Sicherheitseigenschaften zentral sind \cite{Ruehrmair2019,ruhrmair2022secret}.  Dennoch unterscheiden sie sich in anderen wichtigen Punkten, z.B.\ in der Anzahl der Challenges, dem von dieser Arbeit, nicht aber von PUFs unbedingt benötigten externen Abfragemechanismus, und den angestrebten Sicherheitseigenschaften und Einsatzgebieten \cite{ruhrmair2022secret,ruhrmair2020sok}.  Vorrausblickend sei zu betonen, dass diese Unterscheidung im Zusammenhang mit der vorliegenden Arbeit zentral ist.   

Eine wichtige Angriffsform für Strong PUFs sind sogenannte Modellierungsattacken \cite{ruhrmair2010modeling,ruhrmair2013puf,ruhrmair2009foundations,sehnke2010policy}.  In diesen wird kein physikalischer, sondern ein digitaler Klon der PUF-Strukturen erstellt.  Dieser digitale Klon hat dasselbe (digitale) Challenge-Response Verhalten wie das PUF-Original, kann aber im Gegensatz zu letzterem beliebig vervielfältigt werden, was die PUF-Sicherheit bricht. Diese Modellierungsattacken haben im Bereich Strong PUFs zu einer Fülle von neuen, angeblich sichereren Architekturen  geführt und waren für die Entwicklung des Gebiets von zentraler Bedeutung
\cite{chen2009analog,vijayakumar2016machine,sauer2017sensitized,chen2011bistable,tuyls2007strong,csaba2010application,vijayakumar2015novel,kumar2014design,majzoobi2008lightweight,xi2017strong,majzoobi2009techniques,ruhrmair2010applications,xi2022provably,jin2017fpga,csaba2009chip,ruehrmair2012method,lugli2013physical,herder2016trapdoor,majzoobi2010fpga,kappelhoff2022strong}.  Gleichzeitig hat die PUF-Forschung des letzten Jahrzehnts gezeigt, dass sich die genannten Modellierungsattacken sogar noch weiter verbessern lassen, indem Seitenkanalinformationen hinzugefügt werden  \cite{ruhrmair2014efficient,ruhrmair2013power,ruhrmair2014special,tajik2014physical}.
Während die besagten Modellierungsattacken und Gegenmaßnahmen ein zentrales Agens der Strong PUF Forschung der letzten Jahrzehnte darstellen, spielen sie allerdings für die in dieser Arbeit untersuchten fälschungssicheren Labels keine Rolle. Dies illustriert noch einmal die technischen Unterschiede klassischer  PUFs und der in dieser Arbeit behandelten Themen und Primitive. 

Ein weiterer Angriffsvektor, der sich innerhalb des letzten Jahrzehnts herauskristallisiert hat, sind Strong PUF Angriffe auf Protokollebene \cite{van2014protocol,ruhrmair2016security,ruhrmair2013pufs,ruhrmair2012practical,ruhrmair2013practical}.   Im Zusammenhang mit diesen Protokollanalysen müssen auch die generelle Verwendung von Strong PUFs in fortgeschrittenen kryptographischen Protokollen, ihre Formalisierung und formale Sicherheitsbeweise genannt werden \cite{ruhrmair2011physical,ostrovsky2013universally,dachman2014feasibility,ruhrmair2010strong,ruhrmair2010oblivious}.  Wie bereits oben beschrieben unterscheiden all diese Aspekte die PUF-Forschung von dieser Arbeit und von fälschungssicheren Labels, bei denen die genannten Aspekte keine direkte Rolle spielen. 

Falls das Interesse des Lesers an PUFs durch diesen kurzen Überblick geweckt ist, werden zum Abschluss dieser Sektion nur einige wenige weiterführende Beispiele genannt, die für den Leser auch als Startpunkt für eigene Forschungsaktivitäten von Interesse sein könnten;  diese schließen   \cite{rajendran2012nano,horstmeyer2015physically,eliezer2022exploiting,pavanello2021recent,langhuth2011strong,ruhrmair2013optical,arapinis2021quantum,ruhrmair2015virtual,jaeger2010random,gao2018efficient,jin2020erasable,jin2015playpuf,jin2022programmable} mit ein.  Andere, verwandte Arbeiten sind \cite{zalivaka2018reliable,dejean2007rf,lakafosis2010rf,koeberl2013memristor,chatterjee2018rf,mazady2015memristor,zhang2014survey,shi2019approximation,katzenbeisser2011recyclable,delvaux2019machine,nedospasov2013invasive,liu2017acro,john2021halide,rosenfeld2010sensor,rose2013hardware,tang2016securing,majzoobi2012slender,rostami2014robust,maes2009soft,maes2008intrinsic,yu2016lockdown,delvaux2013side,grubel2017silicon,horstmeyer2013physical,buchanan2005fingerprinting}. Besonderes Interesse gilt aus dem Blickwinkel der vorliegenden Arbeit unter anderem sogenannten SIMPL Systemen/SIMPL PUFs \cite{ruhrmair2009simpl,ruhrmair2011simpl,ruhrmair2012simpl,ruhrmair2010towards,chen2011circuit}, die ebenso wie die in dieser Arbeit untersuchten einzigartigen und nicht-fälschbaren Labels ein prinzipiell geheimnisfreies (``secret-free'') Sicherheitsprimitiv darstellen \cite{ruhrmair2022secret}.  Die meisten anderen bekannten PUFs besitzen diese Eigenschaft nicht \cite{ruhrmair2022secret}.

\section{Grundlagen und Nomenklatur}\label{section:Grundlagen}
Im Kern der Antwort auf die Frage, ob es sich bei einem Produktlabel um ein Original oder ein Imitat handelt, befindet sich ein Algorithmus, welcher dabei hilft, den Grad der Gleichheit zweier Punktwolken zu bestimmen.
Wie beschrieben, handelt es sich hierbei um eine Problemstellung aus dem Bereich der Point-Set-Registration, in welchem dafür eine Vielzahl an Algorithmen entwickelt wurde \cite{zhu2019review}.
Tatsächlich sind im Kontext dieser Arbeit sogar nur die Algorithmen von Interesse, die eine Lösung für das (in der Literatur nicht einheitlich so bezeichnete) Simultaneous-Pose-and-Correspondence-Problem liefern.
Hierbei geht es darum, für gegebene Punktwolken eine höchstens affine Transformation und eine Korrespondenzmenge zu finden, sodass der Abstand der korrespondierenden Punkte minimal ist.
Formaler ausgedrückt:
Seien ${X = \{X_1,...,X_N \}}$ und ${Y = \{Y_1,...,Y_M \}}$ die zwei gegebenen Punktwolken, wobei ${N,M \in \mathbb{N}}$ und ${\forall i \in \{1,...,N \}} ,{ \forall j \in \{1,...,M \}}$ gilt ${X_i, Y_j \in \mathbb{R}^D}$ sind jeweils Vektoren gleicher Dimension ${D \in \mathbb{N}}$.
Des Weiteren sei\\
${d:\mathbb{R}^D\times\mathbb{R}\to\mathbb{R}_{\geq0}}$ eine Metrik.
Zur Lösung des Simultaneous-Pose-and-Correspondence-Problem müssen nun eine affine Funktion ${T : \mathbb{R}^D \to \mathbb{R}^D}$ und eine Konsensmenge\\
${C \subseteq X \times Y}$ gefunden werden, die das in \ref{Gleichung:PSR-Problem-Allgemein} gezeigte Optimierungsproblem lösen. \cite{zhu2019review,Besl_1992}
\begin{flalign}\label{Gleichung:PSR-Problem-Allgemein}
  \min_{T, C} \sum_{x,y \in C} d(x, T y)
\end{flalign}

Durch die Verwendung der Produktlabel sind bereits einige zusätzliche Eigenschaften bekannt, wodurch sich das in \ref{Gleichung:PSR-Problem-Allgemein} gezeigte Optimierungsproblem vereinfachen lässt.
Die Struktur der jeweiligen Punktwolken, also die zueinander relativen Positionierungen der einzelnen Punkte einer Punktwolke, soll sich im Kontext dieser Arbeit nicht verändern. 
Somit ist es ausreichend, die affine Funktion $T$ durch eine Translation ${t \in \mathbb{R}^D}$, auch Parallelverschiebung genannt, einen Skalierungsparameter ${s \in \mathbb{R}}$ und eine Rotation ${R \in \mathbb{R}^D \times \mathbb{R}^D}$ im dreidimensionalen Raum {($D = 3$)} zu ersetzen.
Daraus ergibt sich das in \ref{Gleichung:Optimierungs-Problem-Spezifisch} beschriebene vereinfachte und präzisere Optimierungsproblem für den Vergleich der Punktwolken im Zusammenhang der fälschungssicheren Produktlabel.
\begin{flalign}\label{Gleichung:Optimierungs-Problem-Spezifisch}
  \min_{R,t,s,C} \sum_{x,y \in C} d(x,s R y + t)
\end{flalign}

In dieser Arbeit wird im Folgenden die Punktwolke \textbf{$X$} als \textbf{Referenz} und die Punktwolke \textbf{$Y$} als \textbf{Messung} bezeichnet.
Zudem ist die Referenz stets die vom Hersteller eines Produkts erstellte und auf dem Barcode hinterlegte Punktwolke.
Folglich ist die durch den Anwender während des Vorgangs der Produktauthentifizierung gemessene Punktwolke des physischen Labels die Messung.

Wie bereits in Kapitel \ref{chapter:Einleitung} dargelegt, treten bei der Vermessung eines Produktlabels naturgemäß Messfehler auf.
Dadurch kommt es vor, dass einzelne Punkte in einer der beiden Mengen, Referenz oder Messung, enthalten sind, in der anderen jedoch nicht existieren.
In der folgenden Arbeit werden Punkte der Messung, die nicht Teil der Referenz sind, als \textbf{Artefakte} und Punkte der Referenz, welche wiederum bei der Messung fehlen, als \textbf{verlorene Punkte} bezeichnet.
Formaler ausgedrückt: Die Elemente aus $Y \setminus X$ werden Artefakte, die Elemente aus $X \setminus Y$ verlorene Punkte genannt.

Wie sich in Abschnitt \ref{section:EinordnungErgebnisse} zeigen wird, ist der CPD für das zugrundeliegende Problem der vorliegenden Arbeit im Kontext der Entwicklung einer mobilen Applikation der geeignetste Algorithmus. 
Hierbei handelt es sich um einen probabilistischen Ansatz, bei dem die Menge $Y$ als Gauß’sche Mischverteilung angesehen wird.
Dabei stellt die Position eines Punktes den jeweiligen Mittelwert einer Normalverteilung dar.
Der CPD sucht dann eine Transformation, sodass diese auf die Messung $Y$ angewendet passend zur Referenz $X$ ist.
Hierzu werden die Mischverteilungen kohärent verschoben, also alle Teilverteilungen in gleicher Weise.
Dabei werden abwechselnd die Korrespondenzmenge und die Transformationen, also $R$, $s$ und $t$, bestimmt.
Für diese alternierende Bestimmung wird wiederum der Expectation-Maximation (EM) -Algorithmus eingesetzt \cite{Dempster_1977}.
Dies ist ein Verfahren zur Bestimmung der Parameter eines statistischen Modells.
Da allerdings nicht alle wesentlichen Informationen über das statistische Modell existieren, schätzt der EM-Algorithmus die am wahrscheinlichsten erscheinenden Werte für die nötigen Parameter. 
Als Ergebnis liefert der CPD sowohl die Transformation bestehend aus $R$, $s$ und $t$ als auch eine Korrespondenz-Wahrscheinlichkeit $P$ für alle Punkte Paare ${(X_i, Y_j) \in X \times Y}$. \cite{Myronenko2010}

Bei der Verwendung des CPDs werden die in der Praxis auftretenden Fehlerbereiche der einzelnen Punkte nicht berücksichtigt.
Der Fehlerbereich eines Punktes ist hierbei der von seinem dreifachen Fehlervektor aufgespannte Quader, bei dem das Zentrum durch die Position des Punktes beschrieben wird. Für einen Punkt mit der Position ${(x,y,z) \in \mathbb{R}^3}$ und einen Fehler ${(\sigma_x,\sigma_y,\sigma_z) \in \mathbb{R}^3_{\geq 0}}$ wird der Fehlerbereich wie in Gleichung \ref{Gleichung:Fehlerbereich} definiert.
\begin{flalign} \label{Gleichung:Fehlerbereich}
  \mathbb{E}_3((x,y,z)) = 
  [x-3\sigma_x, x+3\sigma_x]
  \times [y-3\sigma_y, y+3\sigma_y]
  \times [z-3\sigma_z, z+3\sigma_z]
\end{flalign}
Im Zusammenhang der fehlerhaften Messung eines Punktes werden $x$ und $\sigma_x$ als die Parameter einer Gauß'schen Normalverteilung mit Mittelwert $x$ und Standardabweichung $\sigma_x$ betrachtet.
Anhand dieser lässt sich die Wahrscheinlichkeit ermitteln, den Punkt mit einem speziellen Wert für die $x$-Koordinate zu messen.
Analoges gilt für $y$ und $z$ mit $\sigma_y$ und $\sigma_z$. \cite{Lankheit2020}

Die Fehlerbereiche der einzelnen Punkte werden im Zuge der abschließenden Gleichheitsuntersuchung herangezogen.
Dabei wird anhand von $P$ für jeden Punkt der Referenz der am wahrscheinlichsten erscheinende korrespondierende Partner-Punkt der transformierten Messung ermittelt.
Im Folgenden sei $C_{Korr} \subseteq X \times Y$ die Menge aller vermuteten Korrespondenz-Paare.
Final wird dann gezählt, bei wie vielen der gebildeten Paare sich beide Punkte im Fehlerbereich des jeweiligen Partners befinden.
Dieses Resultat wird ins Verhältnis zur Gesamtzahl der Punkte der Referenz gesetzt und dadurch die Güte des Punktwolkenvergleichs angegeben.
Das Beschriebene ist in Gleichung \ref{Gleichung:AnzahlErkannt} dargestellt.
\begin{flalign}  \label{Gleichung:AnzahlErkannt}
  P_{Fehlerbereich} &= P_{Fehl} =
  \frac{\abs{
    \left\lbrace 
      (X_i,Y_j) \mid (X_i,Y_j) \in C_{Korr}, X_i \in \mathbb{E}_3(Y_j), Y_j \in \mathbb{E}_3(X_i)
    \right\rbrace}
  }{\abs{X}}
\end{flalign}

In der vorangegangenen Untersuchung hat sich jedoch gezeigt, dass der CPD häufig schlechte Ergebnisse liefert, also in einer geringen Zahl erkannter Punkte resultiert, falls die Messung in Relation zur Referenz stark rotiert ist.
Zur Lösung des Problems wird, bezogen auf die Rotation, ein ähnlicher Schritt wie bei anderen Algorithmen vollzogen: Vor dem Start des CPDs wird der Suchraum, der Raum aller
Rotationen, in gleich große Würfel unterteilt \cite{Campbell2016}.
Hierbei wurde in der Vorarbeit je Achse eine Dreiteilung empfohlen, sodass sich 27 sogenannte Unterwürfel ergeben. 
Für jeden von diesen wird dann ein eigenständiger CPD ausgeführt. \cite{Lankheit2020}
Es ist davon auszugehen, dass ein Anwender die Kamera des Smartphones in etwa senkrecht und korrekt ausgerichtet über dem Produktlabel platzieren kann und folglich nur vergleichsweise kleine Rotationen auftreten werden.
Allerdings hat sich die Aufteilung des Suchraums trotzdem als hilfreich erwiesen und wird daher auch weiterhin beibehalten.

Nach der Vollendung aller CPDs wird, wie in Gleichung \ref{Gleichung:AnzahlErkannt} beschrieben, für jeden einzelnen der prozentuale Wert erkannter Punkte ermittelt und der größte aller als das Gesamtergebnis des Punktwolkenvergleichs ausgewählt.

\section{Ergebnisbeurteilung der Untersuchung von Algorithmen der Point-Set-Registration}\label{section:EinordnungErgebnisse}
In diesem Abschnitt werden die wesentlichen Ergebnisse der vorangegangenen Untersuchung von Lankheit \cite{Lankheit2020} vorgestellt.
Zudem werden die grundsätzlichen und konzeptionellen Unterschiede im Vergleich zu der vorliegenden Arbeit aufgezeigt.
Das sind beispielsweise Annahmen über die Eigenschaften des Labels und die Art der finalen Bewertung des Punktwolkenvergleichs.

Wie bereits dargestellt, musste vor der tatsächlichen Entwicklung einer mobilen Applikation zur Validierung von fälschungssicheren Produktlabeln zuerst ein geeigneter Algorithmus aus dem Bereich der Point-Set-Registration ausgewählt werden, welcher die zwei Punktwolken zufriedenstellend aufeinander abbildet.
Dazu wurden in der vorangegangenen Analyse verschiedene Algorithmen in der Programmiersprache Python, unter Zuhilfenahme von NumPy, implementiert und miteinander verglichen.
NumPy ist eine Bibliothek, die insbesondere bei der Implementierung nummerisch rechenintensiver Verfahren, wie das hier zugrundeliegende, eine bedeutende Unterstützung darstellt.

Bei dem Vergleich der unterschiedlichen Algorithmen hat sich der CPD für die Umsetzung im Kontext einer mobilen Applikation, als geeignetster Algorithmus erwiesen.
Zwar erzielte bei der reinen Punkte-Zuordnung eine Komposition aus CPD und Gaussian Mixture Model Registration (GMMReg) \cite{jian2010GMMReg}, ein Algorithmus, welcher sich so modifizieren lässt, dass die Fehler der einzelnen Punkte schon in die Berechnung der Transformation miteinfließen, minimal bessere Ergebnisse, allerdings ging dies mit einem deutlichen Anstieg der Rechenzeit einher.
Daher eignet sich die zusätzliche Verwendung des GMMRegs nicht, und es wird ausschließlich der CPD eingesetzt. \cite{Lankheit2020}

Des Weiteren liefert die vorangegangene Untersuchung insbesondere zwei interessante Erkenntnisse über die Resultate des Punktwolken-Vergleichs:
\begin{enumerate}
\item Es kommt vor, dass für zwei ursprünglich fast gleiche Punktwolken kaum korrespondierende Punkte gefunden werden, obwohl dies theoretisch möglich sein sollte. 
\item Selbst, wenn alle Korrespondenzen korrekt ermittelt werden, befinden sich häufig nur $50-70~\%$ aller korrespondierenden Punkte im Fehlerbereich ihres Partners. 
\end{enumerate}

Im Gegensatz zu dieser Arbeit wurde in der vorangegangenen Abhandlung Wissen über die Punktwolken verwendet, das in der Praxis nicht vorhanden ist.
Dabei wurde in die Bewertung, wie erfolgreich die Zuordnung der Punktwolken verlief, die Kenntnis über in der Realität korrespondierende Punkte mit einbezogen.
Im Detail wurde dazu im Nenner des in Gleichung \ref{Gleichung:AnzahlErkannt} definierten $P_{Fehl}$ anstelle von $\abs{X}$ die Anzahl der wahren Korrespondenzen verwendet.
Derartige Kenntnisse über die echten Korrespondenzen liegen in der Praxis allerdings nicht vor und deshalb wurden in der vorliegenden Arbeit die beschriebenen Anpassungen in der Evaluation des Punktwolken-Vergleichs vollzogen.

Ein weiterer wichtiger Unterschied zwischen der vorangegangenen und dieser Arbeit liegt darin begründet, dass sich detaillierte, praxisbezogene Eigenschaften, wie zum Beispiel die Größe eines Produktlabels, erst parallel zu dieser Arbeit herauskristallisiert haben.
Mit dem Wissen über die Größe eines Produktlabels können eindeutige Grenzen definiert werden, innerhalb derer sich alle Punkte befinden müssen.
Entsprechend basieren die synthetischen Testdatensätze in dieser Arbeit auf der rein zufälligen Verteilung der Punkte im Bereich der vorgegebenen Begrenzungen.
Da derartige Erkenntnisse zur Erstellung von Testdatensätzen in der vorangegangenen Arbeit allerdings aus Mangel an Existenz nicht verwendet werden konnten und um sicherzustellen, dass die einzelnen Punkte der Punktwolke in einem realistischen Abstand zueinander sind, wurden die Punktwolken basierend auf einer Gauß’schen Normalverteilung erzeugt.
Dadurch wurden zwar relativ wirklichkeitsgetreue Punktwolken erstellt, allerdings wiesen diese zwei unerwünschte Eigenschaften auf.
Erstens hatten sie einen festen Mittelpunkt, an welchem das Vorkommen von Punkten am wahrscheinlichsten war.
In der Praxis sollte es einen derartigen wahrscheinlichsten Punkt jedoch nicht geben.
Und zweitens konnte es passieren, dass einzelne Punkte der Punktwolke deutlich abseits angesiedelt waren, also weit entfernt von allen anderen Punkten.
Auch wenn dies eher unwahrscheinlich war, konnte ein klar von den anderen abgegrenzter Punkt als eine Art Leuchtturm zur Orientierung dienen, da dessen Charakteristik, die große Entfernung zu den anderen Punkten, sowohl in der Referenz als auch in der Messung einfach auszumachen ist.
Dies ist aufgrund der Eigenschaften eines Labels in der Praxis allerdings nicht mehr nur unwahrscheinlich, sondern schlicht unmöglich.
Folglich müssen im Zuge dieser Arbeit neue und vor allem leicht andersartige Punktwolken als Testdatensätze erstellt werden.

Der letzte Unterschied zwischen der Umsetzung des Algorithmus in der vorangegangenen Arbeit und seiner Entwicklung im Kontext einer App besteht neben der Verwendung unterschiedlicher Programmiersprachen und damit einhergehend auch anderer Bibliotheken in der Differenz der zugrundeliegenden Hardware.
Diese hat vor allem auf die Dauer der Berechnung des Punktwolkenvergleichs einen entscheidenden Einfluss.
    \chapter{Unfälschbarkeit der Produktinformationen}\label{chapter:unfälschbarkeitProduktInfos}
Folgende Überlegung liegt dieser Arbeit zugrunde: \textbf{Stimmt die Referenz-Punktwolke mit der Messung des Labels überein, ist das Produkt ein Original.}
Um diesen theoretischen Grundsatz auch in der Praxis gewährleisten zu können, bedarf es noch eines weiteren Sicherheitsmechanismus.
Denn setzt ein Betrüger ein fälschungssicheres Produktlabel ein, fügt dem Barcode aber unechte Informationen über das Produkt an, würde dennoch seine Ware als Original angesehen werden.
Folglich kann es sich auch bei Übereinstimmung eines Labels mit den dazu gespeicherten Referenzdaten um eine Fälschung handeln.
Deshalb muss dem gesamten Verfahren noch ein Schritt hinzugefügt werden, welcher potenziellen Fälschern die Möglichkeit nimmt, fingierte Angaben über das Produkt zu machen.
Dazu unterschreibt jeder Hersteller, dass Label, Produkt und Barcode von ihm stammen.
Solange diese Unterschrift wiederum nicht fälsch- oder kopierbar ist, kann sich niemand als der Hersteller des Originals ausgeben.
Unechte Angaben über das Produkt werden damit erkannt und die Unfälschbarkeit ist wiederhergestellt. \cite{Ruehrmair2021}
Um dies zu erreichen, kommen, wie in Abschnitt \ref{section:digitaleSignatur} beschrieben, digitale Signaturen zum Einsatz.
Wie sich zeigen wird, sind dazu verschiedene algorithmische Verfahren nötig, für die jeweils eine Vielzahl an möglichen Umsetzungen existiert.
Deshalb werden vor der finalen Auswahl der Algorithmen in Abschnitt \ref{subsection:AuswahlDerAlgorithmen} zunächst in Abschnitt \ref{subsection:KriterienDigitaleSignatur} die für diese Arbeit bedeutendsten Kriterien erläutert.
Zudem wird in \ref{section:Elliptisch-Hash} ein kurzer Überblick über die anschließend eingesetzten elliptischen Kurven und Hashfunktionen gegeben.

\section{Digitale Signatur}\label{section:digitaleSignatur}
Um das beschriebene Szenario umzusetzen, benötigt es Verfahren der Kryptographie.
Damit wird die Erforschung mathematischer Techniken bezeichnet, die zum Schutz von Informationen eingesetzt werden können.
Ursprünglich diente die Kryptographie vor allem der Verschlüsselung geheimer Informationen, welche zwischen zwei Parteien ausgetauscht werden.
Dabei einigen sich Sender und Empfänger auf einen geheimen Schlüssel, mit dem die Nachricht ver- und entschlüsselt wird.
Kommen unberechtigte Personen in den Besitz der Nachricht, können sie diese ohne Kenntnis des geheimen Schlüssels nicht dechiffrieren.
Diese Art der Verschlüsselung ist ein Verfahren der symmetrischen Kryptographie. \cite{werner2002problem}
Es ist allerdings offensichtlich, dass der Austausch des geheimen Schlüssels zwischen Hersteller und jedem Anwender in der Praxis nicht sinnvoll umsetzbar ist.
Stattdessen bedarf es Verfahren der asymmetrischen Kryptographie.
Dabei wird statt eines geheimen Schlüssels, den sowohl der Sender als auch der Empfänger verwenden, ein Schlüsselpaar, bestehend aus einem privaten Schlüssel und einem öffentlichen Schlüssel, verwendet \cite{werner2002problem}.

Zudem müssen die auf dem Barcode gespeicherten Informationen nicht chiffriert, sondern nur signiert werden.
Denn durch eine sogenannte digitale Signatur lässt sich, erstens die Integrität des Barcodes nachweisen, also dass dessen Daten seit der Erstellung nicht mehr verändert wurden.
Zweitens zeigt sie die Authentizität der Daten, also dass der Besitzer des geheimen Signierschlüssels der Ersteller des Barcodes ist.
Drittens erfüllt die digitale Signatur den Aspekt der Nicht-Abstreitbarkeit, das heißt, der Hersteller kann nicht dementieren, der Urheber des Barcodes zu sein. \cite{buchmann2008einfuhrung}
Der Hauptgrund für die Verwendung digitaler Signaturen in dieser Arbeit besteht in der Gewährleistung der Integrität und Authentizität der Daten und damit des gesamten Produktes.
Weitere Aspekte der Informationssicherheit, wie zum Beispiel die Vertraulichkeit der Daten, müssen im Kontext fälschungssicherer Produktlabel nicht sichergestellt werden.
Folglich ist eine digitale Signatur ausreichend, und es bedarf keiner Verschlüsselung der gesamten Informationen. 

Im praktischen Einsatz erstellt ein Hersteller mit Hilfe seines geheimen, privaten Schlüssels eine Signatur und speichert diese ebenfalls auf dem Barcode.
Der öffentliche Schlüssel wird für jeden frei zugänglich gemacht und zum Beispiel in einer online Datenbank gespeichert.
Der Anwender prüft dann im Zuge der Produktvalidierung die Daten des Barcodes mit der enthaltenen Signatur und dem öffentlichen Schlüssel auf Korrektheit.
Somit kann jeder die Integrität und Authentizität des Barcodes kontrollieren.
Aufgrund der Analogie einer händischen Unterschrift spricht man von einer digitalen Signatur.
Allerdings wird hierbei keine traditionelle Unterschrift mit einem Stift auf einem Papier getätigt, sondern eine digitale Authentisierung der Daten vollzogen. 

Das gesamte Verfahren der digitalen Signatur umfasst laut des Bundesamtes für Sicherheit in der Informationstechnik (BSI) drei verschiedene Algorithmen:
\begin{enumerate}
\item \glqq Ein Algorithmus zur Generierung von Schlüsselpaaren.\grqq \cite[S. 46]{BSI-TR-02102-1}
\item  \glqq Eine Hashfunktion, die die zu signierenden Daten auf einen Datenblock fester Bitlänge abbildet.\grqq \cite[S. 46]{BSI-TR-02102-1}
\item \glqq Ein Algorithmus zum Signieren der gehashten Daten und ein Algorithmus zum Verifizieren der Signatur.\grqq \cite[S. 46]{BSI-TR-02102-1}
\end{enumerate}
Die beiden unter Punkt 3. genannten Algorithmen werden als nur ein einziger Algorithmus aufgefasst.
Dieser bietet dann sowohl die Funktionalität des Signierens von Daten als auch des Verifizierens einer Signatur.
Deshalb werden in dieser Arbeit, im Kontext der digitalen Signatur, stets nur drei verschiedene Algorithmen betrachtet.
Für jeden dieser Algorithmen gibt es eine Vielzahl an verschiedenen Berechnungsverfahren, welche wiederum sehr differente Ansätze als Grundlage haben und in ihren Ergebnissen unterschiedliche Eigenschaften aufweisen.
Entsprechend müssen vor der finalen Auswahl der Algorithmen zuerst die wichtigsten Kriterien erläutert werden.

\section{Kriterien für die Auswahl der Algorithmen} \label{subsection:KriterienDigitaleSignatur}
Bei der Entwicklung einer neuen Anwendung müssen zuerst die benötigten Kriterien, die von den Algorithmen der digitalen Signatur erfüllt werden sollen, definiert werden.
Anhand dieser lassen sich dann geeignete Algorithmen auswählen.
Im Rahmen der vorliegenden Arbeit ist die wichtigste Eigenschaft eine kurze Signaturlänge, da die Menge der auf dem Barcode gespeicherten Daten möglichst gering sein sollte.
Des Weiteren muss die verwendete digitale Signatur eine hohe Sicherheit bieten.
Denn sobald bei einem Hersteller die digitale Signatur imitiert oder sein privater Schlüssel anderweitig berechnet werden können, sind seine Produkte fälschbar.
Die Beurteilung des Grades der Sicherheit eines kryptographischen Verfahrens ist äußerst komplex und erfordert zahlreiche detaillierte Untersuchungen.
Deshalb werden in dieser Arbeit, im Kontext der digitalen Signatur, ausschließlich Algorithmen mit Empfehlungen des BSI oder des National Institute of Standards and Technology (NIST) verwendet.
Weitere Merkmale der algorithmischen Verfahren einer digitalen Signatur, wie zum Beispiel die Schlüssellänge, spielen für die vorliegende Arbeit nur eine nebensächliche Rolle.

\section{Elliptische Kurven und Hashfunktionen}\label{section:Elliptisch-Hash}
Bevor auf die Auswahl der eingesetzten Algorithmen eingegangen wird, ist die kurze Erklärung einiger mathematischer Grundbegriffe im Kontext der digitalen Signaturen notwendig.
Der Grundsatz des in dieser Arbeit eingesetzten asymmetrischen Verfahrens der Kryptographie beruht auf dem diskreten Logarithmus-Problem auf elliptischen Kurven, genannt Elliptic Curve Discrete Logarithm Problem (ECDLP).
Elliptische Kurven sind geometrische Objekte der Mathematik, auf denen die Eigenschaften der Addition definiert sind.
Das ECDLP wiederum basiert auf den Besonderheiten von Einwegfunktionen. 
\begin{definition}
\textbf{Einwegfunktion (nach  Lenze)}\\
\glqq Es seien $D, W$ zwei beliebige nicht leere Mengen und $f \colon D \to W$ sei eine injektive Funktion. Dann heißt $f$ \textbf{Einwegfunktion} [...], wenn gilt:
\begin{itemize}
\item $f(x)$  ist für alle   $x\in D$  \textbf{sehr effizient} berechenbar,
\item $f^{-1}(y)$  ist für alle   $y\in f(D)$  \textbf{sehr schwer} berechenbar.
\end{itemize}
Dabei bezeichnet $f(D)$ ,
\begin{description}
\item $f(D) \colon=\{ f(x) \mid x \in D \} \subseteq W$,
\end{description}
genau die Teilmenge des Wertebereichs   $W$, deren Elemente als Funktionswerte von   $f$  angenommen werden.\grqq \cite[S.254]{lenze2020einwegfunktionen}
\end{definition}
Einwegfunktionen sind also Funktionen, die sich effizient auswerten lassen, aber nur mit großem Aufwand invertierbar sind.
Wenn man nur $y$ kennt, dann ist es im ECDLP nahezu unmöglich, den Wert von $x$ zu bestimmen. 
Die einzige Chance auf Erfolg bietet nach aktuellem Stand lediglich ein Brute-Force-Angriff, also ein Durchprobieren aller möglichen Werte von $x$.
Das ist für die in dieser Arbeit verwendeten Parameter aber nicht sinnvoll durchführbar. \cite{lenze2020einwegfunktionen,swoboda2008kryptographie}

Eine digitale Signatur für die gesamten Daten anzufertigen, stellt in der Praxis einen enormen, umständlichen und überflüssigen (Rechen-)Aufwand dar.
Deshalb wird tatsächlich nur eine Art Fingerabdruck der ursprünglichen Daten als Eingabe für die digitale Signatur verwendet.
Zur Erstellung repräsentativer und einmaliger Fingerabdrücke werden Hashfunktionen verwendet. 
\begin{definition}
\textbf{Hashfunktion (nach Wätjen)}\\
\glqq Eine Hashfunktion ist eine Funktion $h$, welche die folgenden beiden Eigenschaften besitzt:
\begin{itemize}
\item $h$ bildet Eingaben $x$ einer beliebigen Bitlänge auf Ausgaben $h(x)$ einer festen Bitlänge ab. Es wird $h(x)$ auch Fingerabdruck (fingerprint) von $x$ genannt.
\item Es seien $x$ und $h$ gegeben. Dann ist $h(x)$ leicht zu berechnen.\grqq \cite[S.93]{watjen2018hashfunktionen}
\end{itemize}
\end{definition}
Da es ab einer gewissen Bitlänge der Ausgabe einer Hashfunktion praktisch unmöglich wird, für einen gegebenen Fingerabdruck $z$ einen Inputwert $x$ mit $h(x) = z$ zu berechnen, gelten Hashfunktionen auch als Einwegfunktionen \cite{watjen2018hashfunktionen}.

Eine tiefergehende Erklärung zu elliptischen Kurven, Hash- und Einwegfunktionen würde den Rahmen dieser Arbeit sprengen.
In der Literatur zu den jeweiligen Themen finden sich detailliertere Ausführungen \cite{werner2013elliptische,silverman2009arithmetic,koblitz1987elliptic,koblitz2012introduction,rosing1999implementing,watjen2018hashfunktionen}.

\section{Auswahl der Algorithmen} \label{subsection:AuswahlDerAlgorithmen}
Sowohl das BSI als auch das NIST publizieren regelmäßig Vorschläge, welche grundsätzlichen Kryptographischen Verfahren für eine digitale Signatur verwendet und welche Algorithmen dafür im Detail eingesetzt werden sollten.
Zwar bewerten beide Institute in ihren Publikationen nicht alle bekannten Verfahren und Algorithmen der digitalen Signatur, trotzdem bilden sie die am weitesten verbreiteten, besten erforschten und vermutlich sichersten ab.
Deshalb orientiert sich die vorliegende Arbeit an den Empfehlungen des BSI und NIST. 

Die verschiedenen Varianten der digitalen Signatur, die auf elliptischen Kurven basieren, unterscheiden sich primär in der Größe des Parameters $p$, welcher die Ordnung des verwendeten endlichen Körpers angibt.
Dabei muss gelten: $p \in \{192, 224, 256, 384, 512\}$.
Das BSI empfiehlt, aus Gründen der Sicherheit $p \geq 250$ zu wählen \cite{BSI-TR-02102-1}.
Da die Signaturlänge direkt von $p$ abhängt, wird folglich der kleinste, noch empfohlene Wert gewählt: $p = 256$.
Auf dieser Grundlage und unter Berücksichtigung der unter \ref{subsection:KriterienDigitaleSignatur} genannten Kriterien - möglichst kurze Signaturlänge und hohe Sicherheit - werden folgende drei Algorithmen für die digitale Signatur ausgewählt:

\begin{description}
\item[Algorithmus zur Generierung von Schlüsselpaaren: secp256r1 \cite{brown2010sec2}]
\hfill \\
Zwar benötigen die verwendeten Algorithmen noch weitere unter bestimmten Vorgaben frei wählbare, kryptographische Systemparameter, aber auf die tatsächliche Auswahl dieser wird bei der Entwicklung neuer Systeme meist verzichtet und stattdessen auf standardisierte Empfehlungen zurückgegriffen.
Falls nicht über Zusatzkonstrukte auf andere Programmiersprachen zurückgegriffen werden soll, stehen für die Entwicklung einer mobilen Applikation für Android ausschließlich die Programmiersprachen Kotlin und Java zur Verfügung.
Javas sogenannte Security Bibliothek bietet Möglichkeiten, die gewünschten Algorithmen nach Wahl von $p$ direkt in der App einzubinden.
Folglich muss sich nach der Entscheidung, welche algorithmischen Verfahren zum Einsatz kommen, nicht mehr um die tatsächliche Implementierung dieser gekümmert werden.
Da aber nicht alle bekannten kryptographischen Algorithmen der digitalen Signatur in Javas Security Bibliothek zur Verfügung stehen, wurde bei der finalen Wahl der Algorithmen auch auf die Verwendbarkeit in Java geachtet.
Für das Logarithmus-Problem auf elliptischen Kurven existieren verschiedene Standards zur Schlüsselgenerierung mit jeweils unterschiedlichen kryptographischen Systemparametern \cite{ANSIX9.62,FIPS186-4,FIPS186-5,brown2010sec2}.
Allerdings sind diese in den grundlegenden Eigenschaften - Signaturlänge und Sicherheitsniveau - gleichwertig und es bedarf daher keiner weiteren Unterscheidungen.
Die in Javas Security Bibliothek bereits bestehende Implementierung für den gewählten Algorithmus zur Erstellung von Schlüsselpaaren, secp256r1,  basiert dabei auf dem Standard SEC 2 \cite{brown2010sec2}.

\item[Hashfunktion: SHA256 \cite{FIPS180-4}]
\hfill \\
Bei der Auswahl der Hashfunktion ist vor allem auf die Bitlänge der Ausgabe zu achten.
Denn diese sollte im Idealfall gleich dem Parameter $p$ sein.
Da $p = 256$ gilt und die Empfehlungen des BSI und NIST berücksichtigt werden, wurde SHA256 als Hashfunktion ausgewählt \cite{FIPS180-4,BSI-TR-02102-1,FIPS186-4,FIPS186-5}.

\item[{\parbox[t]{\textwidth}{Algorithmus zum Erstellen und Verifizieren einer Signatur: Elliptic Curve Digital Signature Algorithm (ECDSA) \cite{BSI-TR-03111}}}]
\hfill \\
Im Vergleich zu anderen asymmetrischen kryptographischen Verfahren wie dem Digital Signature Algorithm (DSA) und dem nach seinen Entwicklern Rivest, Shamir und Adleman benannten RSA erzeugen Algorithmen, die auf elliptischen Kurven basieren, deutlich kürzere Signaturen bei gleichem Sicherheitsniveau.
Um beispielsweise ein Sicherheitsniveau von $120$ Bit zu erreichen, benötigen der DSA und RSA jeweils eine Schlüssellänge von $2800$ Bit, der ECDSA eine Schlüssellänge von $240$ Bit.
Die Länge der Signatur würde dann beim RSA $2800$ Bit betragen, hingegen beim DSA und ECDSA nur $480$ Bit. \cite{BSI-TR-03111, BSI-TR-02102-1, FIPS186-4, FIPS186-5}
Somit sind DSA und ECDSA dem RSA klar vorzuziehen.
Da auch ausgewählte öffentliche Schlüssel lokal auf dem Gerät des Anwenders gespeichert werden sollen, wird hier das Verfahren mit der kürzeren Schlüssellänge ausgewählt.
Daraus ergibt sich ein Vorteil bei der Verwendung des ECDSA gegenüber dem DSA.
Zwar existieren noch weitere Methoden, eine digitale Signatur zu erstellen, allerdings werden diese vom BSI \mbox{und/oder} NIST entweder als unsicher eingestuft oder aufgrund fehlender Popularität und geringer Verbreitung nicht begutachtet \cite{BSI-TR-03111, BSI-TR-02102-1, FIPS186-4, FIPS186-5}.
Folglich eignet sich im Zusammenhang der Konzeption und Umsetzung einer mobilen Applikation zur Validierung von fälschungssicheren Produktlabeln der ECDSA als Algorithmus zur Erstellung und Verifizierung von digitalen Signaturen am besten.
\end{description}
Aus der oben beschriebenen Wahl der Parameter und Algorithmen ergibt sich eine finale Schlüssellänge von 256 Bit, ein Sicherheitsniveau von 128 Bit und eine Signaturlänge von 512 Bit.
    \chapter{Auswahl des Barcodes}\label{chapter:auswahlBarcode}
Die als 3D-Punktwolke extrahierte digitale Version des Labels, die digitale Signatur und weitere Produktinformationen, wie zum Beispiel die Seriennummer, der Name des Herstellers und die exakte Produktbezeichnung, müssen mit dem Label auf dem vor Fälschung zu schützenden Produkt angebracht werden.
Denn im Zuge der Validierung des Produktlabels sollen diese Informationen mit einem Smartphone, unter Zuhilfenahme der mobilen Applikation, direkt offline, also ohne die Notwendigkeit einer bestehenden Netzwerkverbindung, auslesbar sein.
Somit wäre es etwa falsch, alle entsprechenden Daten auf einer Website zu hinterlegen.
Des Weiteren müssen die im Zuge der Speicherung und Anbringung der Daten entstehenden Zusatzkosten minimal sein.
Beispielsweise radio frequency identification (RFID)-Tags scheiden damit aufgrund ihres Preises aus.

Im Kontext dieser Arbeit sind Barcodes am besten zum Aufbewahren und späteren Auslesen der zusätzlichen Informationen geeignet.
Ein Barcode ist eine Maschinen-lesbare Repräsentation von Informationen.
Diese wird durch Kombination von stark und schwach reflektierenden Flächen geschaffen. 
Die einzelnen Flächen, meist schwarz und weiß, lassen sich dann zu \glq1\grq en und \glq0\grq en konvertieren.
Durch das Aneinanderreihen dieser ergibt sich dann eine Bitfolge, aus der die enthaltenen Daten berechnet werden können.
Ursprünglich wurden Informationen in einem Feld benachbarter Spalten als lineare eindimensionale (1D-) Barcodes gespeichert.
Da mittels 1D-Barcodes allerdings nur geringe Datenmengen kodiert werden können, wurden die Spalten durch kleine Flächen ersetzt.
Zum Beispiel durch Quadrate oder Kreise, die dann in mehr als nur einer Dimension angeordnet die Weiterentwicklung zu zweidimensionalen (2D-) Barcodes darstellen. \cite{kato2010barcodes, gao2007understanding}

\begin{figure}
  \centering
  \includegraphics[width=0.8\textwidth]{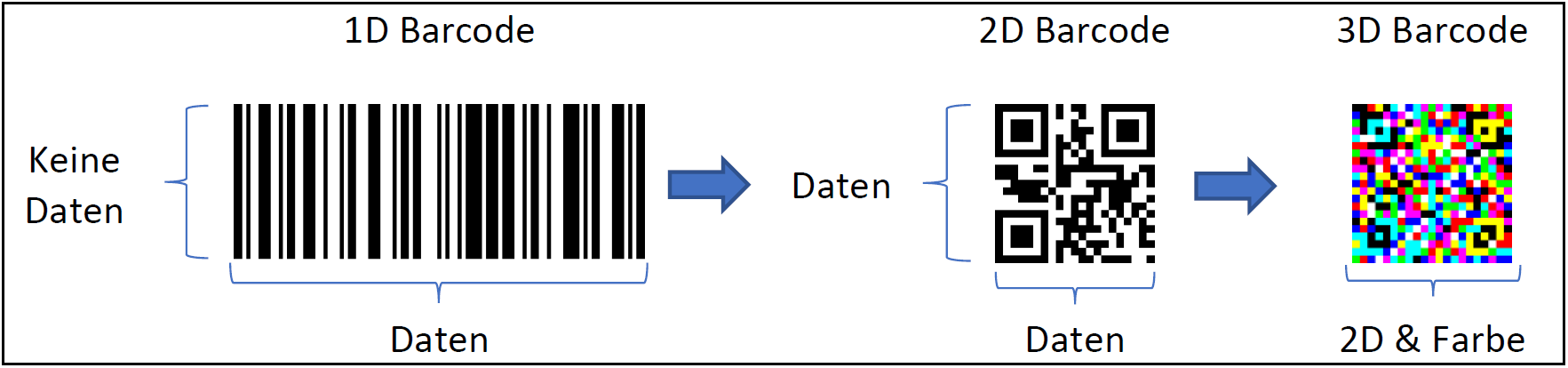}
  \caption{Entwicklung der Barcodes}
  \label{figure:EntwicklungBarcodes}
\end{figure}

Weil aber auch die traditionellen, meist schwarz-weißen, 2D-Barcodes nur eine begrenzte Datenmenge aufnehmen können, wurden diese zu bunten, also aus mehr als zwei Farben bestehenden, Barcodes weiterentwickelt.
Durch die Hinzunahme der Farben entsteht quasi eine dritte Dimension, weshalb die bunten Versionen auch als 3D-Barcodes bezeichnet werden \cite{uitz2012qr}.
In Abbildung \ref{figure:EntwicklungBarcodes} ist die Entwicklung der Barcodes symbolisch anhand ausgewählter Beispiele dargestellt.

Aufgrund der geringen Datenkapazität von 1D-Barcodes werden diese im Folgenden nicht weiter betrachtet und ausschließlich 2D- und 3D-Barcodes für die Überlegungen zur finalen Umsetzung herangezogen \cite{kato2010barcodes}.
Um den für dieses Projekt sinnvollsten Barcode zu bestimmen, werden erst in Kapitel \ref{section:BarcodeKriterien} die zu erfüllenden Kriterien definiert.
Anschließend wird anhand dieser Kriterien in Kapitel \ref{section:BesteBarcodes} jeweils der am ehesten in Frage kommende schwarz-weiße 2D- und bunte 3D-Barcode ermittelt.
Die beiden geeignetsten Kandidaten, Quick-Response-Code (QR-Code) und Just Another Bar Code (JAB Code), werden dann in Abschnitt \ref{section:VergleichJABvsQR} einem direkten Vergleich unterzogen.
Daraus resultierend ergibt sich, dass QR-Codes eine bessere Lösung als JAB Codes darstellen.
Allerdings wird sich dabei zeigen, dass der Ansatz, alle Daten in nur einem Barcode zu kodieren, in der Praxis nicht wie gewünscht umsetzbar ist.
Abschließend wird deshalb in Kapitel \ref{section:QRZweiteilung} das hier auftretende Hauptproblem von QR-Codes, der hohe Flächenbedarf, durch die Aufteilung der Daten in zwei einzelne QR-Codes gelöst.

\section{Barcode Kriterien}\label{section:BarcodeKriterien}
Zur Auswahl eines geeigneten Barcodes müssen vorerst die zu erfüllenden Kriterien festgelegt werden.
Dabei gilt, dass der verwendete Barcode mit einem standardmäßigen Smartphone unkompliziert und schnell lesbar sein muss.
Denn es soll kein spezielles Lesegerät, also eine zusätzliche Hardware, oder ein mehrminütiger Zeitaufwand zum Auslesen nötig sein.
Zudem soll der finale Ausdruck des Barcodes nur wenige Quadratzentimeter groß sein.
Das hat den simplen Grund, dass auf einigen Produkten nicht mehr Fläche zur Verfügung steht.
Auch sollen die Kosten zur Erstellung neuer Barcodes möglichst gering sein.
Da Barcodes aber mit handelsüblichen Druckern auf Papier gedruckt werden können und es sich zudem nur um kleine Flächen handelt, wird dieser Aspekt in der weiteren Betrachtung vernachlässigt.
Außerdem muss für die Verwendung der Barcodes bereits eine frei verfügbare Implementierung vorhanden sein.
Denn im Zuge dieser Arbeit sollen keine neuen Algorithmen zum Erstellen und Auslesen von Barcodes entwickelt werden.
Des Weiteren muss die Datenkapazität des Barcodes groß genug sein, um alle nötigen Informationen aufnehmen zu können.
Fast alle Barcodes, die uns im Alltag begegnen, ob auf Lebensmittelverpackungen, Werbeanzeigen oder Eintrittskarten, kodieren nur einen sehr kleinen Datensatz.
Meist handelt es sich dabei um eine Seriennummer, die Adresse einer Webseite oder einen kurzen Text.
Im Fall der Produktlabel müssen allerdings deutlich größere Datenmengen mit einem Barcode kodiert werden.
Neben dem Label müssen folgende Daten mit auf dem Produkt angebracht werden: 
\begin{itemize}
    \item Repräsentation der 3D-Punktwolke zur eindeutigen Beschreibung des Labels
    \item Fehlerbereich jedes Punktes
    \item Produktinformationen wie der Name des Herstellers, die Seriennummer und die exakte Produktbezeichnung
    \item Digitale Signatur
\end{itemize}
Um eine Aussage über die benötigte Datenkapazität des Barcodes treffen zu können, werden im Folgenden grobe Abschätzungen zur jeweiligen Datenmenge gegeben.

\begin{description}
\item [3D-Punktwolke]\hfill \\
Zur anschaulicheren Erklärung sei der Raum im Folgenden durch ein dreidimensionales Koordinatensystem mit den Achsen $x$, $y$, und $z$ beschrieben.
Die Position eines Punktes der 3D-Punktwolke kann also durch das Tripel $(x,y,z)$ repräsentiert werden. 
Das gesamte Label ist in $x$- und $y$-Dimension maximal einen Millimeter (mm) und in $z$-Dimension maximal $0,1~mm$ groß \cite{Ruehrmair2021}.
Die Positionierungen der Goldnanokügelchen beziehungsweise -stäbchen können auf Nanometer (nm) genau gemessen werden \cite{ZachMarin2021}.
Da sich alle Elemente des Labels innerhalb dieses befinden, sind diese in $x$- und $y$-Dimension maximal einen Millimeter, in $z$-Dimension maximal $0,1~mm$, voneinander entfernt.
Folglich bedarf es zur exakten Repräsentation eines Punktes sechs Ziffern für die $x$- und $y$-Koordinate sowie fünf Ziffern für die $z$-Koordinate, denn $1~mm = 1*10^{6}~nm$.
Daraus ergeben sich für eine Punktwolke, welche aus $p$ Punkten besteht $(6 + 6 + 5) * p = 17*p$ mit dem Barcode zu kodierende Ziffern.
Für die Komplexität der Punktwolke kann angenommen werden, dass diese im Normalfall aus circa $50$ Punkten besteht \cite{Ruehrmair2021}.
Wobei intuitiv gilt, je größer $p$ ist, desto schwerer ist es, das Label zu duplizieren.
Um auf eine zukünftig erwünschte Erhöhung des Sicherheitsniveaus vorbereitet zu sein, werden in dieser Arbeit 3D-Punktwolken verwendet, die aus bis zu $100$ Punkten bestehen.

\item [Fehlerbereich]
Die Messung eines Punktes erfolgt zwar auf Nanometer genau, allerdings kommt es trotzdem zu Messungenauigkeiten.
Zu jedem Punkt wird dabei ein Fehlerbereich angegeben.
Dieser liegt durchschnittlich im niedrigen zweistelligen Nanometerbereich. \cite{ZachMarin2021}
Im Kontext dieser Arbeit wurde festgelegt, für den Fehlerbereich stets zwei Ziffern pro Punkt und Koordinate zu speichern \cite{Ruehrmair2021}.

\item[Produktinformationen]
Wie groß die benötigte Datenmenge für die Produktinformationen in der Praxis sein wird, ist bisher noch nicht absehbar.
Deshalb werden hierfür im folgenden Platzhalter-Texte mit einer Länge von $400$ bis $900$ alphanumerischen Zeichen verwendet.

\item[Digitale Signatur]
Die Länge der digitalen Signatur beträgt in der Theorie, wie in \ref{subsection:AuswahlDerAlgorithmen} beschrieben, $512$ Bit ($64$ Byte).
Um die digitale Signatur in der Praxis aber auch verwenden zu können, wird diese nach der Erstellung standardmäßig noch um weitere Daten ergänzt.
Hierbei handelt es sich zum Beispiel um Informationen über den verwendeten Algorithmus und die dazugehörigen Parameter.
Das ist ein nötiger Schritt, welcher den Daten zur digitalen Signatur weitere sechs bis acht Bytes hinzufügt.
Folglich müssen im Zusammenhang der digitalen Signatur $70$ bis $72$ Byte mit dem Barcode kodiert werden.
\end{description}

\section{Auswahl des besten 2D- und 3D-Barcodes} \label{section:BesteBarcodes}
Sowohl 2D- als auch 3D-Barcodes weisen untereinander jeweils eine direkt wahrnehmbare Ähnlichkeit auf.
Dabei bestehen die erstgenannten nur aus zwei Farben, meist schwarz-weiß, die zweitgenannten stets aus mehr als zwei Farben, meist sogar aus vier bis neun Farben.
Innerhalb der jeweiligen Kategorie sind viele Eigenschaften der Barcodes gleichartig.
Im Vergleich mit Barcodes der anderen Gruppe sind allerdings häufig große Unterschiede zu erkennen.
Daher wird im Folgenden je ein 2D-Barcode, Abschnitt \ref{section:Bester2DBarcode}, und ein 3D-Barcode, Abschnitt \ref{section:Bester3DBarcode}, ausgewählt, die jeweils am besten für die Verwendung im Kontext dieser Arbeit geeignet sind.

\subsection{Auswahl des besten 2D-Barcodes} \label{section:Bester2DBarcode}
\begin{figure}
  \centering
  \includegraphics[width=0.8\textwidth]{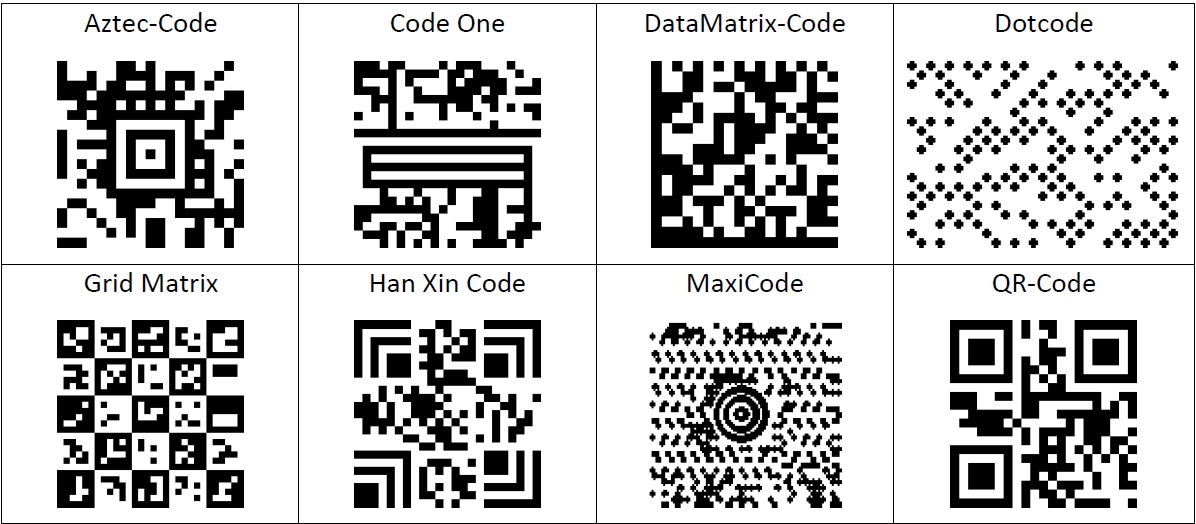}
  \caption{Beispiele 2D-Barcodes}
  \label{figure:BeispieleZweifarbigerBarcodes}
\end{figure}

Zwar existieren viele 2D-Barcodes, wie einige ausgewählte Beispiele in Abbildung \ref{figure:BeispieleZweifarbigerBarcodes} belegen; für die Verwendung im Kontext fälschungssicherer Produktlabel eignen sich allerdings nur wenige. 
Häufig entwickeln und optimieren Unternehmen neue 2D-Barcodes, die primär für den Einsatz in einem spezifischen Fall gedacht sind.
Dies führt dazu, dass erstens zahlreiche Barcodes ausschließlich für bestimmte Personengruppen zugänglich, nur mit speziellen Lesegeräten auslesbar oder Lizenz-pflichtig sind, und dass zweites bei der Verwendung der Barcodes mit andersartigen Eingabe-Daten häufig positive Eigenschaften verloren gehen.
Darüber hinaus existieren bei einigen Barcodes keine öffentlichen Vorschläge für die Implementierung der benötigten Algorithmen zur Generierung und zum Auslesen. 

Letztendlich erfüllen der DataMatrix-Code, standardisiert in \cite{ISODataMatrix}, Aztec-Code, standardisiert in \cite{ISOAztecCode}, und QR-Code, standardisiert in \cite{ISOQRCode}, die in Kapitel \ref{section:BarcodeKriterien} geforderten Kriterien am ehesten.
Von diesen weisen QR-Codes mit bis zu $23648$ Bits die höchste maximal kodierbare Datenmenge auf und werden deshalb als am geeignetsten betrachtet \cite{tiwari2016introduction, warasart2012based}.
Zudem sind QR-Codes bereits erfolgreich in der Praxis erprobt, zum Beispiel in Bereichen des mobilen Marketings, der Online-Werbung und der elektronischen Tickets und Zahlungen \cite{dorado2016mobile, uitz2012qr}.
Dementsprechend sind viele Personen mit der Verwendung von QR-Codes vertraut, wodurch es vermutlich zu weniger Anwenderfehlern beim Auslesen der Daten kommen wird \cite{tiwari2016introduction}.

\subsection{Auswahl des besten 3D-Barcodes} \label{section:Bester3DBarcode}
Die Entwicklung neuer 3D-Barcodes erfolgt im Gegensatz zu ihren zweidimensionalen Vorgängern mehrheitlich im Kontext wissenschaftlicher Arbeiten.
Dabei ist häufig die maximal kodierbare Datenmenge oder die Datendichte von 2D-Barcodes zu gering, so dass deshalb ein neuer 3D-Barcode entwickelt wird.
In der Literatur finden sich einige vielversprechende Vorschläge für 3D-Barcodes \cite{querini2014reliability, grillo2010high, yang2016towards, yang2018robust, melgar2012cqr, melgar2016high, melgar2019high}.
Allerdings scheinen diese meist nur als theoretische Argumentationsgrundlage zu dienen.
Häufig erfolgt keine sinnvolle praktische Umsetzung.
Es existieren weder Anwendungen zum Generieren noch zum Auslesen solcher 3D-Barcodes.
Derartige Anwendungen wären beispielsweise Webseiten oder mobile Applikationen mittels welcher ein praxisbezogener Test möglich wäre.
Entsprechend fehlen auch die jeweils nötigen Algorithmen zur Verwendung der 3D-Barcodes.

Ein 3D-Barcode, der auch in der Praxis umgesetzt und verwendet wurde, ist der bereits 2007 von Microsoft Research entwickelte High Capacity Color Barcode (HCCB) \cite{microsoftHCCB,parikh2008localization}. 
Zwar war die Verwendung des HCCBs sowohl nur für spezielle Anwendungen vorgesehen als auch ausschließlich mit einer bestehenden Verbindung zu einer online Datenbank möglich, aber es wurde immerhin eine App zum Lesen der Barcodes bereitgestellt \cite{Querini2DColor,yang2018robust}.
Allerdings wurde die Weiterentwicklung des Barcodes inzwischen beendet und die App wieder vom Markt genommen.
Es finden sich keine öffentlichen Implementierungen für die Verwendung des HCCBs.
Folglich eignet sich der HCCB nicht für die Nutzung in dieser Arbeit.

\begin{figure}
  \centering
  \includegraphics[width=2cm,height=2cm,keepaspectratio]{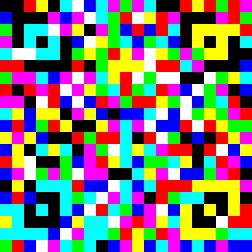}
  \caption{Beispiel eines JAB Codes}
  \label{figure:BeispielJABCode}
\end{figure}

Ein weiteres Beispiel eines 3D-Barcodes ist der 2018 vom Fraunhofer-Institut für Sichere Informationstechnologie entworfene, in Abbildung \ref{figure:BeispielJABCode} beispielhaft dargestellte, JAB Code \cite{berchtold2020jab}.
Hierbei handelt es sich um eine neue und junge Technologie, die im Auftrag des BSI entwickelt wurde.
Sowohl die Algorithmen zur Erstellung neuer als auch die Algorithmen zum Auslesen bereits bestehender JAB Codes sind öffentlich verfügbar. \cite{BSI-JAB}
Des Weiteren stellen die Entwickler sowohl eine Webseite als auch eine mobile Applikation bereit, mit denen die Generierung und das Auslesen der JAB Codes erprobt werden kann \cite{jabcode.org,jabcodeGit}.
Zudem erfüllen JAB Codes in der Theorie alle in Kapitel \ref{section:BarcodeKriterien} definierten Kriterien und verfügen darüber hinaus sogar noch über eine höhere Datenkapazität und Datendichte als traditionelle 2D-Barcodes.

\section{Vergleich zwischen QR- und JAB Code}\label{section:VergleichJABvsQR}
Vermeintlich scheinen JAB Codes für die Nutzung im Kontext dieser Arbeit am besten geeignet, da sie in der theoretischen Untersuchung alle anderen in Frage kommenden Barcodes überragen.
Sie verfügen laut der Entwickler über eine ungefähr dreimal höhere Datendichte als QR-Codes \cite{berchtold2020jab}.
Das heißt, mit einem JAB Code können auf gleicher Fläche in etwa dreimal so viele Daten kodiert werden.
Das macht intuitiv auch Sinn, denn durch die Hinzunahme der Farben als eine quasi dritte Dimension erscheinen JAB Codes ihren 2D Vorgängern überlegen.
Für die exakt gleichen Daten als Eingabe ergibt sich daraus in der Theorie folgende These: Ein JAB Code kann im Vergleich zu einem QR-Code in kleinerer Größe gedruckt und anschließend ausgelesen werden.

Um diese These zu untersuchen, wurde ein ausführlicher, anwendungsorientierter Vergleich vollzogen.
Dazu wurden mehrere beispielhafte Datensätze auf Grundlage der tatsächlichen Produktlabel, wie in \ref{section:BarcodeKriterien} beschrieben, erstellt.
Dabei blieben die digitale Signatur und die zusätzlichen Produktinformationen während des gesamten Vergleichs unverändert, einzig die Größe der 3D-Punktwolke, welche aus $p$ Punkten besteht, gestaltete sich variabel. 
Die Datensätze bildeten dann die Grundlage zur Generierung mehrerer JAB und QR-Codes.

Bis zu einem gewissen Grad können beschädigte JAB und QR-Codes noch erfolgreich dekodiert werden.
Dieser Grad, das Fehlerkorrektur-Level, kann bei der Erstellung selbständig gewählt werden.
Je höher dieser ist, desto geringer ist allerdings die maximal kodierbare Datenmenge. \cite{ISOQRCode, BSI-JAB}
Dies liegt auf der Hand, da zur potenziellen Fehlerkorrektur redundante Informationen gespeichert werden müssen.
Im Kontext dieser Arbeit werden nur JAB und QR-Codes mit dem jeweils niedrigsten Fehlerkorrektur-Level erstellt.
Dadurch wird bei beiden Barcodes die maximale Datenkapazität erreicht.

Zum direkten Vergleich wurden JAB Codes, genau wie QR-Codes, in der standardmäßig quadratischen Form erstellt.
Anschließend wurden die Barcodes mit dem Drucker Aficio MP C3002 der Firma Ricoh, mit einer Qualität von $1200$ Punkten pro Zoll (engl. dots per inch) auf handelsüblichem weißen Papier in verschiedenen Größen mit Seitenlängen zwischen $10$ und $0,5~cm$, ausgedruckt. 

Das Auslesen und Dekodieren von QR-Codes lässt sich bei modernen Smartphones direkt mit der integrierten Kamera, ohne zusätzliche Software, durchführen.
Zwar gibt es auch noch eine Vielzahl weiterer Methoden zur Dekodierung von QR-Codes, aber diese spiegelt die gewünschte Umsetzung in der finalen Applikation zur Validierung der Produktlabel am besten wider.
Bei JAB Codes gibt es erstens die Möglichkeit, diese mit einer von den Entwicklern bereitgestellten App direkt mit dem Smartphone auszulesen \cite{jabcodeGit}, oder zweitens ein Bild des JAB Codes aufzunehmen und dieses dann auf einer von den Entwicklern ebenfalls zur Verfügung gestellten Webseite auswerten zu lassen \cite{jabcode.org}.
Wobei mittels der zweiten Methode deutlich bessere Ergebnisse erzielt wurden.
Zudem hat es sich als hilfreich erwiesen, wenn der gesamte Bildinhalt nur aus dem JAB Code bestand und somit nichts anderes auf dem Bild zu sehen war.
Beispielsweise wurde das Bild nach der Aufnahme händisch zugeschnitten, um tatsächlich nur den JAB Code zu enthalten.
Da für den Anwender kein zusätzlicher Aufwand entstehen soll, ist das manuelle Zuschneiden im Zuge der Validierung von fälschungssicheren Produktlabeln zwar nicht erwünscht, weil sich der Vergleich jedoch auf die Untersuchung der Dekodierbarkeit der Barcodes konzentriert wird diese Problematik toleriert.
Das Auslesen der Barcodes wurde mit drei verschiedenen Smartphones, Samsung Galaxy S20 FE, Samsung Galaxy S9 und Huawai P30 vollzogen.
Trotz unterschiedlich qualitativer Kameras wurden dabei allerdings keine signifikanten Differenzen wahrgenommen.

Da sowohl QR-Codes als auch JAB Codes mit zunehmender Datenmenge an Komplexität gewinnen, also aus immer mehr Reihen und Spalten bestehen, stellt sich die Frage, bis zu welcher minimalen Größe die Barcodes druck- und auslesbar sind.
Denn eines der in Kapitel \ref{section:BarcodeKriterien} beschriebenen Kriterien ist, dass nur eine Fläche von wenigen Quadratzentimetern für den Barcode verwendet werden soll.
Zur Ermittlung der Grenze, wie klein die Barcodes gedruckt werden können und weiterhin trotzdem noch dekodierbar sind, wurde für alle Datensätze jeweils der kleinste noch auslesbare Barcode eruiert.
Die Daten der Punktwolke machen dabei im Vergleich zu den übrigen Informationen den größten Teil des gesamten zu kodierenden Datensatzes aus. 
Die Zunahme der Komplexität der Barcodes lässt sich in etwa linear zur Größe von $p$, der Anzahl der Punkte in einer Punktwolke, beschreiben.
Damit kann die minimale Druckgröße der jeweiligen Barcodes in Abhängigkeit von $p$ angegeben werden.
Für JAB Codes gilt, dass die Seitenlänge mindestens $\frac{p}{10}~cm$ lang sein muss.
Also bei einer Punktwolke, die aus nur $30$ Punkten besteht, muss der JAB Code schon mindestens $3~cm*3~cm$ groß sein.
JAB Codes mit Seitenlängen von unter $\frac{p}{10}~cm$ zeigten sich als so gut wie nicht mehr dekodierbar.
Zudem mussten selbst für JAB Codes mit einer Seitenlänge von mehr als $\frac{p}{10}~cm$ mehrere Versuche zur Dekodierung unternommen werden.
Folglich bedurfte es pro Barcode eines mehrminütigen Zeitaufwands zum Auslesen der Daten.
Da dies offensichtlich nicht benutzerfreundlich ist, empfiehlt sich der Grenze von $\frac{p}{10}~cm$ sogar noch hinzuzufügen, dass der JAB Code eine Seitenlänge von $5~cm$ niemals unterschreiten sollte.
Benutzerfreundlichkeit bedeutet hierbei, dass es ohne großen Aufwand und relativ einfach möglich sein sollte, in wenigen Sekunden den jeweiligen Barcode auszulesen.
Selbst für $p = 30$ sollte der JAB Code also mindestens in einer Größe von $5~cm*5~cm$ gedruckt werden.

QR-Codes erzielten im Vergleich deutlich bessere Ergebnisse.
Hier gilt als Grenze, innerhalb derer sie noch benutzerfreundlich dekodierbar sind, $\frac{p}{10}~cm$ pro Seite.
QR-Codes dieser Größe ließen sich in wenigen Sekunden mit der integrierten Kamera des Smartphones auslesen.
Die Grenze des maximal Dekodierbaren beträgt bei QR-Codes, unter optimalen Bedingungen und mit Vernachlässigung der Benutzerfreundlichkeit, sogar $\frac{p}{20}~cm$ pro Seitenlänge.
QR-Codes unterhalb dieser Grenze konnten mit dem bloßen Auge schon als vermutlich nicht dekodierbar wahrgenommen werden, da es sich nicht mehr um klare Kanten und eindeutig voneinander getrennte schwarz-weiße Quadrate handelte.
Die Druckqualität nimmt bei zu kleiner Größe entscheidend ab und stellt entsprechend den limitierenden Faktor dar. 

Auch bei JAB Codes nimmt die Druckqualität ab einer gewissen Größe merklich ab.
Dabei wirkte es, als würden die Grenzen zwischen den einzelnen unterschiedlich farbigen Quadraten verschwimmen.
Allerdings ist die Druckqualität hier nicht der alleinige limitierende Faktor.
Denn selbst JAB Codes, die scheinbar eine ausreichende Qualität mit voneinander abtrennbaren farbigen Quadraten aufwiesen, ließen sich teilweise nicht mehr dekodieren.
Möglicherweise liegt es daran, dass es sich bei JAB Codes um eine vergleichsweise junge Technologie handelt und die Algorithmen noch nicht final ausgreift sind.
Falls mit fortschreitender Zeit und steigendem Bekanntheitsgrad die Algorithmen zur Erstellung, Erkennung und Dekodierung weiterentwickelt werden, bieten JAB Codes eine vielversprechende Option.
Bisher erfüllen JAB Codes die im Kontext dieser Arbeit geforderten Kriterien jedoch nicht zufriedenstellend. 

Zusammenfassend gilt, dass QR-Codes im praktischen Vergleich mit den in der Theorie überlegenen JAB Codes wider Erwarten bessere Ergebnisse erzielten.
Womöglich liegt dies an bereits optimierten Algorithmen oder den höheren Kontrasten zwischen den jeweils eingesetzten Farben.
Alle weiteren in Kapitel \ref{section:BarcodeKriterien} definierten Kriterien erfüllen beide Barcodes in etwa gleichermaßen.
Dementsprechend bedarf es dahingehend keiner detaillierten Unterscheidung.
Folglich eignen sich QR-Codes im Kontext dieses beschriebenen Anwendungsfalls am besten.

\section{QR-Code Zweiteilung}\label{section:QRZweiteilung}
Wie sich gezeigt hat, erfüllen QR-Codes die in Kapitel \ref{section:BarcodeKriterien} beschriebenen Kriterien besser als alle anderen untersuchten Barcodes.
Allerdings gilt trotzdem, dass die Barcodes für eine benutzerfreundliche Verwendung mindestens mit einer Größe von $\frac{p}{10}~cm$ Seitenlänge gedruckt werden müssen.
Die Kodierung einer Punktwolke mit $p = 50$ Punkten würde also eine Fläche von $5~cm * 5~cm = 25~cm^2$ beanspruchen.
Das ist für die Verwendung im Kontext fälschungssicherer Produktlabel zu groß, und es muss eine Lösung mit einem geringeren Flächenbedarf gefunden werden.

Die in einem QR-Code gespeicherte Information ist eine Bitfolge, die aus den Eingabedaten generiert wird.
Je länger diese ist, umso größer muss der QR-Code ausgedruckt werden.
Daher gilt es, die Daten mit einer möglichst kurzen Bitfolge darzustellen. 

Für die Erstellung der QR-Codes und damit für die Generierung der Bitfolge aus den Eingabedaten stehen verschiedene Modi zur Verfügung.
Von diesen sind im Kontext der vorliegenden Arbeit zwei von Interesse.
Zum einen der nummerische Modus, bei dem die Eingabedaten nur aus Ziffern bestehen, wobei je drei zu einem zehn Bit langen Block zusammengefasst werden.
Zum anderen ein Modus, bei dem die Daten neben Großbuchstaben, Ziffern und wenigen ausgewählten weiteren Symbolen auch andere Zeichen, wie zum Beispiel Kleinbuchstaben, enthalten. \cite{ISOQRCode}
Dabei wird aus den Daten nach ISO/IEC 8859-1 \cite{ISOLatin1}, auch bekannt als Latin-1, eine Bitfolge generiert und jedes Zeichen durch eine Folge von acht Bits beschrieben.
Weitere Modi spielen im Kontext dieser Arbeit keine Rolle, weil sie für die Entwicklung der mobilen Applikation nicht verwendbar sind oder das zugrundeliegende Alphabet der Daten ungeeignet ist.
Der eingesetzte Modus wird dabei stets zur Kodierung jedes einzelnen Zeichens der gesamten Eingabedaten verwendet.
Da die hier zustande kommenden Datensätze nur dem zweiten Modus zugeordnet werden können, weil beispielsweise Kleinbuchstaben in den Produktinformationen enthalten sind, wird für jedes Zeichen eine Bitfolge der Länge acht generiert.
Dies ist für einen Teil der Daten allerdings äußerst ineffizient.
Die Informationen der 3D-Punktwolke und die dazugehörigen Fehlerbereiche machen stets den größten Teil des gesamten Datensatzes aus.
Und obwohl diese nur aus Ziffern bestehen, werden sie trotzdem nicht im effizienteren Modus kodiert.
Dadurch ist die resultierende Bitfolge dieses Teils der Daten um fast $60~\%$ länger als sie im nummerischen Modus wäre.

Das Problem der ungünstigen Kodierung des Großteils der Informationen lässt sich jedoch erfolgreich umgehen.
Dazu werden statt nur eines alle Daten enthaltenen QR-Codes zwei QR-Codes erstellt.
Der erste QR-Code erhält bei der Erstellung die Informationen der 3D-Punktwolke und die Fehlerbereiche.
Dadurch wird für den größeren Anteil der Daten die Kodierung im effizienten nummerischen Modus vollzogen.
Für den zweiten QR-Code bleibt der Modus der Kodierung zwar unverändert, dafür wird für diesen mit den Produktinformationen nur ein kleiner Anteil der Daten als Eingabe verwendet.
Die digitale Signatur lässt sich den Eingabedaten beider QR-Codes effizient hinzufügen.
Um diese allerdings nicht zu zertrennen, wird sie mit Blick auf die Größe des Ausdrucks dem QR-Code angehängt, der bereits die Produktinformationen enthält. 

Die Zweiteilung der Daten hat zur Folge, dass die beiden resultierenden Bitfolgen jeweils deutlich kürzer sind und der Flächenbedarf dadurch verringert wird.
Für den ersten QR-Code bestehend aus den Daten der Punktwolke und den dazugehörigen Fehlerbereichen gilt, dass dieser für $p$ Punkte mit einer minimalen Seitenlänge von bis zu $\frac{p}{100}~cm$ gedruckt werden kann.
Um allerdings eine gewisse Benutzerfreundlichkeit zu erreichen, empfiehlt es sich, diesen wenigstens mit einer Seitenlänge von $\frac{p}{60}~cm$ und in einer Mindestgröße von $1~cm * 1~cm$ zu drucken.
Im Folgenden sei $z$ die Anzahl der Zeichen der Produktinformationen.
In der Praxis hat sich gezeigt, dass der zweite QR-Code, bestehend aus den Produktinformationen und der digitalen Signatur, in etwa mit einer Seitenlänge von $\frac{z}{800}~cm$ gedruckt und anschließend erfolgreich ausgelesen werden kann.
Allerdings sollten aus Gründen der Benutzerfreundlichkeit auch bei diesem die Maße erhöht werden.
Empfehlenswert erscheint hierbei eine Seitenlänge von $\frac{z}{500}~cm$ und zudem eine Mindestgröße von $~1cm * 1~cm$ zu sein.

Für den standardmäßig zu erwartenden Fall gilt, dass beide QR-Codes jeweils auf einer Fläche von $1~cm * 1~cm$ gedruckt und anschließend erfolgreich ausgelesen werden können.
Folglich wurde der Flächenbedarf für eine Punktwolke mit $50$ Punkten von inakzeptablen $25~cm^2$ auf akzeptable $2~cm^2$ gesenkt.
Durch die Aufteilung der Daten auf zwei QR-Codes erhält man zudem Spielraum für etwaige zukünftige Erhöhungen der Datenmengen.
Denn bisher benötigen beide Datensätze jeweils nur circa $30~\%$ der maximalen Datenkapazität der QR-Codes.

Zwar muss der Benutzer zur Validierung des Produktlabels zwei QR-Codes auslesen, aber aufgrund der enormen Flächenersparnis wird dies in Kauf genommen.
Zudem ist das Dekodieren von QR-Codes allgemein mit sehr wenig Aufwand verbunden und dem Benutzer entsprechend zumutbar.
    \chapter{Erklärung zur Implementierung und Verwendung der mobilen Applikation} \label{chapter:EntwicklungDerApp}
\begin{figure}[htb]
  \centering
  \includegraphics[width=0.32\textwidth]{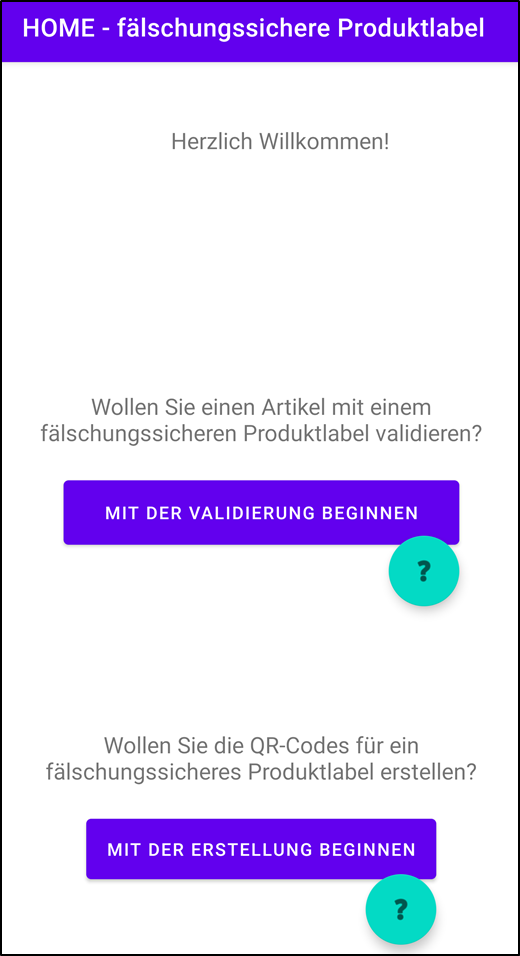}
  \caption{Bildschirmaufnahme der entwickelten mobilen Applikation. Dem Anwender wird zu Beginn je Anwendungsfall eine Auswahlmöglichkeiten geboten.}
  \label{figure:Screen-Home}
\end{figure}

Nachdem in Kapitel \ref{chapter:unfälschbarkeitProduktInfos} bereits die Notwendigkeit der Verwendung einer digitalen Signatur aufgezeigt und zudem in Abschnitt \ref{section:QRZweiteilung} die Zweiteilung der QR-Codes als Lösung für die Speicherung der Referenzdaten neben dem Label vorgestellt wurde, soll in diesem Kapitel die tatsächliche Entwicklung und Verwendung der mobilen Applikation erläutert werden.
Da es sich, wie bereits in Kapitel \ref{chapter:Einleitung} erklärt, bei der Umsetzung der App ausschließlich um einen Machbarkeitsnachweis handelt wird diese zum einen nur für Android entworfen (die Entwicklung für weitere Betriebssysteme ist für das angestrebte Ziel nicht nötig), zum anderen sind nicht alle der zukünftig geplanten Funktionalitäten Teil der Implementierung.
Deshalb erfolgt zuerst in Kapitel \ref{section:FehlendeFunktionalitäten} eine Erklärung, um welche Funktionalitäten es sich handelt, wie ihre zukünftige Umsetzung geplant ist und wie mit ihrem Auslassen in der Implementierung umgegangen wird.
Anschließend werden die zwei unterschiedlichen Szenarien der mobilen Applikation, erstens die Sicherung eines neuen Produktes durch den Hersteller in \ref{section:SicherungEigenerProdukte}, und zweitens die Validierung eines fälschungssicheren Produktlabels in \ref{section:ValidierungBestehenderProduktlabel}, symbolisch durch die Benutzung eines Anwenders vorgestellt und bei Bedarf mit zusätzlichen Implementierungsdetails ergänzt.
Zur Veranschaulichung der beiden Anwendungsfälle dient Abbildung \ref{figure:Screen-Home}.
Hierbei handelt es sich um eine Bildschirmaufnahme der App.
Dem Anwender wird zu Beginn die Wahl zwischen den zwei Anwendungsfällen gelassen.

\section{Ausgelassene Funktionalitäten}\label{section:FehlendeFunktionalitäten}
Im Folgenden werden zwei für den praktischen Einsatz des Verfahrens der Produktauthentifizierung zwar essenzielle, bisher allerdings nicht verwendbare respektive überflüssige Funktionalitäten vorgestellt.
Dabei handelt es sich erstens um das Scannen der Punktwolke eines physischen Produktlabels (Abschnitt \ref{subsection:ScannenLabel}) und zweitens um die Einrichtung einer Public-Key-Infrastruktur (PKI) (Abschnitt \ref{subsection:EntwicklungPKI}).
Beide sind bisher noch nicht Teil der mobilen Applikation.
Trotzdem werden die Resultate der Verwendung dieser bereits benötigt.
Deshalb wird im Folgenden ein kurzer Überblick über diese Funktionalitäten gegeben und unter anderem dargestellt, wie mit ihrer Vakanz umgegangen wird.

\subsection{Scannen der Produktlabel}\label{subsection:ScannenLabel}
Wie bereits erklärt, können die im Produktlabel enthaltenen Goldnanokügelchen bzw. -stäbchen in digitaler Form durch eine 3D-Punktwolke repräsentiert werden.
Um die Eigenschaften des Produktlabels auszulesen und die benötigten Informationen zu extrahieren, soll zukünftig eine von Marin neue, dafür speziell entwickelte Technologie eingesetzt werden \cite{ZachMarin2021}.
Allerdings lag diese, genauso wie physische Produktlabel, zum Zeitpunkt der Entwicklung der App nicht vor.
Folglich wird anstelle tatsächlicher Messungen der Label mit synthetischen Datensätzen gearbeitet.
Für den bisherigen Stand der Entwicklung der Validierung von fälschungssicheren Produktlabeln ist das jedoch nebensächlich, da die App in dieser Arbeit vor allem als ein konzeptioneller Nachweis der Umsetzbarkeit des algorithmischen Verfahrens dienen soll.
Statt der Vermessung eines realen Labels mit der Smartphone-Kamera wird dem Anwender die Möglichkeit gegeben, synthetische Daten aus externer Quelle zu importieren.
Eine detaillierte Erläuterung zur Erstellung der künstlichen Datensätze findet sich in Kapitel \ref{chapter:Evaluation}.

\subsection{Entwicklung einer Public-Key-Infrastruktur}\label{subsection:EntwicklungPKI}
In Kapitel \ref{chapter:unfälschbarkeitProduktInfos} wurde die bedeutende Rolle der digitalen Signatur im Kontext der Produktauthentifizierung dargestellt.
Um diese aber tatsächlich auch wie in der Theorie geplant verwenden zu können, bedarf es weiterer Komponenten zusätzlich zu dem in Kapitel \ref{subsection:AuswahlDerAlgorithmen} beschriebenen Signaturverfahren. 
Denn falls es einem Produktfälscher gelänge, dass der Prüfer des Produktes und damit der Prüfer der Signatur anstelle des vom Hersteller des Originals generierten öffentlichen Schlüssels den öffentlichen Schlüssel des Fälschers zur Prüfung der Signatur verwendet, würde der Prüfer alle Produkte des Fälschers positiv verifizieren.
Es ist offensichtlich, dass dies eine große Gefahr darstellt und dadurch das gesamte Verfahren des Echtheitsnachweises überflüssig werden würde. 

Um zu verhindern, dass derartige Täuschungsversuche erfolgreich sind, werden sogenannte Zertifikate eingesetzt.
Diese verbinden insbesondere den Namen des Schlüsselinhabers, hier also den Namen des Herstellers eines Produktes, mit seinem öffentlichen Schlüssel zu einer festen Einheit.
Ein Zertifikat wird dann wiederum von einer vertrauenswürdigen Zertifizierungsinstanz signiert.
Damit ist sichergestellt, dass die Einheit, der Name des Schlüsselinhabers und sein öffentlicher Schlüssel korrekt gebildet wurden.
In der Praxis wird neben der Zertifizierungsinstanz noch eine Reihe weiterer Komponenten benötigt, um digitale Signaturen, wie theoretisch angedacht, einsetzen zu können.
Die Gesamtheit aller benötigten Komponenten und die damit einhergehenden Prozesse wird als Public-Key-Infrastruktur bezeichnet. \cite{BSI-GrundlagenDigSig}
Die Erschaffung einer passenden PKI ist laut des BSI allerdings eine aufwendige und anspruchsvolle Aufgabe, welche nur unter der Aufsicht entsprechender Sicherheitsexperten geschehen sollte \cite{BSI-TR-02102-1}.
Da die Entwicklung einer PKI allerdings sowohl den Rahmen der vorliegenden Arbeit sprengen würde als auch vor allem für den konzeptionellen Nachweis der Validierung von fälschungssicheren Produktlabeln nicht von Interesse ist, wird dies Aufgabe zukünftiger Projekte sein.
Wenn es sich im Folgenden um das Signieren der Daten und die Verifikation einer Signatur handelt, wird folglich nur symbolisch eine lokal auf dem Smartphone gespeicherte Datei verwendet, die alle benötigten Schlüssel enthält.

\section{Sicherung eigener Produkte vor Fälschung} \label{section:SicherungEigenerProdukte}
\begin{figure}[htb]
  \centering
  \includegraphics[width=0.32\textwidth]{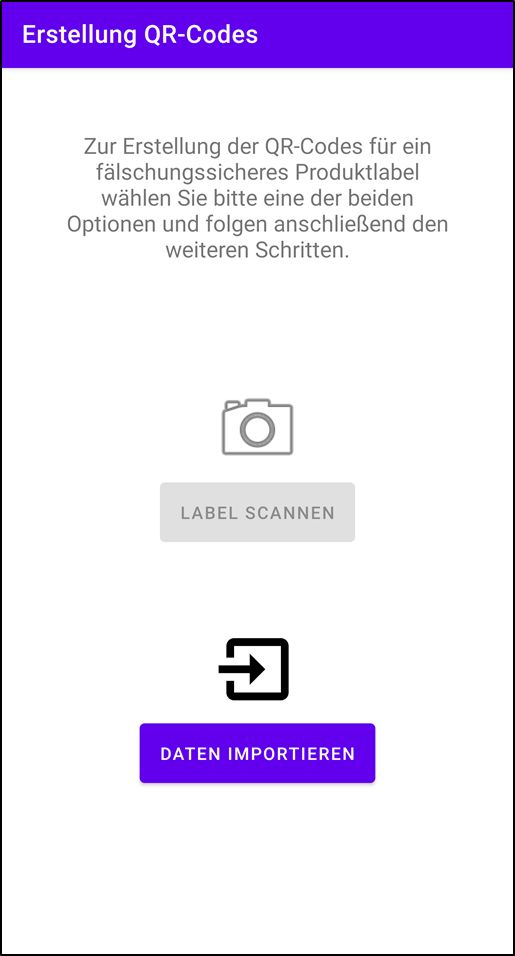}
  \caption{Bildschirmaufnahme der entwickelten mobilen Applikation. Der Anwender hat auf Grund fehlender Technologien nur die Möglichkeit die benötigten Daten zu importieren.}
  \label{figure:Screen-Import}
\end{figure}

Ein Hersteller beginnt den Prozess der Sicherung seines Produktes damit, auf diesem ein fälschungssicheres Label anzubringen.
Ob dieses selbstständig erzeugt oder von einer anderen Quelle bezogen wird, spielt für die vorliegende Arbeit keine Rolle.
Nach der Anbringung des Labels sollen dessen Eigenschaften zukünftig mit der integrierten Kamera des Smartphones als 3D-Punktwolke extrahiert werden.
Da dies aber, wie oben unter \ref{subsection:ScannenLabel} beschrieben, bisher nicht umsetzbar ist, wird dem Anwender stattdessen die Möglichkeit gegeben, eine lokal gespeicherte Datei, welche die Informationen der Punktwolke enthält, auszuwählen.
Die Abbildung \ref{figure:Screen-Import} stellt die entsprechende Umsetzung in der mobilen Applikation dar.
Die graue, nicht funktionsfähige Schaltfläche zum Scannen eines Labels drückt das Fehlen der entsprechenden Technologie aus.
Für die Daten der Punktwolke, deren dazugehörigen Fehlerbereiche und die Produktinformationen wird dann mit dem privaten Schlüssel des Herstellers eine digitale Signatur erstellt.
Die Signatur und Produktinformationen wie auch die Punktwolke und deren Fehlerbereiche werden dann, wie in Abschnitt \ref{section:QRZweiteilung} beschrieben, mit je einem QR-Code kodiert.
Nachdem diese automatisch auf dem Smartphone gespeichert wurden, erhält der Anwender abschließend eine Anweisung zur korrekten Anbringung der QR-Codes auf dem Produkt.

\section{Validierung bestehender Produktlabel} \label{section:ValidierungBestehenderProduktlabel}
Der zweite Anwendungsfall der mobilen Applikation ist die Validierung eines Produktlabels, also die Überprüfung des Ursprungs eines Produktes.
Hierbei müssen zuerst die beiden QR-Codes dekodiert werden.
Dazu wurde der App, unter Verwendung der Kamera des Smartphones, die Funktionalität des Extrahierens der Informationen eines QR-Codes hinzugefügt.
Es bedarf also weder zusätzlicher Hardware, wie zum Beispiel eines Warenscanners, noch zusätzlicher Software.

Nach der Dekodierung der beiden QR-Codes wird unmittelbar die Korrektheit der digitalen Signatur überprüft.
Auch hier wird, wie in Abschnitt \ref{subsection:EntwicklungPKI} dargestellt, der benötigte öffentliche Schlüssel des Herstellers nur symbolisch einer lokalen Datei entnommen.
Falls die Verifikation der Signatur negativ ist, wird die Echtheitsuntersuchung des Produktes umgehend abgebrochen und dem Anwender mitgeteilt, dass der angegebene Hersteller nicht der Ersteller der Signatur ist und es sich bei dem Produkt folglich um ein Imitat handelt.

Liefert die Überprüfung der digitalen Signatur allerdings ein positives Ergebnis, wird der Anwender im nächsten Schritt dazu aufgefordert, die Eigenschaften des Produktlabels auszulesen.
Da auch hier wieder, wie in Abschnitt \ref{subsection:ScannenLabel} erläutert, die entsprechende Technologie nicht verfügbar ist, wird dem Anwender die Möglichkeit geboten, eine Datei, die die Punktwolke der Referenzmessung enthält, auszuwählen.
Enthalten die beiden Punktwolken ähnlich viele Elemente, wird anschließend mit einer in der Programmiersprache Kotlin \cite{samuel2017programming} verfassten Version des CPDs die Messung so gut wie möglich auf die Referenz abgebildet.
Würde die Anzahl der Punkte zwischen Referenz und Messung nicht verglichen, könnte das Verfahren theoretisch manipuliert werden.
Denn dem Label könnten maximal viele Goldnanokügelchen respektive -stäbchen hinzugefügt werden, sodass mit hoher Wahrscheinlichkeit jeder Punkt der Referenz im Label vorhanden ist.

Wie erläutert, existiert durch die Aufteilung des Raums in sogenannte Unterwürfel nicht nur ein CPD, der ein Ergebnis berechnet, sondern $27$ voneinander unabhängige CPDs, die entsprechend $27$ unterschiedliche Resultate erzielen.
Damit sich dieser Zuwachs an Berechnungen nicht in einem enormen Anstieg der Laufzeiten widerspiegelt, werden so viele CPDs wie möglich parallel ausgeführt.
Die unpräzise Aussage \textit{\glqq so viele CPDs wie möglich\grqq{}} beruht auf den unterschiedlichen Hardware-Eigenschaften der verschiedenen Smartphones, die bedingen, dass der Parallelisierungsgrad variiert.

\begin{figure}[b!]
  \begin{subfigure}{0.45\textwidth}
  \centering
    \includegraphics[width=0.72\textwidth]{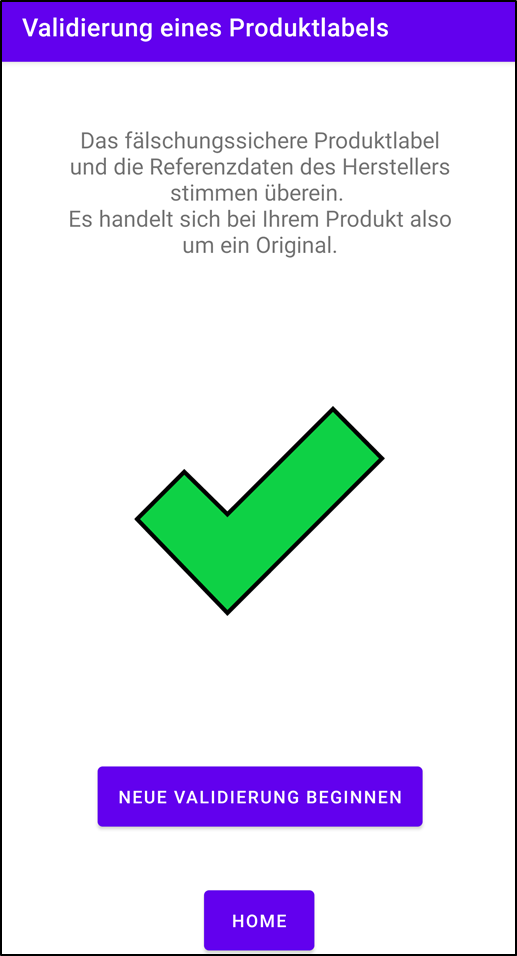}
    \subcaption{Positives Resultat}
    \label{figure:Screen-Pos}
  \end{subfigure}
  \hspace{0.1\textwidth}
  \begin{subfigure}{0.45\textwidth}
   \centering
    \includegraphics[width=0.72\textwidth]{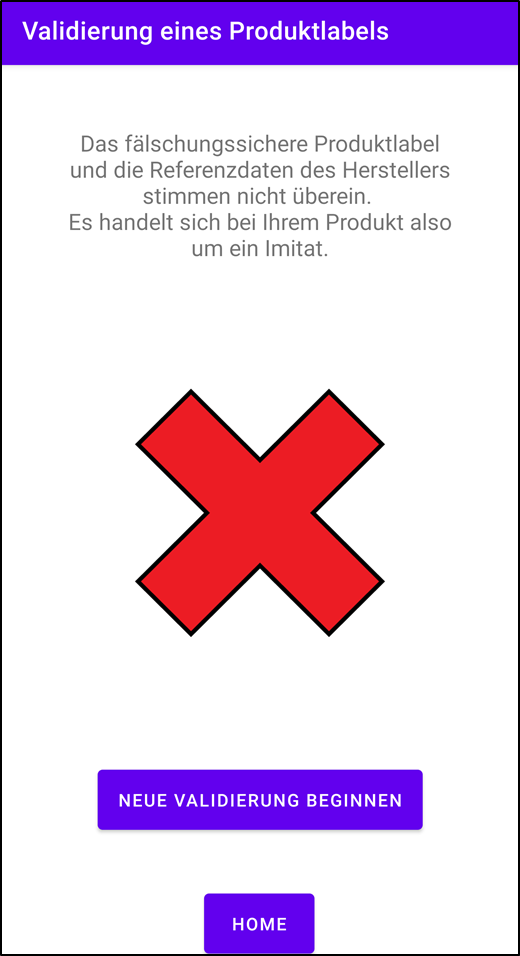}
    \subcaption{Negatives Resultat}
    \label{figure:Screen-Neg}
  \end{subfigure}
  \caption{Zwei Bildschirmaufnahmen der entwickelten mobilen Applikation. Dem Anwender wird jeweils das Resultat der Validierung eines Produktlabels angezeigt.}
  \label{figure:Screen-Ergebnisse}
\end{figure}

Wie erläutert, ist das Ergebnis des Punktwolkenvergleichs eine Prozentzahl der korrekt zugeordneten Punkte.
Um den Anwender nicht mit der Bewertung dieses Resultats zu konfrontieren, wird ein Schwellenwert definiert und in Abhängigkeit zu diesem ein klares \textit{\glq Ja\grq{}} oder \textit{\glq Nein\grq{}} als Antwort auf die Frage, ob es sich bei dem Produkt um ein Original handelt, gegeben.
Die Abbildung \ref{figure:Screen-Ergebnisse} zeigt jeweils eine Aufnahme des Bildschirms der entwickelten App für die beiden möglichen Fälle.
Dabei ist in \ref{figure:Screen-Pos} ein positives, und in \ref{figure:Screen-Neg} ein negatives Resultat der Validierung eines Produktlabels zu sehen.

Für eine Zusammenfassung des beschriebenen Verfahrens zur Ermittlung der Gleichheit zweier Punktwolken sei der in Pseudocode dargestellte Algorithmus \ref{algorithm:Ermittlung-Gleichheit} im Anhang der vorliegenden Arbeit zu betrachten.

Welcher Wert sich im finalen praktischen Einsatz der Validierung von fälschungssicheren Produktlabeln als Schwelle zwischen \textit{\glq Gleich\grq{}} und \textit{\glq Ungleich\grq{}} eignet, wird sich erst mit dem tatsächlichen Testen mehrerer physischer Label zeigen.
Anhand der Tests mit synthetischen Daten (siehe Kapitel \ref{chapter:Evaluation}) lässt sich erkennen, dass dabei insbesondere die Anzahl verlorener Punkte, die Messungenauigkeiten je Punkt sowie die Genauigkeit einer potenziellen Fälschung eine entscheidende Rolle spielen werden.

    \chapter{Ergebnisse und Evaluation}\label{chapter:Evaluation}
Da sowohl die Verwendung der zwei QR-Codes, wie in Kapitel \ref{chapter:auswahlBarcode} aufgezeigt, als auch der Einsatz der digitalen Signatur, wie in Kapitel \ref{chapter:unfälschbarkeitProduktInfos} erläutert, in sich geschlossene, eigenständige und ihren geplanten Zweck erfüllende Themen darstellen werden zu diesen keine weiteren Ergebnisse präsentiert.
Zudem handelt es sich bei beiden ausschließlich um ergänzende und insbesondere auch in gewissen Maßen austauschbare Technologien.
Hingegen stellt das Produktlabel und damit verbunden der Vergleich von 3D-Punktwolken im Kontext dieser Arbeit den unverzichtbaren Kern des Verfahrens dar.

Nachdem in den vorherigen Kapiteln sowohl die verwendeten Algorithmen (Abschnitt \ref{section:Grundlagen}) als auch die Umsetzung dieser in der mobilen Applikation (Kapitel \ref{chapter:EntwicklungDerApp}) erläutert wurden, werden nun die Ergebnisse des Punktwolkenvergleichs evaluiert.
Zwar wurde in der dieser Arbeit vorangegangenen Untersuchung (siehe \cite{Lankheit2020})
bereits teilweise eine Evaluation des CPDs vorgenommen, allerdings diente diese primär dem Vergleich mit anderen Algorithmen, so dass essenzielle Kernaussagen bezüglich der Resultate nicht ausreichend beleuchtet wurden.
Außerdem weisen die Testdaten, wie in Abschnitt \ref{section:EinordnungErgebnisse} aufgezeigt, feine, aber doch relevante Unterschiede auf.
Aufgrund dieser zwei Sachverhalte und weiterer neuer Erkenntnisse bezüglich der physischen Produktlabel ist die Evaluation des Vergleichs der zwei Punktwolken notwendig. 

Dementsprechend wird im Folgenden in Abschnitt \ref{section:Testumgebung} zunächst die Testumgebung vorgestellt.
Anschließend findet eine schrittweise Annäherung der Punktwolken an den in der Praxis zu erwartenden Normalfall statt.
Damit verbunden erfolgt die Untersuchung der grundlegenden Fragestellung, ob Gleiches als gleich erkannt wird.
Hierbei werden zuerst in Abschnitt \ref{section:ohneZusatz} die Ergebnisse unter Laborbedingungen aufgezeigt.
Im zweiten Schritt werden der Messung in Abschnitt \ref{section:art+lostPoints} Artefakte und verlorene Punkte hinzugefügt und deren Einfluss auf das Resultat dargestellt.
Der abschließende Schritt in Richtung zum erwarteten Normalfall ist dann in Abschnitt \ref{section:UngenauePositionierung} das Hinzufügen von Messungenauigkeiten. 

Um auch die zweite grundlegende Fragestellung, ob Ungleiches tatsächlich als ungleich identifiziert wird, zu beantworten, werden in Abschnitt \ref{section:UngleichePunktwolken} die Ergebnisse des CPDs bei Eingabe zweier unterschiedlicher Punktwolken analysiert.

Im Kontext der ersten Fragestellung, ob Gleiches als solches erkannt wird, stellen Referenz und Messung jeweils eine digitalisierte Version des gleichen physischen Produktlabels dar.
Deshalb ist bis hier hin die Aussage, dass beide Punktwolken ursprünglich gleich sind, gerechtfertigt.
Da aber aufgrund von Artefakten, verlorenen Punkten und verrauschten Messungen die Punktwolken der Referenz und Messung tatsächlich keineswegs gleich, sondern maximal ähnlich zueinander sind, ergibt sich zusätzlich die angepasste Fragestellung, bis zu welchem Grad ähnliches als vermeintlich gleich angesehen wird.
Dazu wird den Testdaten in Abschnitt \ref{section:Fälschungen} neben der standardmäßigen Verrauschung eine zusätzliche Verfälschung je Punkt hinzugefügt.
Zwar stellt dies im Detail nur eine andersartige Messungenauigkeit dar, kann aber vor allem  als ein Fälschungsversuch angesehen werden, bei dem ein Produktfälscher das Produktlabel mit einer gewissen Ungenauigkeit nachbildet.

Abschließend wird in Abschnitt \ref{section:Zeitanalyse} mit der Analyse der Laufzeit des Punktwolkenvergleichs die Frage geklärt, ob sich das eingesetzte Verfahren und die damit verbundene Verwendung der Algorithmen derart in einer mobilen Applikation umsetzen lassen, dass nach maximal einer Sekunde die benötigten Ergebnisse vorliegen.

\section{Testumgebung} \label{section:Testumgebung}
Wie bereits dargestellt, liegen während der Anfertigung dieser Arbeit keine physischen Produktlabel vor und es muss stattdessen mit künstlich erzeugten Daten gearbeitet werden. 
Immerhin können die synthetischen Testdaten auf Grundlage einiger von Marin bereitgestellter Messungen realer Label erstellt werden \cite{Lankheit2020}.
Ein Testdatum besteht im Folgenden aus zwei 3D-Punktwolken: erstens der Referenz $X$ und zweitens der Messung $Y$.

Für die Referenz gilt, dass alle Punkte in einem Quader, also dem Produktlabel, zufällig verteilt sind.
Die Seitenlängen des Quaders sind hierbei (jeweils in Nanometern) $10^6$ entlang der $x$- und $y$-Achse, sowie $10^5$ entlang der $z$-Achse \cite{Ruehrmair2021}.
Zudem wird für jeden Punkt ein Fehlerbereich definiert.
Dieser Fehlerbereich, zusammengesetzt aus dem Fehlerradius je Achse, wird dabei für jeden Punkt einer Gauß'schen Normalverteilung entnommen, mit Erwartungswert ${\mu \approx (6,5,5)}$ und Standardabweichung ${\sigma \approx (10,8,8)}$ für Kügelchen sowie ${\mu \approx (22,19,43)}$ und ${\sigma \approx (11,10,21)}$ für Stäbchen.
Auch diese Werte sind wieder in Nanometern zu interpretieren und wurden mit der Bedingung, dass je Koordinate nur ein zweistelliger Fehlerwert existiert, aus den bereitgestellten Messungen physischer Label berechnet \cite{Ruehrmair2021}.
Um Erkenntnisse darüber zu erlangen, wie viele Kügelchen beziehungsweise Stäbchen ein Produktlabel tatsächlich enthalten sollte, werden Referenzen mit ${\abs{ X } \in \{25,30,35,40,45,50,60,75,100\}}$ für Kügelchen sowie ${\abs{ X } \in \{24,30,34,40,44,50,60,74,100\}}$ für Stäbchen erstellt.
Dies sollte den Bereich ausreichend abdecken, der in der Praxis als sinnvoll erscheint.
Der kleine Unterschied beider Mengen entsteht dadurch, dass ein Stäbchen in digitaler Version immer durch zwei Punkte vertreten wird und somit eine Punktwolke, die eine Repräsentation von Stäbchen darstellt, nur eine gerade Anzahl an Punkten enthalten kann.

Den zweiten Teil eines Testdatums, die Messung, bildet eine transformierte Version der Referenz.
Die Transformation besteht hierbei stets aus einer Rotation und einer Translation.
Da ein Anwender des Verfahrens allerdings in der Lage sein dürfte, die Kamera des Smartphones in etwa korrekt ausgerichtet über dem Label zu platzieren, wird die Rotation in der weiteren Untersuchung auf maximal $20^\circ$ je Achse beschränkt.
Im weiteren Verlauf der Evaluation des Punktwolkenvergleichs werden je nach Sonderfall noch weitere Veränderungen an der Messung vorgenommen, um die praktischen Gegebenheiten präziser nachzubilden.

Um möglichst exakte Erkenntnisse über die realen Anwendungsfälle der Validierung von fälschungssicheren Produktlabeln zu erhalten, wird stets eine Vielzahl an verschiedenen Referenzen mit jeweils einer dazu gehörigen Messung erstellt und evaluiert.
Dies symbolisiert das einmalige Validieren von unterschiedlichen Produktlabeln.
Da die Authentizitätsprüfung eines Produkts allerdings meistens nicht nur einmal, sondern mehrfach erfolgt - ein Geldschein wechselt beispielsweise häufig seinen Besitzer - wird zudem für die Referenzen auch eine Menge unterschiedlicher Messungen angefertigt und untersucht.

Zur Bewertung, wie gut die Ergebnisse im Mittel sind, wird im Folgenden stets der Median verwendet.

\section{Punktwolkenvergleich unter Laborbedingungen}\label{section:ohneZusatz}
Bevor die erzeugten Testdaten in den nachfolgenden Kapiteln noch mehr dem in der Praxis zu erwartenden Fall angeglichen werden, wird zunächst der Nachweis erbracht, dass der CPD unter Laborbedingungen im Stande ist, annähernd optimale Ergebnisse zu erzielen.
Laborbedingungen bedeuten hierbei, dass die Messung ausschließlich eine rotiert und linear verschobene Version der Referenz ist. 

Zwar werden sowohl Label, die Goldnanokügelchen, als auch Label, die Goldnanostäbchen enthalten, in digitalisierter Form durch eine 3D-Punktwolke dargestellt, allerdings werden sich im weiteren Verlauf der Evaluation leichte Unterschiede zwischen den beiden Varianten zeigen.
Selbst unter Laborbedingungen lassen sich die Differenzen bereits erkennen.
Zur Untersuchung wurden hierbei jeweils für Stäbchen und Kügelchen zu allen, unter \ref{section:Testumgebung} genannten, neun verschiedenen Größen von $\abs{X}$ je zehn verschiedene synthetische Messungen zu zehn unterschiedlichen Referenzen erstellt.
Insgesamt wurden also je Label-Variante 900 Tests durchgeführt. 

Bei Punktwolken auf Kügelchen-Basis konnte der CPD die Transformation für $Y$ in $99~\%$ der Testfälle derart präzise berechnen, dass sich abschließend alle Punkte im Fehlerbereich ihres Partners befanden.
Allerdings konnte bei den verbliebenen $1~\%$ der Testfälle kein einziger Punkt korrekt im Fehlerbereich des Partners abgebildet werden.
In solchen Fällen bleibt der CPD meist in lokalen Minima hängen, wodurch unter anderem die Punktepaare-Bildung fehlerhaft verläuft.
Dadurch befindet sich bei der abschließenden Abstandsmessung keiner der Punkte im Fehlerbereich des vermeintlichen Partners.
Auch wenn das hier nur schwach zu erkennen ist, zeigt sich bereits, dass die Ergebnisse meist um zwei Zentren verteilt sind.
Entweder es werden viele, hier sogar alle, Punkte korrekt ihrem Partner zugeordnet und befinden sich auch im jeweiligen Fehlerbereich oder die Zuordnung schlägt gänzlich fehl und es befinden sich nur wenige, hier sogar gar keine, Punkte in ausreichender Nähe ihres Partners.
Das macht intuitiv auch Sinn. 
Denn der CPD geht nicht nach dem Prinzip vor, so viele Punkte wie möglich ihrem Partner korrekt zuzuordnen, sondern versucht, stets die gesamte Punktwolke $Y$, also alle Punkte der Wolke gleichermaßen, auf die Punktwolke $X$ abzubilden.
Dadurch entsteht die Verteilung der Ergebnisse um die zwei Zentren.

Für Punktwolken basierend auf Stäbchen wird sich dieses Phänomen in den folgenden Kapiteln zwar auch zeigen, unter Laborbedingungen ist es allerdings noch nicht zu beobachten.
Dafür lässt sich hier bereits eine weitere Eigenschaft erkennen: Stäbchen liefern konstantere Ergebnisse als Kügelchen.
Es befinden sich zwar seltener alle Punkte im Fehlerbereich des vermuteten Partners, selbst unter Laborbedingungen ist das nur in $38~\%$ aller Tests der Fall, aber es existieren dafür auch deutlich weniger Ausreißer, also Punktwolken, bei denen kaum Punkte im Fehlerbereich des Partners gefunden wurden.
Das Testdatum mit den niedrigsten prozentual erkannten Punkten hatte immer noch einen Wert von $88~\%$.

\begin{figure}[b!]
  \centering
  \fbox{
  \includegraphics[width=0.9\textwidth]{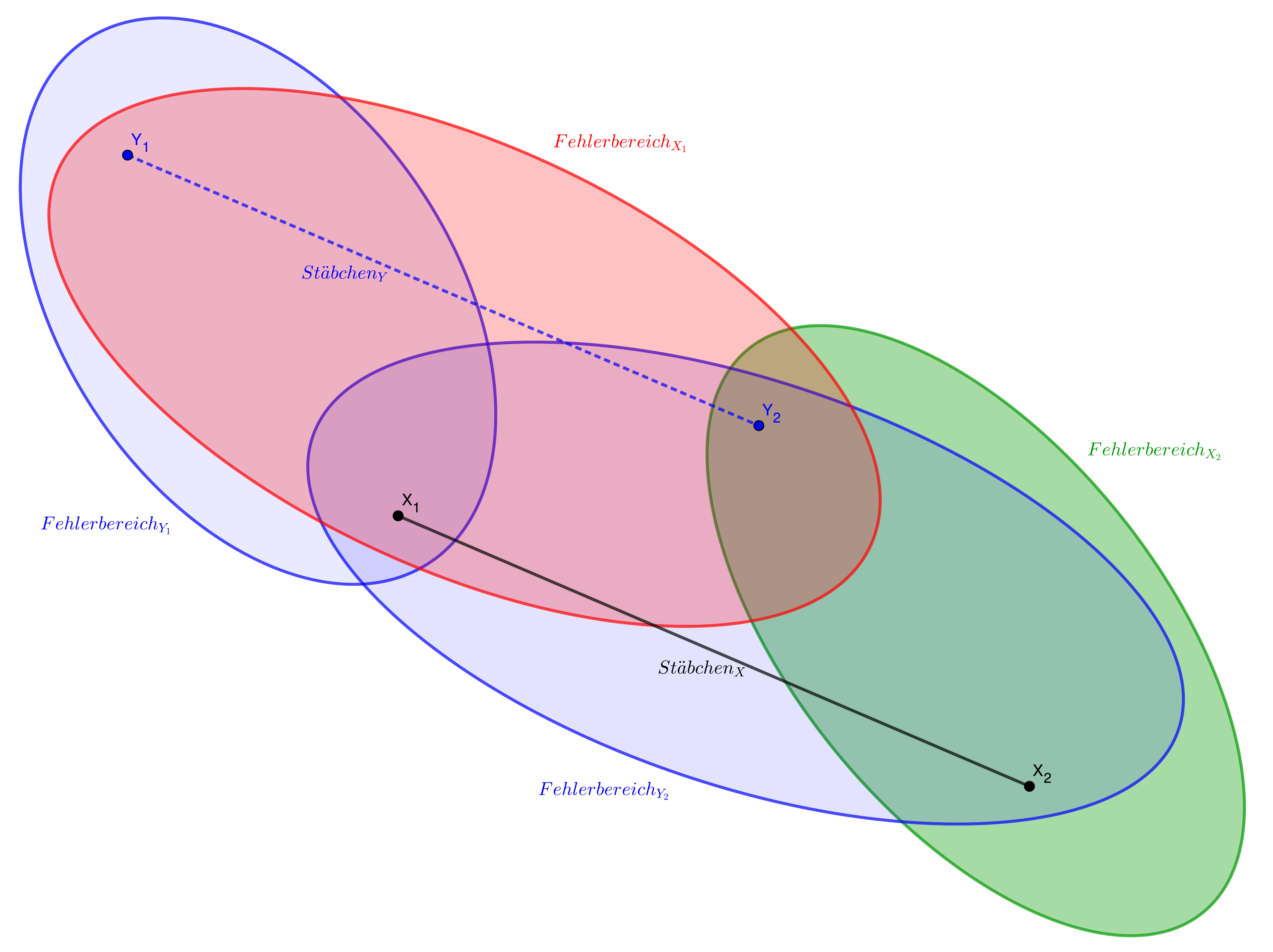}
  }
  \caption{Zweidimensionale symbolische Veranschaulichung eines Stäbchens und dessen transformierter Messung sowie der Fehlerbereiche aller vier Punkte.}
  \label{figure:StäbchenVeranschaulichung}
\end{figure}

Eine Erklärung, weshalb selbst bei sehr guten Ergebnissen nicht immer $100~\%$ der Punkte korrekt erkannt werden (der Mittelwert liegt hier auch nur bei $98~\%$) liefern die besonderen Eigenschaften der Stäbchen.
Dadurch dass die zwei Punkte, welche ein Stäbchen repräsentieren, zueinander meist einen deutlich geringeren Abstand aufweisen als zu allen anderen Punkten der Punktwolke, bildet der CPD in manchen Fällen falsche Korrespondenzpaare, sprich er irrt sich bei der Zuordnung der einzelnen Punkte eines Stäbchens.
Für eine entsprechende visuelle zweidimensionale Veranschaulichung sei die Abbildung \ref{figure:StäbchenVeranschaulichung} zu betrachten.
Hierbei handelt es sich um eine symbolische Darstellung, bei welcher insbesondere die einzelnen Flächen nur der Anschaulichkeit dienen und in der Praxis leicht abweichende Charakteristika aufweisen.
In schwarz ist das physische Stäbchen mit seinen zwei Endpunkten $X_{1}$ und $X_{2}$ dargestellt, in blau die nach Ausführung des CPDs berechneten transformierten Punkte $Y_{1}$ und $Y_{2}$ mit ihren jeweiligen Fehlerbereichen.
Die roten und grünen Flächen stehen hierbei symbolisch für die Fehlerbereiche von $X_{1}$ und $X_{2}$.
Da sich in einer Punktwolke noch weitere Punkte befinden, die das Resultat beeinflussen, kann es vorkommen, dass der CPD Ergebnisse liefert, bei denen einzelne transformierte Punkte zwar gut ausgerichtet wurden, aber trotzdem leicht verschoben zu ihren ursprünglichen Partnern sind.
Dieses Szenario ist in der Abbildung \ref{figure:StäbchenVeranschaulichung} ebenfalls sinnbildlich dargestellt.
Nach der Ausführung des CPDs wird, wie in Kapitel \ref{chapter:Validierung} beschrieben, für jeden Punkt der Referenz, hier also für $X_{1}$ und $X_{2}$, ein Partner-Punkt der transformierten Messung ermittelt.
Zwar erfolgt dies beim CPD auf Basis berechneter Wahrscheinlichkeiten, der Einfachheit halber soll hier nun aber symbolisch der euklidische Abstand zweier Punkte als Korrespondenzkriterium dienen.
Es würde damit in diesem Beispiel für beide Punkte der Referenz jeweils der Punkt $Y_{2}$ als Partner ermittelt werden.
Denn sein Abstand ist sowohl von $X_{1}$ als auch von $X_{2}$ aus betrachtet im Vergleich zu den anderen Punkten der Messung am geringsten.
Die Paare, bei denen abschließend untersucht wird, ob beide Punkte im Fehlerbereich des jeweils anderen liegen, lauten folglich: $(X_{1},Y_{2})$ und $(X_{2},Y_{2})$.
Obwohl die Untersuchung bei beiden Paaren ein positives Ergebnis liefert, kann offensichtlich nur eines der beiden zur finalen Zählung der korrekten Punkte-Erkennungen verwendet werden.
Andernfalls würde ein und derselbe Punkt der Messung unterschiedlichen Punkten der Referenz zugeordnet werden.
Dies darf aus offensichtlichen Gründen aber nicht passieren.
In dieser beispielhaften Darstellung würde final also einer der Punkte nicht korrekt erkannt werden und das Ergebnis des Punktwolkenvergleichs entsprechend keine $100~\%$ aufweisen, obwohl dies mit der Paarbildung  $(X_{1},Y_{1})$ und  $(X_{2},Y_{2})$ möglich wäre.
So kann es vorkommen, dass selbst bei scheinbar guten ermittelten Transformationen nicht alle Punkte korrekt zugeordnet werden.

\section{Artefakte und verlorene Punkte}\label{section:art+lostPoints}
Um die Testumgebung einen Schritt mehr den tatsächlichen Gegebenheiten anzupassen, fließen in die Datensätze nun sowohl Artefakte als auch verlorene Punkte mit ein.
In Messungen physischer Label beträgt der Anteil an Artefakten und verlorenen Punkten laut bisherigen Erkenntnissen in etwa $15~\%$ \cite{Lankheit2020}.
Im Folgenden bezeichne $p_{Art}$ den Anteil an Artefakten und $p_{Ver}$ den Anteil an verlorenen Punkten.
Des Weiteren gilt in der Praxis ${p_{Art} \le 0.2}$ und ${p_{Ver} \le 0.2}$ \cite{Lankheit2020}.
Entsprechend bewegt sich der Anteil an Artefakten und verlorenen Punkten in den weiteren Tests stets im Bereich zwischen $10-20~\%$.

\begin{figure}
  \centering
  \fbox{
  \includegraphics[width=0.7\textwidth]{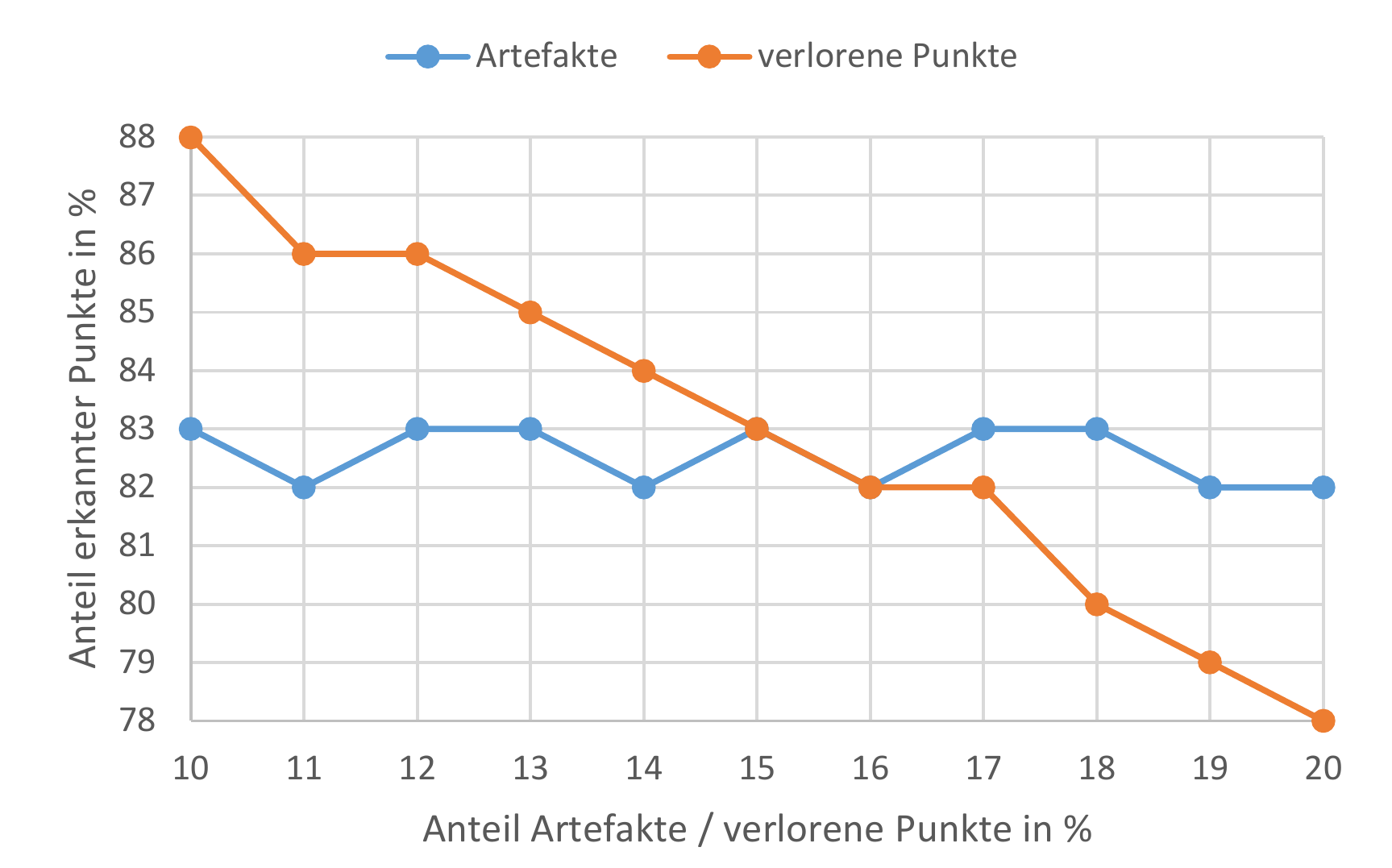}
  }
  \caption{Anteil erkannter Punkte in Abhängigkeit der Anzahl an Artefakten und verlorenen Punkten. Testdaten basieren auf Stäbchen.}
  \label{figure:Einzeln-Art-Ver}
\end{figure}

Für den Punktwolkenvergleich unter Einfluss von Artefakten und verlorenen Punkten gilt: Artefakte wirken sich nur minimal auf das Resultat aus, da sie den Algorithmus lediglich potenziell irritieren, aber vor allem an keinem Punkt der ursprünglichen Punktwolke eine Veränderung vornehmen.
Verlorene Punkte dagegen haben deutlich mehr negativen Einfluss auf das Ergebnis.
Denn ein nicht existenter Punkt kann unmöglich korrekt zugeordnet werden und seine Abwesenheit in der Messung schlägt sich somit bei der finalen Auszählung der erkannten Punkte unmittelbar auf das Resultat nieder.
Dieser naheliegende und einleuchtende Zusammenhang lässt sich in Abbildung \ref{figure:Einzeln-Art-Ver} einwandfrei erkennen.
Die verwendeten Testdaten sind zugunsten der Übersichtlichkeit zwar nur auf Stäbchen basierend, aber auch Daten auf Kügelchen-Basis liefern vergleichbare Ergebnisse.
Bei der Beurteilung der einzelnen Graphen darf nicht vergessen werden, dass bei jedem Testdatum nicht nur eine der beiden Verunreinigungen aufgetreten ist, sondern sowohl Punkte verlorengegangen als auch Artefakte hinzugekommen sind.
Folglich wird derjenige (blaue) Graph, der den Erfolg in Abhängigkeit des Anteils an Artefakten darstellt, auch von den verlorenen Punkten beeinflusst.
Man erkennt jedoch, dass dieser konstant in etwa $15~\%$ niedriger ist, als es die Ergebnisse unter Laborbedingungen waren.
Da durchschnittlich aber auch $15~\%$ der Punkte verlorengegangen sind, stellt ein mittlerer Erkennungsanteil von $83~\%$ ungefähr das Optimum dar.
Betrachtet man nun zusätzlich den (orangenen) Graphen, der die prozentuale Anzahl erkannter Punkte in Abhängigkeit des Anteils verlorener Punkte abbildet, wird noch deutlicher, dass primär die Zahl verlorengegangener Punkte Einfluss auf das Resultat hat.

\begin{figure}
  \centering
  \fbox{
  \includegraphics[width=0.8\textwidth]{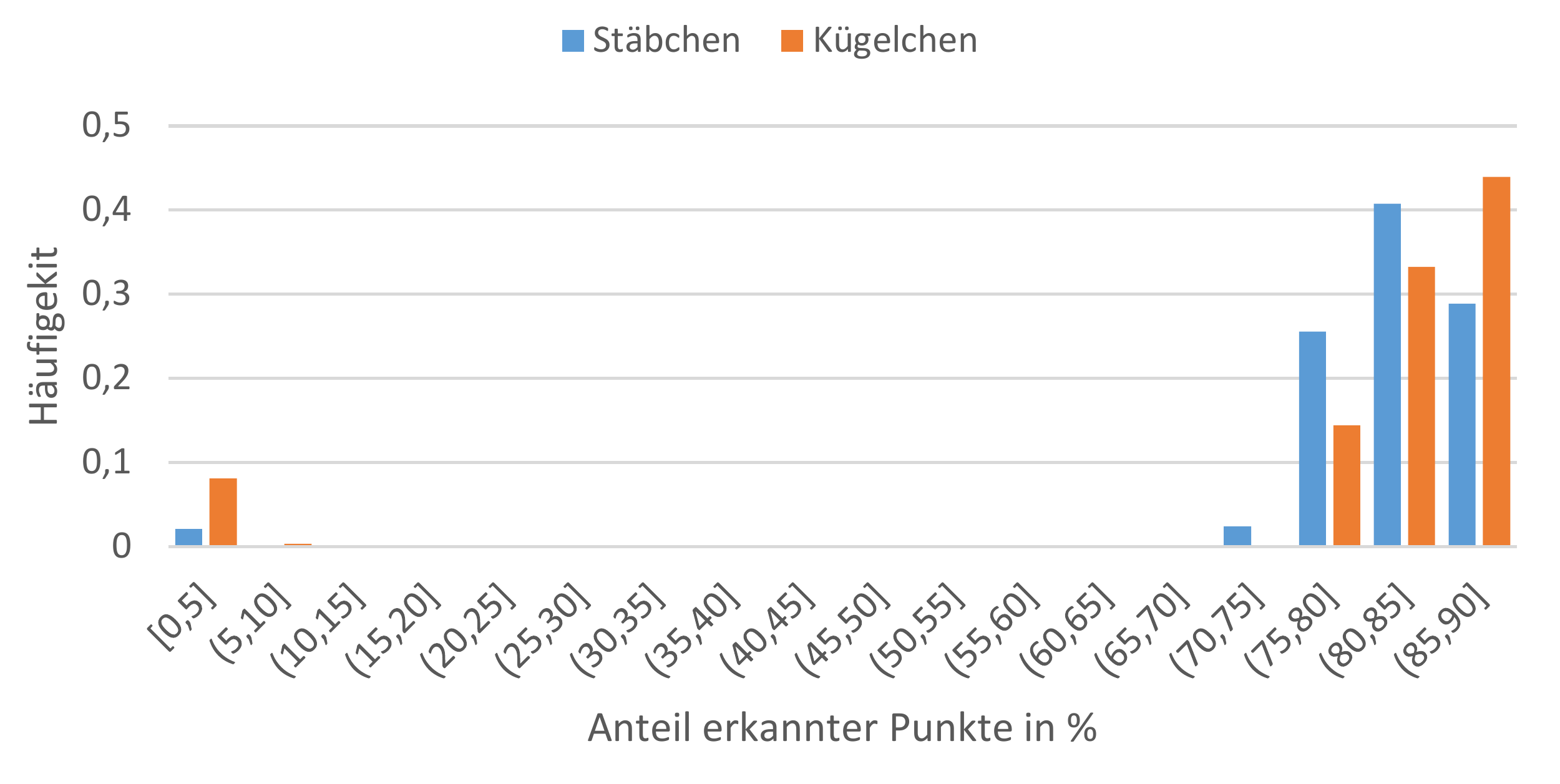}
  }
  \caption{Häufigkeit der Testergebnisse im jeweiligen Intervall an prozentual erkannten Punkten. Testdaten beinhalten verlorene Punkte und Artefakte.}
  \label{figure:Intervalle-Art-Ver}
\end{figure}

Die Graphen der Abbildung \ref{figure:Einzeln-Art-Ver} verlaufen fast konstant, respektive linear, was sich dadurch erklären lässt, dass jeder der einzelnen Datenpunkte den Median einer Menge von mehreren Hundert Elementen darstellt.
Dies liefert in etwa die statistisch zu erwartenden Ergebnisse, welche von einzelnen Ausreißern nur minimal beeinflusst werden.
Da es allerdings, wie bereits selbst unter optimalen Laborbedingungen bei Kügelchen geschehen, vorkommt, dass bei einzelnen Testdaten nur wenige Punkte korrekt erkannt werden, ist des Weiteren Abbildung \ref{figure:Intervalle-Art-Ver} zu betrachten.
Diese stellt dar, wie oft Testergebnisse im jeweiligen Intervall an prozentual erkannten Punkten vorkommen.
Da pro Testdatum stets mindestens $10~\%$ verlorene Punkte enthalten sind und somit maximal $90~\%$ aller Punkte korrekt erkannt werden können, ist bei diesem Diagramm nur der entsprechende Abschnitt der x-Achse dargestellt.
Es lässt sich erkennen, dass der Anteil an Testfällen, bei denen mindestens $70~\%$ aller Punkte korrekt erkannt werden, für Punktwolken auf Kügelchen-Basis bei $92~\%$ und für Punktwolken basierend auf Stäbchen sogar bei $98~\%$ liegt.
Allerdings zeigt sich auch, dass bei $2~\%$ aller Tests auf Stäbchen-Basis nur wenige Punkte korrekt erkannt werden.
Auf Kügelchen-Basis liegt dieser Wert sogar bei $8~\%$.
Obwohl in die Testdaten bisher noch keinerlei ungenaue Positionierung der einzelnen Punkte mit eingeflossen ist, würde in der Praxis für Produktlabel basierend auf Stäbchen in etwa bei einem von $50$ Validierungsversuchen ein fälschlicherweise negatives Ergebnis berechnet werden.
Für Produktlabel, die Kügelchen enthalten, würde dies sogar in vier von $50$ Fällen passieren.
Auf diesen Umstand wird in den folgenden Kapiteln noch detaillierter eingegangen.

\section{Punktwolken unter realistischen Bedingungen mit ungenauer Positionierung}\label{section:UngenauePositionierung}
In der Praxis tritt neben verlorenen Punkten und Artefakte auch eine leichte Verrauschung der Messung auf.
Entsprechend wird in diesem Abschnitt jedem Punkt in jeder Koordinate eine leichte Messungenauigkeit hinzugefügt, wodurch jeder Punkt der Testdaten, wie in Abschnitt \ref{section:Testumgebung} beschrieben, verschoben wird.
Damit handelt es sich nun um maximal realitätsnahe Datensätze.

Im Vergleich zu den Ergebnissen des vorherigen Abschnitts bleiben die Erkenntnisse im Kern unverändert:
Erstens beeinflussen Artefakte das Resultat nur minimal.
Zweitens führt ein Anstieg an verlorenen Punkten logischerweise direkt zu einem Abfall an prozentual erkannten Punkten.
Drittens wird immer noch bei den meisten Testdaten ein gutes Resultat erzielt.
Und viertens werden weiterhin in seltenen Fällen kaum Punkte korrekt erkannt.

\begin{figure}
  \centering
  \fbox{
  \includegraphics[width=0.8\textwidth]{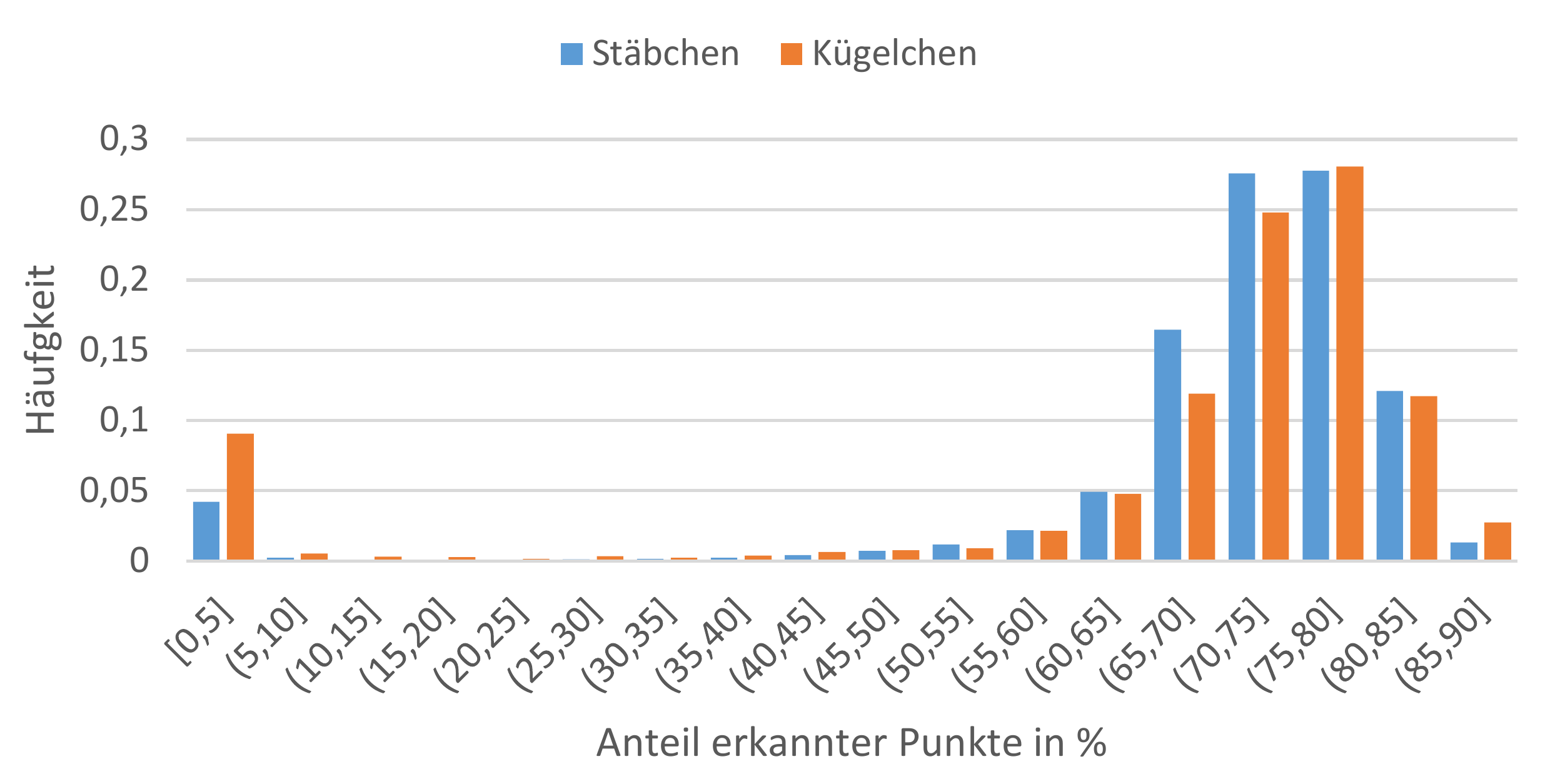}
  }
  \caption{Häufigkeit der Testergebnisse im jeweiligen Intervall an prozentual erkannten Punkten. Testdaten beinhalten verlorene Punkte, Artefakte und eine Messungenauigkeit pro Punkt.}
  \label{fig:Interv-Art-Ver-Noisy}
\end{figure}

Da graphische Veranschaulichungen und tiefergehende Erläuterungen zu den ersten beiden Aussagen keine wesentlichen neuen Informationen enthalten würden, wird an dieser Stelle auf eine weitere Grafik und zusätzliche Erklärungen verzichtet und stattdessen erneut auf Abschnitt \ref{section:art+lostPoints} und insbesondere die Abbildung \ref{figure:Einzeln-Art-Ver} verwiesen.
Bei der dritten und vierten Aussage sind allerdings leichte und vor allem relevante Veränderungen zu erkennen, weshalb hier Abbildung \ref{fig:Interv-Art-Ver-Noisy} zu betrachten ist.
Dabei handelt es sich erneut um die Verteilung der Testergebnisse in Abhängigkeit der prozentual erkannten Punkte.
Nun unterliegen die Testdaten allerdings zusätzlich dem Einfluss von Messungenauigkeiten.
Wie es intuitiv auch zu erwarten ist, werden die Ergebnisse durch das Auftreten der Verrauschung im Durchschnitt etwas schlechter.
Im Mittel werden sowohl bei Testdaten basierend auf Stäbchen als auch bei Testdaten auf Kügelchen-Basis, pro Label $74~\%$ aller Punkte korrekt erkannt.
Dieser Wert ist in etwa zehn Prozentpunkte niedriger als bei nicht verrauschten Messungen.
Wie sich anhand des Diagramms in \ref{fig:Interv-Art-Ver-Noisy} auch erkennen lässt, steigt insbesondere der Anteil an Testdaten, bei welchen die Punkte-Erkennung keine guten Resultate erzielt.
So werden bei circa $6~\%$ aller Testdaten, basierend auf Stäbchen, weniger als die Hälfte aller Punkte korrekt erkannt.
Auf Kügelchen-Basis beträgt dieser Wert sogar fast $13~\%$.

Angenommen die Schwelle, um von Gleichheit der Punktwolken auszugehen, liegt bei $50~\%$ erkannter Punkte:
Dann würde in der Praxis bei der Validierung eines Produktlabels, falls dieses Stäbchen beinhaltet, in etwa jeder 16. Versuch zu einem fälschlicherweise negativen Resultat führen.
Besteht das Produktlabel aus Kügelchen, so würde im Durchschnitt sogar jeder achte Validierungsversuch irrtümlich fehlschlagen.

\begin{figure}
  \centering
  \fbox{
  \includegraphics[width=0.5\textwidth]{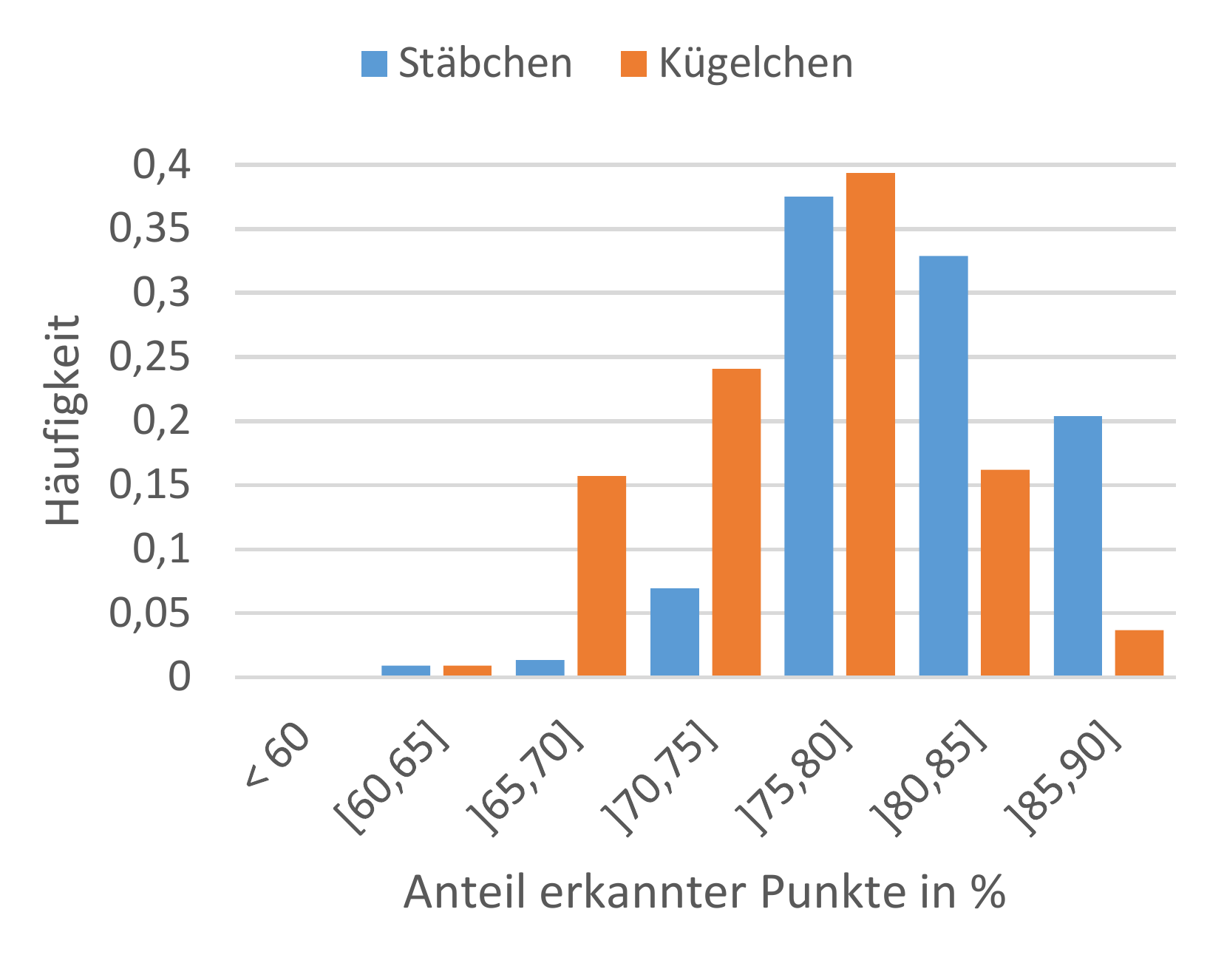}
  }
  \caption{Häufigkeit der Testergebnisse verschiedener Messungen zu einer Referenz im jeweiligen Intervall an prozentual erkannten Punkten. Die Ergebnisse stellen das Optimum dar.}
  \label{fig:Eine_Ref-Optimum}
\end{figure}

\begin{figure}
  \centering
  \fbox{
  \includegraphics[width=0.8\textwidth]{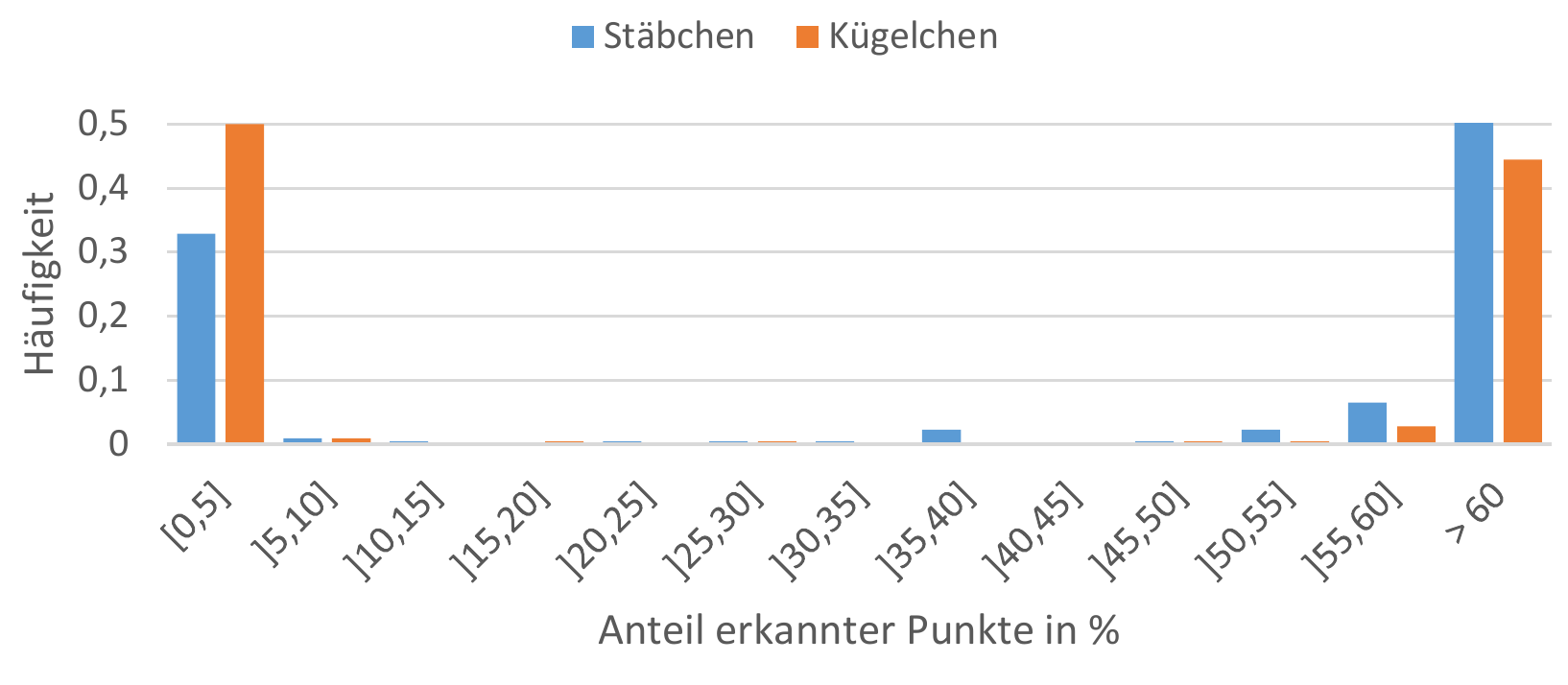}
  }
  \caption{Häufigkeit der Testergebnisse verschiedener Messungen zu einer Referenz im jeweiligen Intervall an prozentual erkannten Punkten. Die Ergebnisse stellen das Pessimum dar.}
  \label{fig:Eine_Ref_Pessimum}
\end{figure}

Um diesen Umstand etwas detaillierter zu beleuchten, seien die Abbildungen \ref{fig:Eine_Ref-Optimum} und \ref{fig:Eine_Ref_Pessimum} zu betrachten.
Hierbei handelt es sich erneut um die Häufigkeit der Testergebnisse im jeweiligen Intervall an prozentual erkannten Punkten.
Im Gegensatz zu vorherigen Grafiken basieren die Daten hier allerdings nicht auf einer Vielzahl unterschiedlicher Referenzen mit verschiedenen Messungen, sondern auf der Validierung jeweils einer einzigen Referenz mit mehreren Hundert Messungen. 

In der Grafik \ref{fig:Eine_Ref-Optimum} ist der optimale und erwünschte Extremfall dargestellt.
Für das untersuchte Label, sowohl auf Stäbchen- als auch auf Kügelchen-Basis, werden bei jeder Messung mindestens $60~\%$ aller Punkte korrekt erkannt.
Insbesondere tritt kein einziger Ausreißer auf.
Entsprechend ist in der Grafik nur ein Teil der x-Achse zu sehen. 

In Grafik \ref{fig:Eine_Ref_Pessimum} ist der dazu gegensätzliche und unerwünschte Extremfall abgebildet.
Um bei der Grafik den Blick auf das Wesentliche zu lenken, wurde auch hier wieder eine Anpassung der x-Achse vorgenommen.
Denn es soll nicht erneut aufgezeigt werden, wie häufig einzelne gute Ergebnisse im Vergleich zueinander auftreten, sondern dass bei einer Vielzahl der Messungen schlechte Resultate erzielt werden.
Sowohl für das auf Stäbchen als auch das auf Kügelchen basierende Label liefern vergleichsweise wenige Messungen ein gutes Ergebnis.
Bei verschiedenen Messungen des Labels auf Stäbchen-Basis werden in fast $40~\%$ der Tests weniger als die Hälfte aller Punkte korrekt erkannt.
Für das auf Kügelchen basierende Label tritt dieser Fall sogar in über $50~\%$ aller Messungen ein.

In den beiden Diagrammen \ref{fig:Eine_Ref-Optimum} und \ref{fig:Eine_Ref_Pessimum} sind jeweils nur die Testergebnisse verschiedener erstellter Messungen zu einer einzelnen Referenz dargestellt.
Zudem wurde in beiden Fällen eine Referenz ausgewählt, die überdurchschnittlich gute respektive unterdurchschnittlich schlechte Ergebnisse erzielte.
Zwar spiegeln die ausgewählten Beispiele damit nicht den durchschnittlichen Normalfall wider, verdeutlichen aber eine andere wichtige Erkenntnis: Scheinbar ist die Güte des Punktwolkenvergleichs zu einem gewissen Maße von der Referenz abhängig, also der realen Anordnung der Stäbchen beziehungsweise Kügelchen im Label.
So liegt der Wert, bei dem für Messungen basierend auf Stäbchen weniger als die Hälfte aller Punkte korrekt erkannt wird, für knapp $50~\%$ aller Referenzen bei unter $5~\%$.
Bei circa $10~\%$ aller Label liefert der Punktwolkenvergleich für jede vierte Messung ein Ergebnis von weniger als der Hälfte an korrekt erkannten Punkten.
Label auf Kügelchen-Basis haben hierbei im Wesentlichen erneut vergleichbare, im Detail allerdings wieder leicht schlechtere Resultate erzielt. 

Eine zusätzliche und für den weiteren Verlauf des Projekts der Validierung von fälschungssicheren Produktlabeln interessante Erkenntnis liefert der Vergleich unterschiedlicher Punktwolkengrößen.
Es hat sich gezeigt, dass bis zu einem gewissen Schwellenwert gilt: Je mehr Punkte in einer Punktwolke enthalten sind, desto konstantere Ergebnisse werden erzielt.
Also je größer $\abs{X}$ ist, desto seltener kommt es vor, dass nur wenige Punkte eines Testdatums korrekt erkannt werden.
Allerdings wirkt sich eine Erhöhung von $\abs{X}$ nur bis zu einem Wert von circa $40$ spürbar positiv auf die Resultate aus.
Zwischen Werten von $40$ bis $100$ Punkten pro Punktwolke sind kaum Veränderungen wahrzunehmen.

\section{Unterscheidung ungleicher Produktlabel} \label{section:UngleichePunktwolken}
Nachdem bisher untersucht wurde, ob Gleiches als gleich erkannt wird, soll in diesem Abschnitt überprüft werden, ob Ungleiches als solches identifizierbar ist.
Aus Sicht der praktischen Anwendung wird also nun ein physisches Produktlabel mit der Referenzmessung eines anderen Produktlabels verglichen. 
Zur Untersuchung wurden hierbei jeweils für Stäbchen und Kügelchen zu allen, unter \ref{section:Testumgebung} genannten, neun verschiedenen Größen von $\abs{X}$ für zehn unterschiedliche Referenzen jeweils zehn falsche Messungen erstellt und einem Vergleich unterzogen.
Eine falsche Messung bezeichnet dabei die Messung eines anderen physischen Produktlabels als das der Referenz zugrundeliegende.
Insgesamt wurden für diesen Abschnitt folglich je Label-Variante 900 Tests durchgeführt.

Es zeigt sich dabei für Punktwolken sowohl auf Kügelchen- als auch auf Stäbchen-Basis, dass maximal eine niedrige einstellige Prozentzahl aller Punkte erkannt wird.
Das mag im ersten Moment überraschend klingen, denn bei Ungleichheit könnte man erwarten, dass stets exakt $0~\%$ Übereinstimmung herrscht.
Allerdings lässt sich jede Punktwolke durch Rotation und Translation so darstellen, dass sich mindestens einer der Punkte in ausreichender Nähe zu einem der Punkte einer beliebigen zweiten Punktwolke befindet. 

Da bei keinem einzigen der Testfälle die Anzahl prozentual erkannter Punkte zweistellig war, kann festgestellt werden, dass die verwendeten Algorithmen Ungleiches erfolgreich als ungleich identifizieren.

\section{Verfälschte Punktwolken}\label{section:Fälschungen}
Bisher wurde überprüft, ob Gleiches als gleich und Ungleiches als ungleich erkannt wird.
Dabei wurden die Messung und Referenz trotz naturgemäßer Differenzen aufgrund des gleichen zugrundeliegenden Produktlabels als gleich deklariert.
Folglich werden ungleiche Punktwolken, die einander nur ähnlich sind, bereits als gleich interpretiert.
Nun stellt sich die Frage, bis zu welchem Grad von Ungleichheit noch auf Gleichheit geschlossen wird.
Diese Fragestellung erscheint konstruiert und wenig Praxisbezug zu haben, denn es ist aufgrund der zufälligen Verteilung der einzelnen Stäbchen beziehungsweise Kügelchen äußerst unwahrscheinlich, dass in der Praxis zwei unterschiedliche Produktlabel ausreichend gleichartig sind.
Zudem ist es selbst bei zwei exakt gleichen Produktlabeln unrealistisch, dass deren Gleichartigkeit jemals bemerkt und ausgenutzt werden würde.
Die praktische Grundlage für die Fragestellung beruht auf dem Wunsch, ein aktiv gefälschtes Produktlabel ebenfalls als solches zu erkennen und als ungleich auszuweisen.
Der dieser Arbeit übergeordnete Grundsatz beruht darauf, dass eine Reproduktion eines Labels weniger genau ist als die Vermessung eines Labels.
Auf dieser Grundlage werden die Label als fälschungssicher betrachtet.
Nichtsdestotrotz stellt sich die Frage, wie sich die Resultate des Algorithmus verändern, falls das Label eine Fälschung ist - also falls ein Fälscher in der Lage wäre, jedes Kügelchen respektive Stäbchen des physischen Produktlabels mit einer gewissen Ungenauigkeit an derselben Position wie im Original-Label zu platzieren.
Für den Punktwolkenvergleich bedeutet das, Referenz und Messung sind einander zwar ähnlich, aber resultieren nicht aus der Vermessung ein und desselben physischen Produktlabels.
Denn die Referenz-Punktwolke des Herstellers des originalen Produktes kann aufgrund der eingesetzten digitalen Signatur nicht verändert werden.

\begin{figure}
  \centering
  \fbox{
  \includegraphics[width=0.75\textwidth]{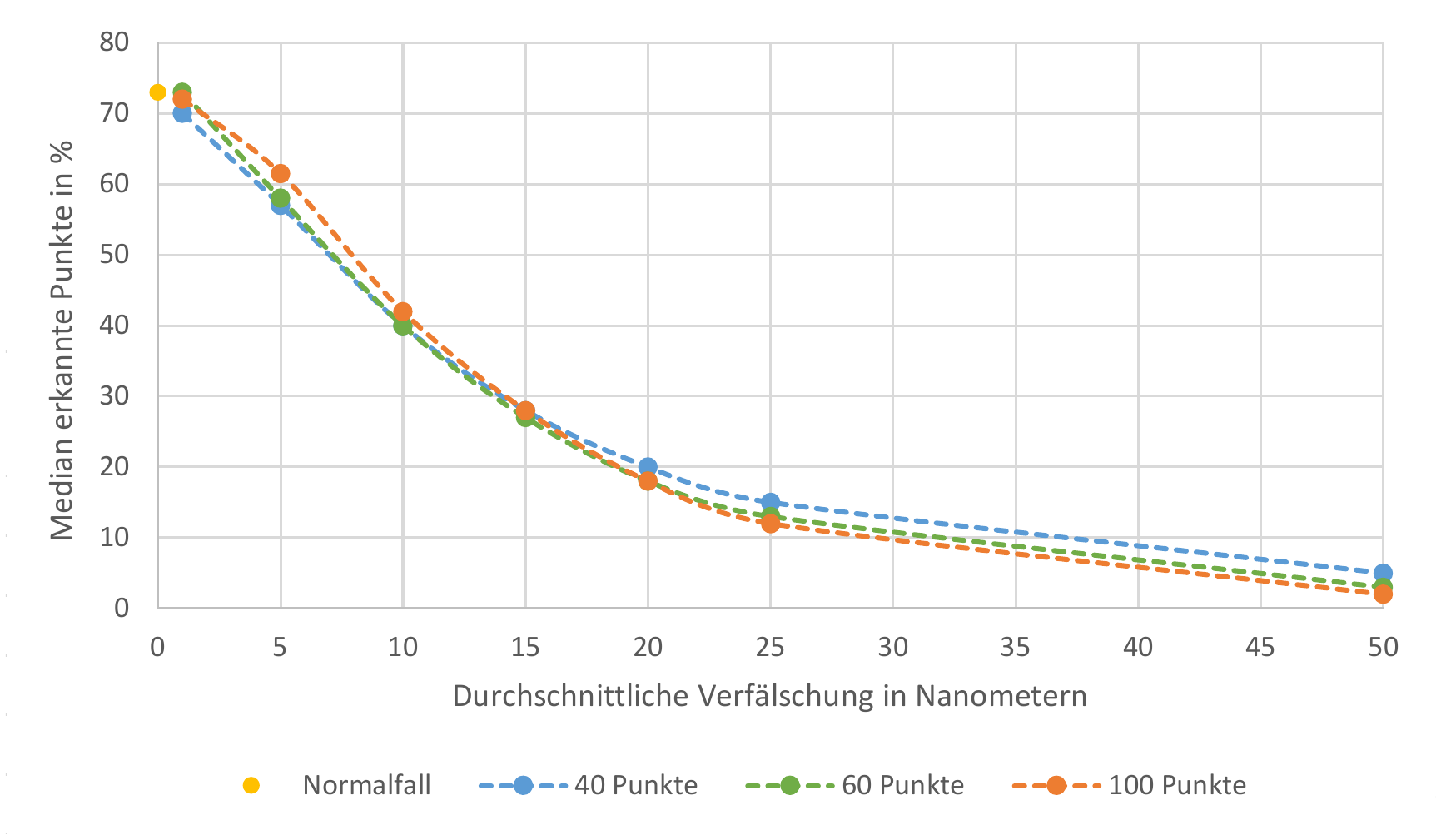}
  }
  \caption{Durchschnittlich erkannte Punkte je zusätzlicher Verfälschung. Testdaten basieren auf Kügelchen.}
  \label{fig:Fake_Median_Beads}
\end{figure}

\begin{figure}
  \centering
  \fbox{
  \includegraphics[width=0.75\textwidth]{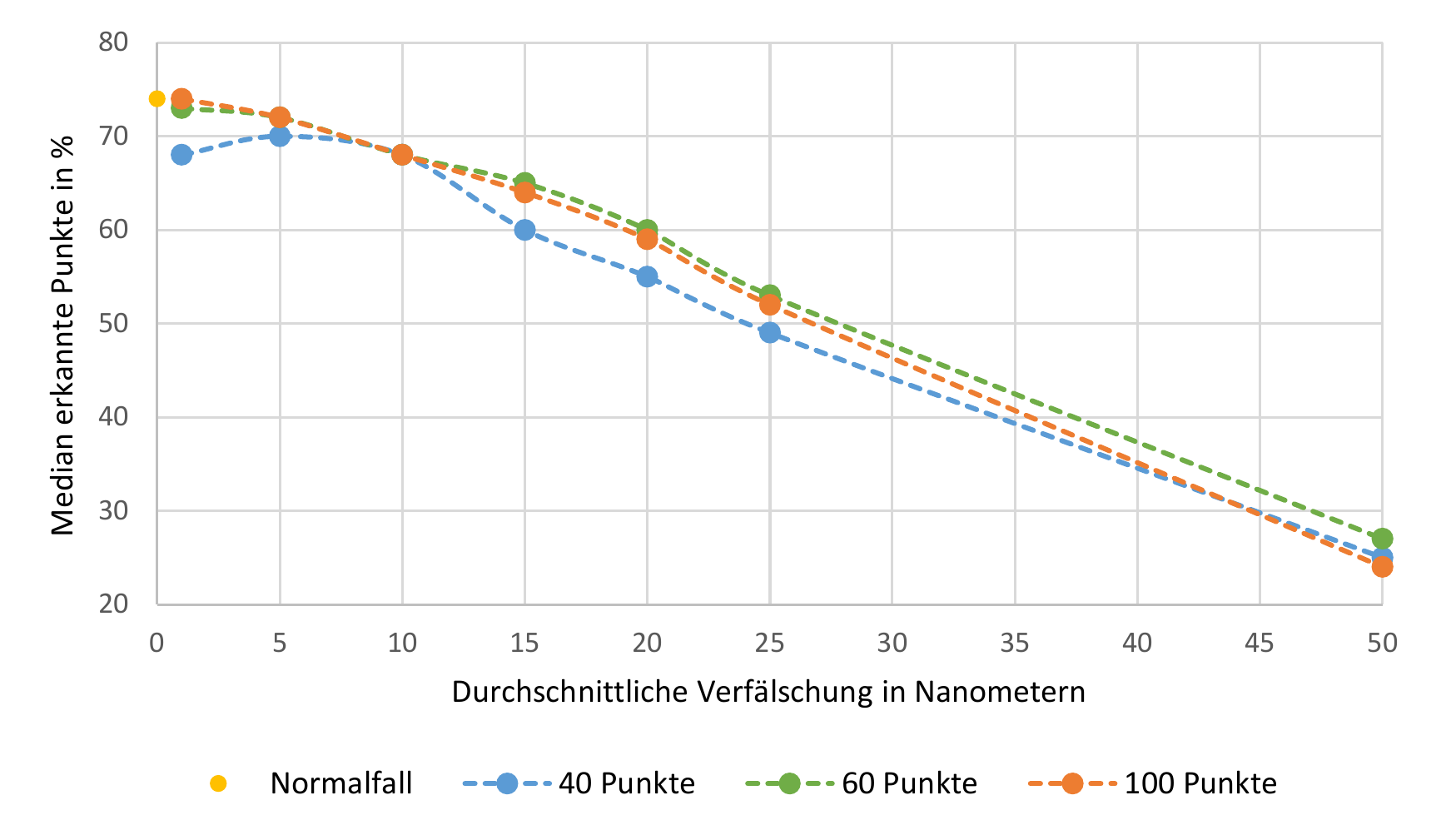}
  }
  \caption{Durchschnittlich erkannte Punkte je zusätzlicher Verfälschung. Testdaten basieren auf Stäbchen.}
  \label{fig:Fake_Median_Rods}
\end{figure}

Zur Analyse des Verhaltens bei verfälschten Punktwolken wurden dieselben Referenzen wie in Abschnitt \ref{section:UngenauePositionierung} verwendet und die dazugehörigen Messungen unterschiedlich stark verfälscht.
Die Verfälschung einer Messung stellt faktisch nichts anderes als eine zusätzliche Verrauschung jedes Punktes dar.
Intuitiv sollte also gelten, je höher der Grad der Verfälschung ist, desto weniger Punkte werden korrekt erkannt.
Ist der Verfälschungsgrad allerdings niedrig, so sollte auch die Veränderung zum Normalfall nur geringfügig erkennbar sein.
Wie sich anhand der Graphen der Abbildungen \ref{fig:Fake_Median_Beads} und \ref{fig:Fake_Median_Rods} erkennen lässt, treffen diese Vermutungen zu.
Der in den Abbildungen gezeigte Normalfall bedeutet, dass sowohl Punkte verlorengegangen und Artefakte hinzugekommen sind als auch dass die in der Praxis standardmäßig auftretende Messungenauigkeit mit eingeflossen ist.
Eine Verfälschung bedeutet dann, dass zusätzlich zum Normalfall noch eine gewisse Ungenauigkeit bei der exakten Positionierung der einzelnen Punkte auftritt.

Anhand der Grafik \ref{fig:Fake_Median_Beads} zeigt sich, dass die Resultate für Label auf Kügelchen-Basis bei einer Verfälschung von nur $1~nm$ kaum vom Normalfall zu unterscheiden sind.
Selbst bei einem Verfälschungsgrad von $5~nm$ ist die Anzahl der durchschnittlich erkannten Punkte nur um etwa zehn Prozentpunkte niedriger.
Wenn die Fälschungen allerdings mit einer Ungenauigkeit von $25~nm$ beziehungsweise $50~nm$ angefertigt sind, werden im Mittel nur in etwa $15~\%$ respektive $5~\%$ aller Punkte korrekt erkannt.

Bei Labeln auf Stäbchen-Basis (Grafik \ref{fig:Fake_Median_Rods}) bleiben die Grundaussagen zwar dieselben, sprich die Verfälschung hat erst ab einem gewissen Grad einen spürbaren Einfluss auf das Ergebnis, aber im Vergleich zu den Kügelchen bedarf es einer höheren absoluten Verfälschung.
Auch wenn das vorerst verwunderlich erscheinen mag, ist es eine schlüssige Erkenntnis.
Denn wie bereits erwähnt ist die standardmäßige Messungenauigkeit bei Labeln auf Stäbchen-Basis deutlich höher als bei Labeln auf Kügelchen-Basis.
Für Details sei Kapitel \ref{section:Testumgebung} zu betrachten.
Folglich sind die absoluten Werte immer in Relation mit der Messungenauigkeit zu setzen.
Und im Vergleich der relativen Verfälschung schneiden Label auf Stäbchen- und Kügelchen-Basis ähnlich ab.
Beispielsweise sind die Ergebnisse für Label auf Stäbchen-Basis bei einer Verfälschung von $50~nm$ in etwa dieselben wie für Label auf Kügelchen-Basis bei einer Verfälschung von $15~nm$.
Die beiden dargelegten Werte von $50~nm$ und $15~nm$ entsprechen in etwa dem Verhältnis der Messungenauigkeit und zeigen damit beispielhaft die Vergleichbarkeit der beiden Label-Varianten auf.

Des Weiteren lässt sich anhand der beiden Grafiken \ref{fig:Fake_Median_Beads} und \ref{fig:Fake_Median_Rods} erneut erkennen, dass die Anzahl der Punkte einer Punktwolke bei ausreichender Größe kaum spürbaren Einfluss auf das Resultat hat.
Wie bereits in Abschnitt \ref{section:UngenauePositionierung} aufgezeigt, gilt auch hier, dass die Resultate ab einer Anzahl von circa $40$ Punkten pro Punktwolke untereinander kaum Unterschiede aufweisen.

\begin{figure}
  \centering
  \fbox{
  \includegraphics[width=0.8\textwidth]{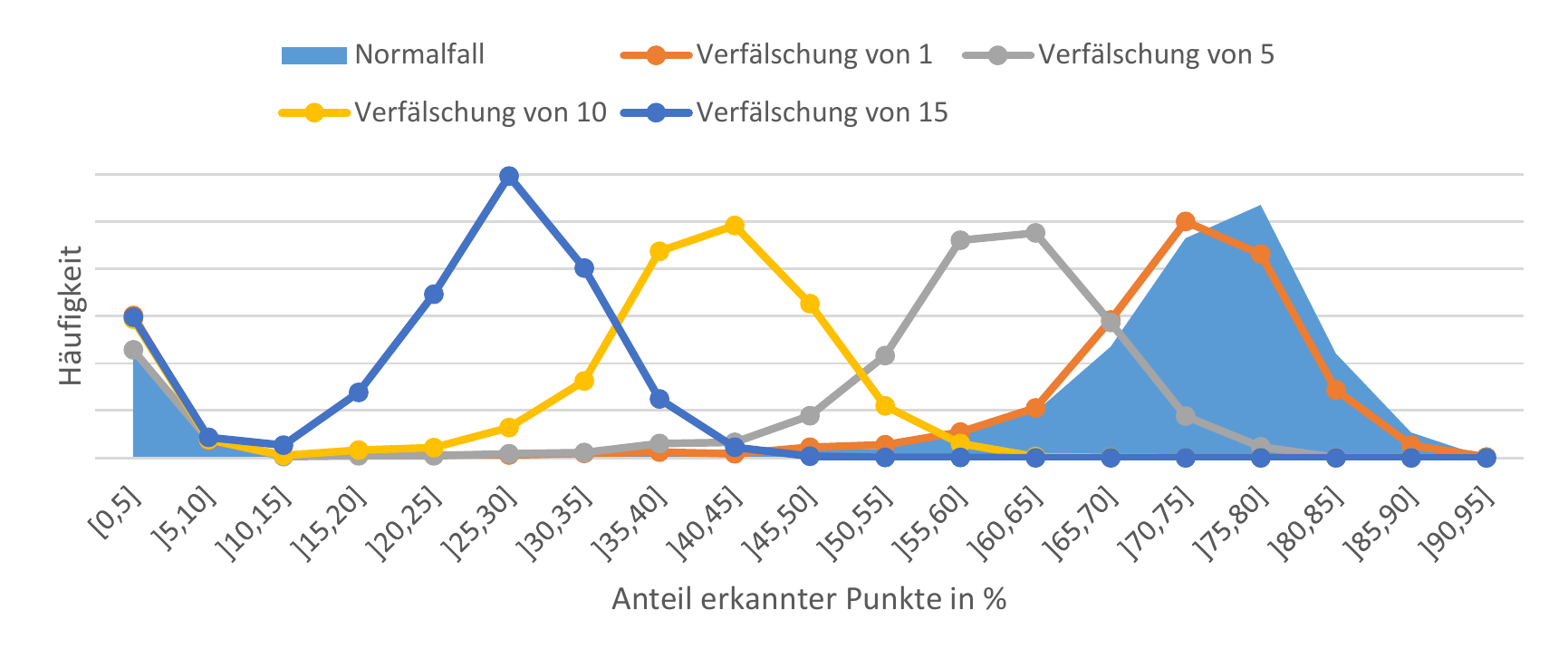}
  }
  \caption{Verteilung prozentual erkannter Punkte mit unterschiedlichen Verfälschungsgraden. Testdaten basieren auf Kügelchen.}
  \label{fig:Fake_Hills_Beads}
\end{figure}

Um den Einfluss der Verfälschung auf das Ergebnis nicht nur, wie in den Abbildungen \ref{fig:Fake_Median_Beads} und \ref{fig:Fake_Median_Rods}, anhand des Medians darzustellen, sei des weiteren Abbildung \ref{fig:Fake_Hills_Beads} zu betrachten.
Hierbei stellt die blaue Fläche die standardmäßige Verteilung der Testergebnisse für Label auf Kügelchen-Basis dar.
Die einzelnen unterschiedlich farbigen Linien zeigen die Verteilungen der Resultate unter verschiedenen Graden der Verfälschungen auf.
Die y-Achse stellt die Häufigkeit dar.
Die exakten Werte dieser spielen bei der Grafik keine Rolle, da ausschließlich die zueinander relative Verteilung aufgezeigt werden soll.

Die oben bereits erläuterten Phänomene sind auch hier bei der Abbildung \ref{fig:Fake_Hills_Beads} wieder zu erkennen.
Bei einer minimalen Verfälschung von nur $1~nm$ (orangene Linie) unterscheiden sich die Ergebnisse kaum spürbar zum Normalfall.
Mit Zunahme des Grades der Verfälschung werden deutlich häufiger nur noch wenige der Punkte korrekt erkannt.
Das ist in der Grafik sichtbar durch die Links-Verschiebung des Kegels der häufigen Ergebnisse unter Zunahme des Grades der Verfälschung.
Weil die Grundaussagen für Label auf Stäbchen-Basis gleichermaßen gelten und sich die entsprechende Grafik ähnlich darstellt, wird an dieser Stelle auf weiterführende Erläuterungen verzichtet.

\section{Laufzeiten}\label{section:Zeitanalyse}
Nachdem in den vorherigen Kapiteln die Ergebnisse des Punktwolkenvergleichs dargelegt wurden, wird im Folgenden die Berechnungsdauer analysiert.
Darunter ist die Zeit zu verstehen, die ab dem Zeitpunkt, zu dem beide Punktwolken in digitalisierter Form vorliegen, bis zur Anzeige des Resultats der Validierung des Produktlabels verstreicht.
Die davorliegende Zeit, die sowohl für das Extrahieren aller Daten aus den beiden QR-Codes als auch für das tatsächliche Auslesen aller Informationen aus dem physischen Produktlabel benötigt wird, spielt keine Rolle.
Denn es handelt sich hier erstens um Schritte, für die der notwendige Zeitaufwand primär von den Handlungen des Anwenders und nur sekundär von der App bestimmt wird.
Zweitens wird sich erst zukünftig mit dem Einsatz der Technologie zum Auslesen eines physischen Produktlabels und der Einführung einer PKI zeigen, mit welcher Zeitspanne zu rechnen ist. 

Eine der Hauptfragen der vorliegenden Arbeit ist, ob im Zuge der Validierung eines Produktlabels das Resultat des Punktwolkenvergleichs von der App in unter einer Sekunde  berechnet werden kann.
Um eine Antwort auf diese Fragestellung zu liefern, wird die Berechnungsdauer des zu erwartenden Normalfalls, wie in \ref{section:UngenauePositionierung} beschrieben, auf drei verschiedenen Smartphones in Abhängigkeit der Punktwolkengröße gemessen.
Dabei werden die nachfolgend genannten Zeiten stets Punktwolken auf Kügelchen-Basis zur Grundlage haben.
Punktwolken basierend auf Stäbchen weisen jedoch keine spürbaren Unterschiede auf.

\begin{figure}
  \centering
  \fbox{
  \includegraphics[width=0.8\textwidth]{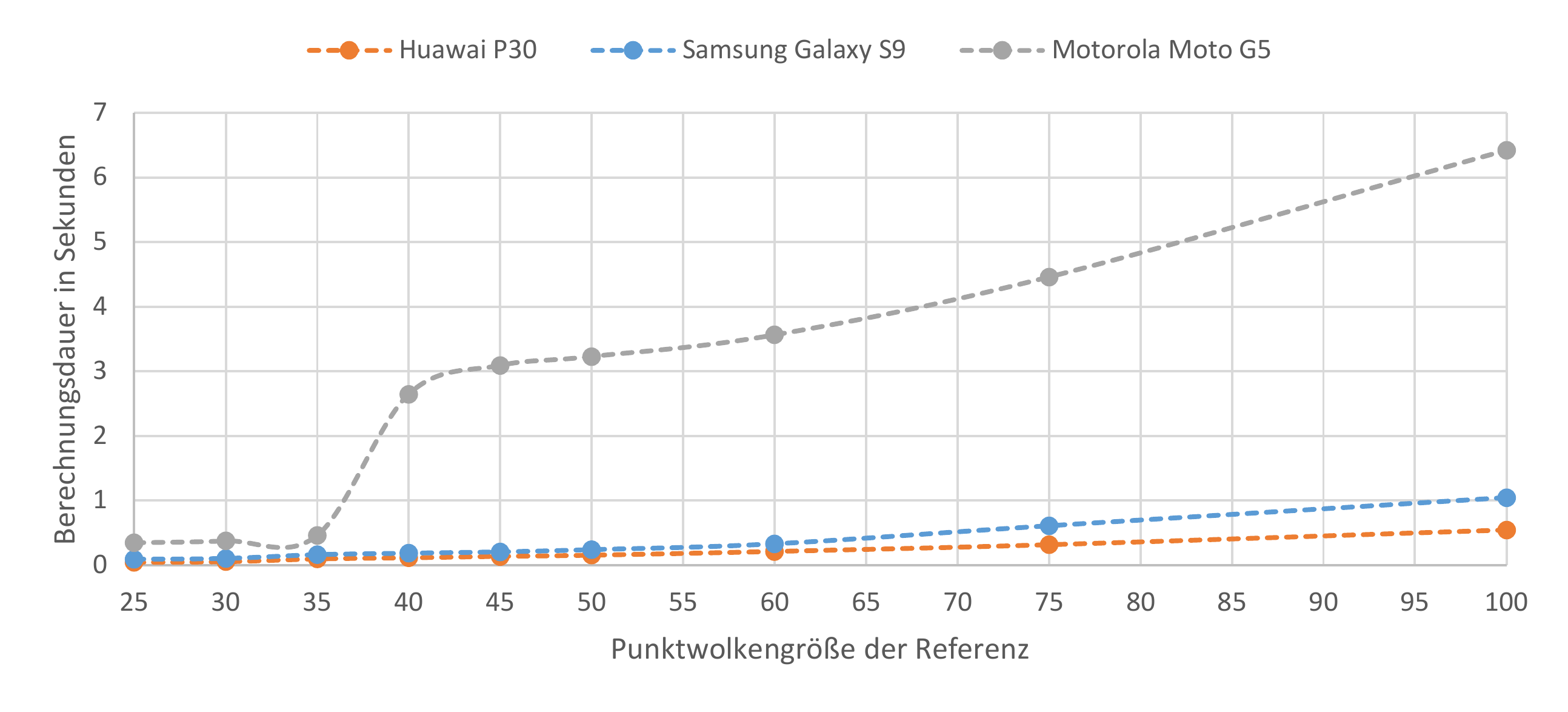}
  }
  \caption{Durchschnittliche Berechnungsdauer verschiedener Smartphones in Sekunden in Abhängigkeit der Punktwolkengröße. Testdaten basieren auf Kügelchen.}
  \label{fig:Zeiten-Sekunden}
\end{figure}

Das älteste eingesetzte Smartphone ist das im März 2017 mit $2$ Gigabyte (GB) Arbeitsspeicher (RAM von engl. Random Access Memory) erschienene Moto G5 der Marke Motorola.
Wie sich anhand der Abbildung \ref{fig:Zeiten-Sekunden} erkennen lässt, wird das Limit von einer Sekunde Berechnungsdauer nur bis zu einer Punktwolkengröße von $35$ Punkten eingehalten.
Sobald $40$ oder mehr Punkte in der Referenz enthalten sind, werden zwischen $2,5$ und $6,5$ Sekunden für die Berechnung des Ergebnisses benötigt.

\begin{figure}
  \centering
  \fbox{
  \includegraphics[width=0.8\textwidth]{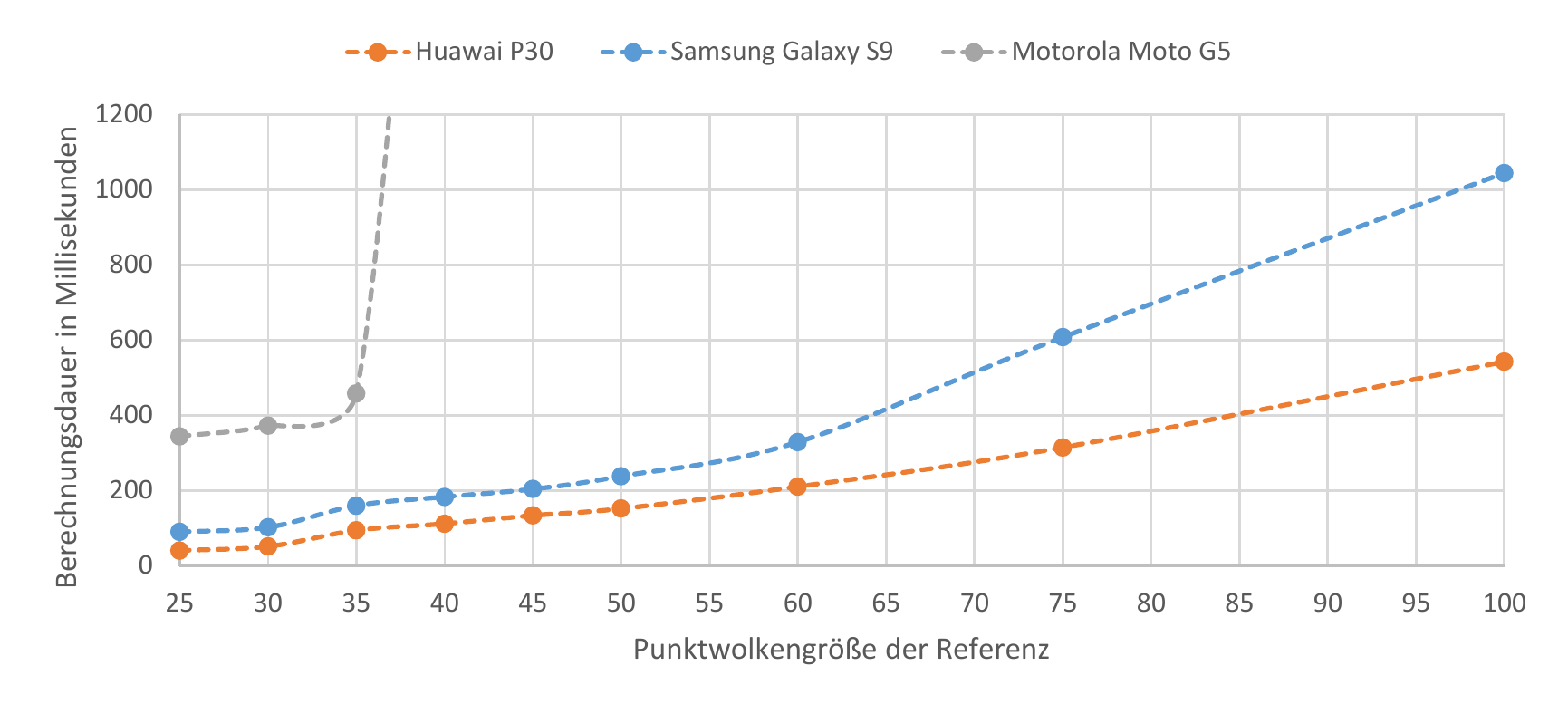}
  }
  \caption{Durchschnittliche Berechnungsdauer verschiedener Smartphones in Millisekunden in Abhängigkeit der Punktwolkengröße. Testdaten basieren auf Kügelchen.}
  \label{fig:Zeiten-Millisekuden}
\end{figure}

Um auch über die jeweilige Berechnungsdauer der zwei weiteren eingesetzten Smartphones detailliertere Aussagen treffen zu können, sei Abbildung \ref{fig:Zeiten-Millisekuden} zu betrachten.
Hierbei handelt es sich um einen Ausschnitt des Diagramms der Abbildung \ref{fig:Zeiten-Sekunden}, wobei die an der y-Achse aufgetragene Berechnungsdauer auf $1,2$ Sekunden begrenzt ist.
Das zweite verwendete Smartphone ist das von Samsung im März 2018 herausgebrachte Galaxy S9 mit $4\,GB$ RAM.
Für dieses zeigt sich, dass die Berechnungsdauer bei nahezu allen untersuchten Punktwolkengrößen durchschnittlich weniger als eine Sekunde beträgt.
Einzig für Referenzen die $100$ Punkte enthalten, liegt die Berechnungsdauer im Mittel mit $1040$ Millisekunden minimal über dem Limit von einer Sekunde.
Das dritte und neueste zur Laufzeitanalyse eingesetzte Smartphone ist das im April 2019 von Huawai herausgebrachte P30 mit $6\,GB$ RAM.
Bei diesem zeigt sich, dass für jede betrachtete Größe der Referenz-Punktwolke das Limit von maximal einer Sekunde nicht nur eingehalten, sondern gar deutlich unterschritten wird.
Selbst bei Referenzen mit $100$ Punkten wird durchschnittlich nur circa eine halbe Sekunde für die Berechnung des Ergebnisses benötigt.
Für Punktwolken mit $60$ oder weniger Punkten beträgt die Berechnungsdauer im Mittel sogar stets weniger als eine viertel Sekunde.

Für die Evaluation der gemessenen Laufzeiten ist erstens anzumerken, dass nach aktueller Planung in der Praxis nur circa $50$ Punkte pro Punktwolke enthalten sein werden.
Speziell der Einsatz von Punktwolken der Größe $100$ oder mehr erscheint bisher unwahrscheinlich. \cite{Ruehrmair2021}
Zweitens vergeht noch eine gewisse Zeit, bis das Verfahren der Validierung von fälschungssicheren Produktlabeln in der Realität eingesetzt werden kann.
Bis zur finalen Verwendung der mobilen Applikation wird sich folglich insbesondere die Hardware der Smartphones weiterentwickeln.
Dementsprechend können die Laufzeiten des Moto G5 bei der Beurteilung der Berechnungsdauer ausgeblendet werden.
Denn es handelt sich bei diesem bereits zum Zeitpunkt der Untersuchung im Zuge der vorliegenden Arbeit um ein überdurchschnittlich altes Smartphone mit unterdurchschnittlich leistungsschwacher Hardware.
Zudem werden sogar mit dem mehrere Jahre alten Samsung Galaxy S9 die gewünschten Ergebnisse erzielt.
Die Verwendung des moderneren Huawai P30 führt im Vergleich zum Galaxy S9 sogar zu einer Halbierung der Berechnungsdauer.
Abschließend kann die Fragestellung, ob das Zeitlimit von maximal einer Sekunde für die Berechnung des Resultats auf einem Smartphone einzuhalten ist, klar positiv beantwortet werden.
    \chapter{Fazit}\label{chapter:Fazit}
Abschließend werden in Abschnitt \ref{section:Konklusion} die erzielten Ergebnisse hinsichtlich der Aufgabenstellung diskutiert und die im Verlauf dieser Arbeit erkannte Problematik, das heißt die teilweise unbefriedigenden Ergebnisse der eingesetzten Algorithmen dargestellt.
Zur Problemlösung bedarf es vermutlich weiterer Untersuchungen.
Dazu wird in Abschnitt \ref{section:Ausblick} ein Ausblick über mögliche Ansätze nachfolgender Arbeiten gegeben.

\section{Konklusion}\label{section:Konklusion}
Das hauptsächliche, übergeordnete Ziel dieser Arbeit war, zu untersuchen, ob sich im Zusammenhang der Validierung von fälschungssicheren Produktlabeln und den damit einhergehenden algorithmischen Verfahren eine mobile Applikation entwickeln lässt, sodass diese den gewünschten Anforderungen entspricht.
Da es sich hierbei um ein umfassendes und unpräzises Ziel handelt, wurde dieses in mehrere Teilziele untergliedert.

Erstens sollte eine digitale Signatur ausgemacht werden, welche die geforderten Kriterien optimal erfüllt.
In diesem Zusammenhang wurden Signaturen basierend auf elliptischen Kurven als die ideale Lösung ermittelt und zudem geeignete Algorithmen vorgestellt.
Es wurde also erfolgreich eine geeignete digitale Signatur eruiert, allerdings konnte sich hierbei verschiedenster internationaler Standards bedient werden, wodurch insbesondere keine neuartige Erkenntnis erzielt wurde.

Das weitere Vorgehen diente dem zweiten Ziel, einen geeigneten Barcode-Typ zur Übermittlung der benötigten Informationen auszumachen.
Im Ergebnis wurden QR-Codes als die im Kontext dieser Arbeit am praktikabelsten Barcodes aufgezeigt.
Aufgrund des enormen Flächenbedarfs bei der Verwendung von nur einem QR-Code, der alle benötigten Daten kodiert, wurde eine Zweiteilung der Informationen und damit einhergehend die Verwendung unterschiedlicher QR-Code-Modi als beste Lösung vorgestellt.
Die zu Beginn dieser Arbeit noch offene Frage, in welcher Form alle Daten sinnvoll auf einem Produkt angebracht werden können, wurde also ebenfalls erfolgreich beantwortet.
Insbesondere die Erkenntnis, dass mehrfarbige und damit quasi dreidimensionale Barcodes - wie zum Beispiel JAB Codes - trotz ihrer theoretischen Vorteile gegenüber schwarz-weißen zweidimensionalen QR-Codes im praktischen Vergleich schlechter abschneiden, ist erstaunlich.
Anzumerken ist allerdings, dass es sich hierbei lediglich um eine aktuelle Bestandsaufnahme handelt.
JAB Codes stellen eine vergleichsweise junge Technologie dar, die sich in Zukunft möglicherweise entscheidend weiterentwickeln wird. 

Resultierend aus den ersten beiden Zielen bestand das dritte Ziel darin, eine mobile Applikation zu entwickeln, die alle erwünschten Funktionalitäten vereint.
Dazu wurde eine Anwendung implementiert, mittels derer sowohl ein Hersteller neue digital signierte QR-Codes für seine Produktlabel erstellen als auch ein Anwender die Authentifizierung des Produktes vollziehen kann.
Teile der durch die mobile Applikation durchgeführten Produktauthentifizierung sind das Auslesen und Dekodieren der Daten der beiden QR-Codes, die Verifizierung der digitalen Signatur mit dem öffentlichen Schlüssel des Herstellers, die Möglichkeit des Hinzufügens einer künstlich erstellten Messung des Produktlabels sowie die abschließende Berechnung des Grades der Übereinstimmung der zwei Punktwolken.
Als Resultat wird dem Anwender abschließend mitgeteilt, ob es sich bei dem Produkt um ein Original oder ein Imitat handelt.
Somit wurden die gewünschten Anforderungen an die mobile Applikation, insbesondere bezogen auf die geringen Laufzeiten des Punktwolkenvergleichs und den Umfang an Funktionalitäten, erfolgreich umgesetzt. 

Die zu Beginn definierten Ziele wurden folglich erreicht und die grundsätzliche Machbarkeit der Verwendung einer mobilen Applikation zur Validierung von fälschungssicheren Produktlabeln nachgewiesen.

Nichtsdestotrotz hat sich im Laufe der detaillierten Auswertung der Testergebnisse mit den, wenn auch selten, auftretenden schlechten Ergebnissen des Punktwolkenvergleichs ein entscheidendes Problem gezeigt.
Zwar ist diese Problematik bereits in der dieser Arbeit zugrundeliegenden Untersuchung der jeweiligen Algorithmen aufgetreten, allerdings wurden die daraus resultierenden Folgen noch nicht erkannt \cite{Lankheit2020}.
Wenngleich sich die Resultate der Berechnung der Anzahl erkannter Punkte im Vergleich zur Vorarbeit verbessert haben - dort wurden im Mittel nur zwischen $50$ und $70~\%$ der Punkte korrekt erkannt \cite{Lankheit2020}, nun liegt der Wert an durchschnittlich richtig erkannten Punkten bei über $70~\%$ - bleibt das problematische Phänomen weiterhin bestehen.
Auf Grundlage der durchgeführten Tests lässt sich erkennen, dass für Produktlabel basierend auf Stäbchen vermutlich bei einem von 16 Validierungsversuchen, also bei circa $6~\%$ aller Fälle, ein fälschlicherweise negatives Ergebnis berechnet wird.
Nun könnte man zwar standardmäßig nach einem negativen Befund eine erneute Überprüfung des Produktlabels durchführen und falls das zweite Resultat positiv ist, wenigstens verspätet von einem Original ausgehen, allerdings stellt auch das keine sinnvolle Lösung des Problems dar.
Zum einen würde mit so einem Ansatz möglicherweise das Vertrauen der Anwender in die Technologie schwinden.
Zum anderen ist die Wahrscheinlichkeit, selbst nach zweifacher Messung einen fälschlicherweise negativen Befund zu bekommen, mit circa $0,4~\%$ noch immer deutlich zu hoch.
Denn für die Praxis ist erwünscht, lediglich in maximal einem von einer Milliarde Versuchen ein irrtümlich negatives Resultat zu erhalten \cite{Ruehrmair2021}.

Abschließend lässt sich also zusammenfassen: Das hauptsächliche übergeordnete Ziel dieser Arbeit, die Entwicklung einer mobilen Applikation im Kontext der Produktauthentifizierung, wurde erreicht.
Die erhofften Ergebnisse wurden erzielt und die Validierung von fälschungssicheren Produktlabeln mit einem Smartphone ist prinzipiell durchführbar.
Allerdings wurde ein essentielles Problem aufgezeigt, das weiterführende Untersuchungen nach sich ziehen dürfte.

\section{Ausblick}\label{section:Ausblick}
Nachdem mit der vorliegenden Arbeit der Nachweis erbracht wurde, dass sich die Validierung von fälschungssicheren Produktlabeln mit einer mobilen Applikation durchführen lässt, sollten sich nachfolgende Untersuchungen der Frage widmen, wie die Ergebnisse des reinen Punktwolkenvergleichs verbessert werden können.
Wobei nicht gemeint ist, die durchschnittliche Anzahl erkannter Punkte zu erhöhen, sondern das Problem zu beheben, dass zu häufig fälschlicherweise negative Resultate erzielt werden.

Ansätze, die möglicherweise zur Lösung des Problems führen, können sowohl durch Anpassungen des physischen Produktlabels und deren Vermessungen als auch durch Veränderungen der algorithmischen Verfahren gefunden werden.
Da Modifikationen des physischen Produktlabels in Bezug auf die erörterte Problemstellung allerdings nicht ohne Weiteres umsetzbar sind, sollte zuerst über eine Anpassung der Algorithmen nachgedacht werden. 

Ein Ansatz könnte darin bestehen, die Punkte einer Punktwolke nicht nur kohärent zu bewegen, sondern neben Rotationen, Skalierungen und Translationen auch weitere Veränderungen wie Verzerrungen und Stauchungen zuzulassen.
Dadurch würden die Punktwolken nicht mehr als ein starres, zusammenhängendes Gebilde angesehen werden und der Punktwolkenvergleich möglicherweise bessere Resultate liefern.
Auch zu dieser etwas toleranteren Art des Punktwolkenvergleichs mit weniger Restriktionen wurde bereits eine Vielzahl an Algorithmen vorgestellt.
Um allerdings nicht zu großzügig verschiedene Punktwolken stets als gleich zu interpretieren und insbesondere ungleiche Punktwolken auch weiterhin als solche zu erkennen, würde sich eine Komposition mehrerer Algorithmen anbieten.
Beispielsweise könnte im ersten Schritt am bisherigen restriktiven Verfahren festgehalten werden, um mit diesem eine vorläufige grobe Ausrichtung zu ermitteln.
Darauf aufbauend würden dann erst im zweiten Schritt weitere großzügigere Algorithmen das finale Ergebnis berechnen. 

Ein zweiter Ansatz könnte sein, stets mehrere voneinander unabhängige Verfahren für den Punktwolkenvergleich einzusetzen.
Zwar liefert der CPD durchschnittlich bessere Resultate als die weiteren untersuchten Algorithmen \cite{Lankheit2020}, aber womöglich würden andere Methoden, die im Gegensatz zum CPD keinen probabilistischen Ansatz verfolgen, genau in den seltenen Fällen eines fälschlicherweise negativen Ergebnisses aufgrund ihrer differenten Herangehensweise zum gewünschten Ausgang führen.
Es könnten also pro Validierung eines Produktlabels beispielsweise zwei eigenständige Algorithmen jeweils ein Resultat berechnen und das bessere wird abschließend als Gesamtergebnis ausgewählt. 

Ein dritter Ansatz, um durch reine Veränderungen des algorithmischen Verfahrens den gewünschten Effekt zu erzielen, könnte sein, die Metrik der finalen Auswertung zu überarbeiten.
Eventuell ist der Grundgedanke, dass sich jeder Punkt im Fehlerbereich seines Partners befinden soll, nicht optimal.
Vielleicht wäre es zum Beispiel erfolgreicher, für einen gewissen Anteil aller Punkte die Summe der Abstände zu ihren Partnern mit einem festgelegten Schwellenwert zu vergleichen.
Auf Grundlage des Vergleichs könnten dann Aussagen über die Gleichheit der Punktwolken getroffen werden.

Abschließende Möglichkeiten zur Verbesserung der Ergebnisse stellen Modifikationen des physischen Produktlabels dar.
Beispielsweise könnte die Einführung eines oder mehrerer Leuchtturmelemente, das heißt Elemente, die bei jeder Messung eindeutig identifizierbar sind und somit stets als Orientierung dienen, zu einer Besserung der Resultate führen. 

Werden die fälschlicherweise negativen Resultate durch eine Modifikation der bisherigen Algorithmen, ein komplett neues algorithmisches Verfahren oder Anpassungen der physischen Produktlabel minimiert, kann die finale praktische Umsetzung des Schutzes vor Produktfälschungen realisiert werden.
Mit der Einführung des geplanten Mechanismus könnten zukünftige Imitationsversuche erkannt und der weltweit steigende Trend an Produktpiraterie umgekehrt werden.
Insbesondere unmittelbare Gefahren für die Gesundheit durch gefälschte Medikamente und andere medizinische Artikel könnten damit abgewendet werden.
%
%
    \appendix
    \chapter{Appendix}

\begin{algorithm}
  \DontPrintSemicolon
  \Daten{\textit{suchraeume\_rotation, maximale\_groessenabweichung, prozent\_mindestens\_erkannt}}
  \usedAlgorithm{\textit{CPD}}
  \Ein{\textit{referenz $= (X, \sigma_X)$, messung $= (X, \sigma_X)$}}
  \Aus{Die Punktwolken sind: \textit{Gleich/Ungleich}}
  \Beginn{
    \tcc{Ein bedeutender Größenunterschied der zwei Punktwolken resultiert in Ungleichheit.}
    \uWenn{
    \textit{ $\frac{\abs{\abs{referenz}-\abs{messung}}}{\abs{referenz}}> $ maximale\_groessenabweichung}}
    {\Zurueck{Ungleich}}
    \BlankLine
    \BlankLine
    \tcc{Das Ergebnis prozentual erkannter Punkte jedes einzelnen CPDs ist zu Beginn $0$.}
    \textit{einzelergebnisse} $\leftarrow$ \textit{initialisiere alle $ \abs{suchraeume\_rotation}$ mit $0$}\;
    \BlankLine
    \FuerAlle{\textit{ $s \in$ suchraeume\_rotation}}{
        initialisiere \textit{CPD} mit Suchraum \textit{s}\;
        \BlankLine
        \BlankLine
        \tcc{Ermittle die Parameter zur Angleichung der Punktwolken.}
        \textit{skalierung, rotation, translation, korrespondenzmenge $\leftarrow$ CPD $(X, Y)$}\;
        \BlankLine
        \BlankLine
        \tcc{Berechne für die Messung die angeglichene Punktwolke.}
        \textit{angeglichene\_messung $\leftarrow$ skalierung $*$ rotation$(Y)$ $+$ translation}\;
        \BlankLine
        \textit{anzahl\_erkannter\_punkte $\leftarrow 0$}\;
        \BlankLine
        \FuerJedes{\textit{$(x,y_{angeglichen}) \in korrespondenzmenge$}}{
        \BlankLine
            \uWenn{\textit{$x$ im Fehlerbereich($y_{angeglichen}$) und $y_{angeglichen}$ im Fehlerbereich($x$)}}
            {\textit{anzahl\_erkannter\_punkte++}}
        \BlankLine
        }
        \BlankLine
        \BlankLine
        \tcc{Die prozentual erkannten Punkte werden pro CPD als Einzelergebnis gespeichert.}
        \textit{einzelergebnisse}[s] $\leftarrow \frac{anzahl\_erkannter\_punkte}{\abs{X}} $\;
        \BlankLine
    }
        \BlankLine
        \tcc{Als Gesamtergebnis wird das beste aller Einzelergebnisse ausgewählt.}
    \textit{gesamtergebnis} $\leftarrow$ maximum(\textit{einzelergebnisse}) \;
     \BlankLine
        \BlankLine
        \tcc{Ist der Prozentsatz erkannter Punkte ausreichend sind die Punktwolken Gleich, andernfalls Ungleich.}
    \uWenn{\textit{gesamtergebnis $>$ prozent\_mindestens\_erkannt}}
    {\BlankLine
    \Zurueck{\textit{Gleich}}}
    \BlankLine
    \Sonst{\BlankLine
    \Zurueck{\textit{Ungleich}}
    \BlankLine
    }
  }
 \caption{Ermittlung der Gleichheit zweier Punktwolken.}
 \label{algorithm:Ermittlung-Gleichheit}
\end{algorithm}

\begin{figure}
  \centering
  \includegraphics[width=0.8\textwidth]{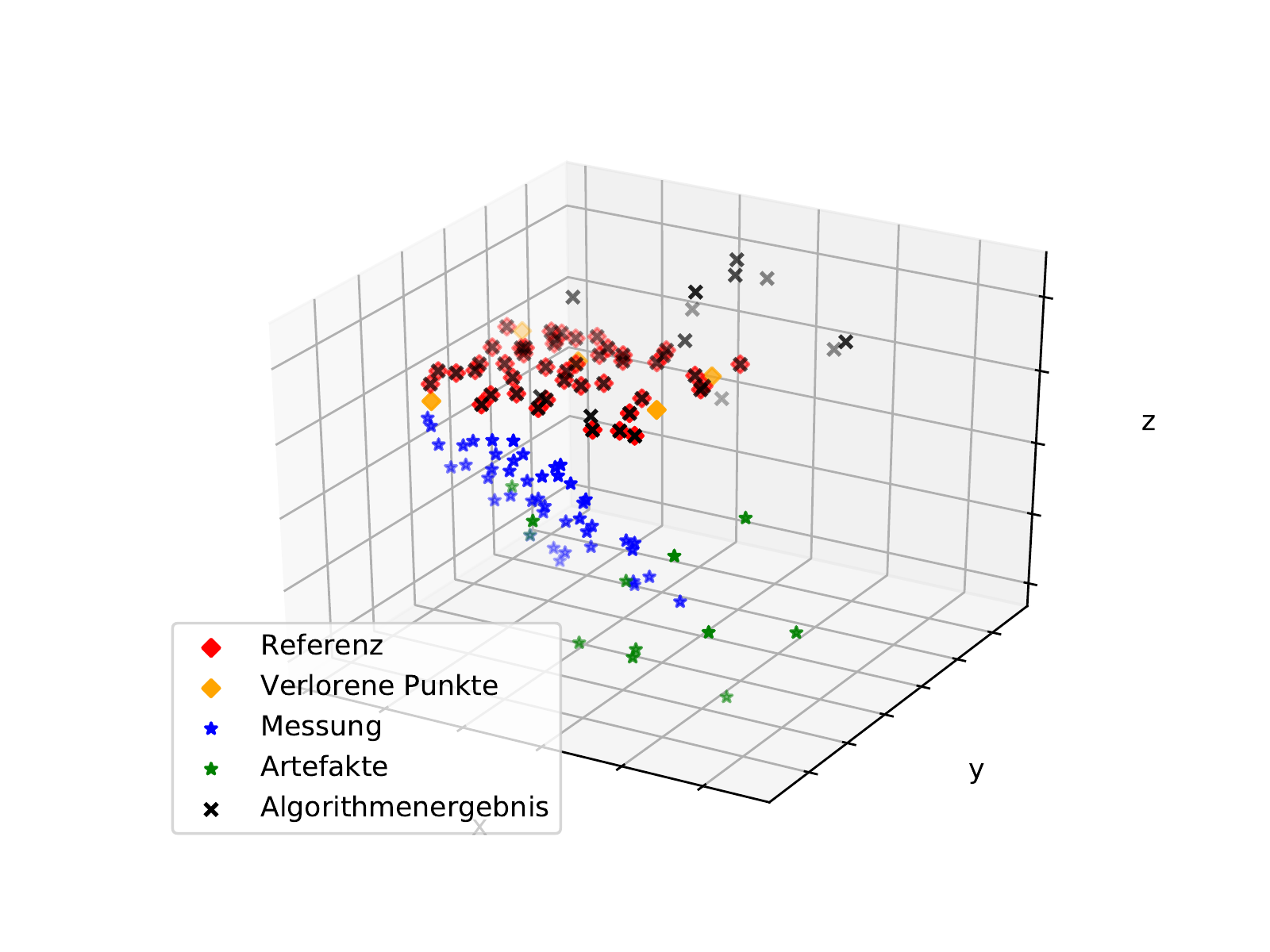}
  \caption{ Beispielhafte 3D-Punktwolken mit Artefakten, verlorenen Punkten und verrauschten Positionierungen sowie den Ergebnissen der Algorithmen. Hierbei ist ein positives Resultat zu sehen. \cite[S. 41]{Lankheit2020}
  }
  \label{figure:3D-Gut}
\end{figure}

\begin{figure}
  \centering
  \includegraphics[width=0.8\textwidth]{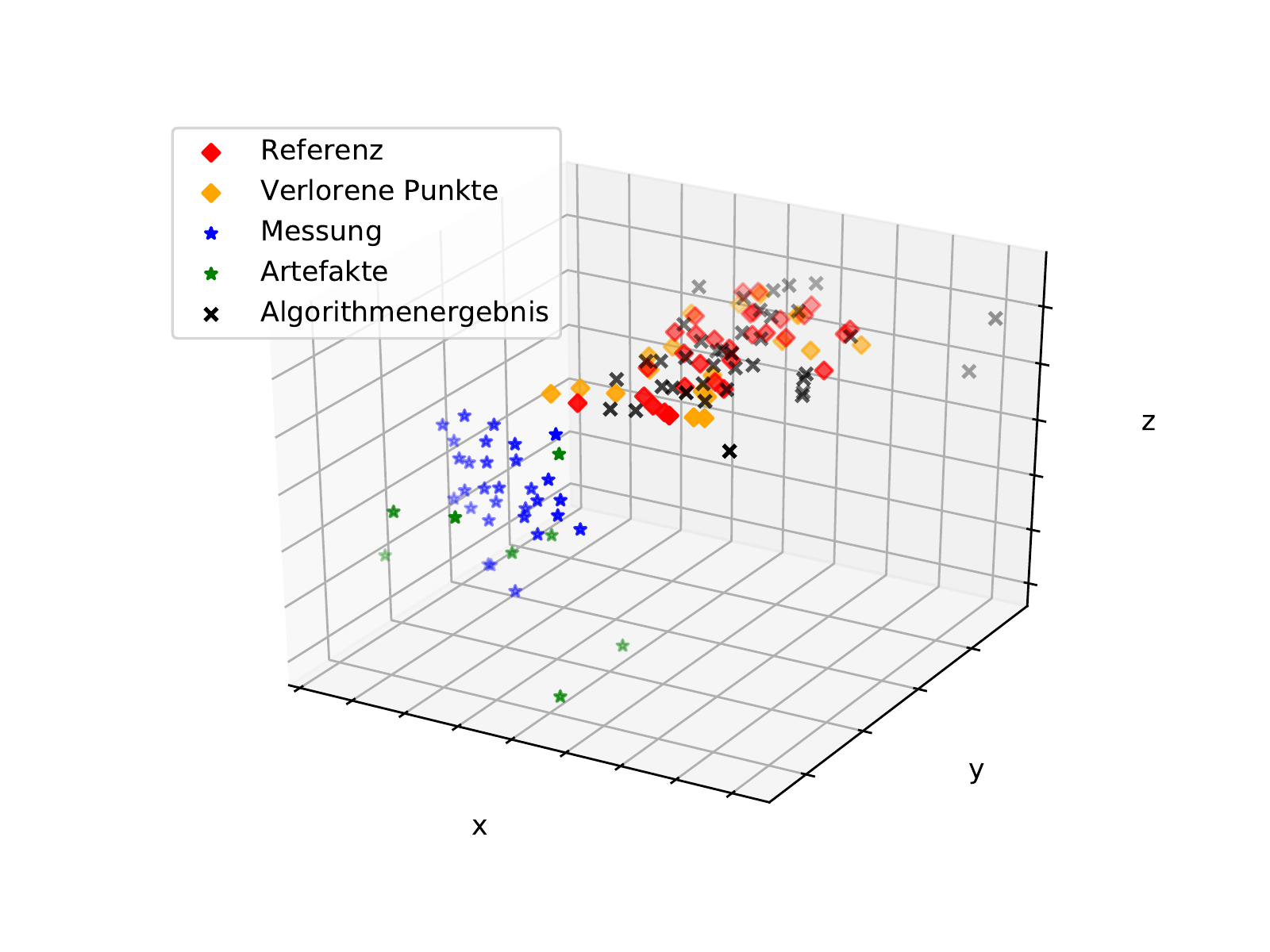}
  \caption{
    Beispielhafte 3D-Punktwolken mit Artefakten, verlorenen Punkten und verrauschten Positionierungen sowie den Ergebnissen der Algorithmen. Hierbei ist ein negatives Resultat zu sehen. \cite[S. 42]{Lankheit2020}
  }
  \label{figure:3D-Schlecht}
\end{figure}
%
%
    \backmatter
    \listoffigures                                
%
%
    \begin{spacing}{0.9}                          
       \bibliographystyle{geralpha}               
       \bibliography{bibliography}                
    \end{spacing}
\end{document}